\documentclass{article}
\usepackage[twocolumn]{emulateapj}
\submitted{To appear in the Astrophysical Journal 2005, January 10}

\def\lsim{\mathrel{\rlap{\lower 4pt \hbox{\hskip 1pt $\sim$}}\raise 1pt \hbox
        {$<$}}}
\def\gsim{\mathrel{\rlap{\lower 4pt \hbox{\hskip 1pt $\sim$}}\raise 1pt \hbox
        {$>$}}}

\righthead{Variations in the Abundance Pattern of Extremely Metal-poor Stars}

\lefthead{Umeda \& Nomoto}

\begin{document}
\title{Variations in the Abundance Pattern of  Extremely Metal-poor Stars
and Nucleosynthesis in Population III Supernovae}

\author{Hideyuki Umeda and Ken'ichi Nomoto}


\affil{
Department of Astronomy, University of Tokyo, Hongo, Bunkyo-ku,
113-0033, Japan \\ 
umeda@astron.s.u-tokyo.ac.jp; nomoto@astron.s.u-tokyo.ac.jp}

\begin{abstract}

 We calculate nucleosynthesis in Population (Pop) III supernovae (SNe)
and compare the yields with various abundance patterns of extremely
metal-poor (EMP) stars.  We assume that the observed EMP stars are the
second generation stars, which have the metal-abundance patterns of
Pop III SNe.  Previous theoretical yields of Pop III SNe cannot
explain the trends in the abundance ratios among iron-peak elements
(Mn, Co, Ni, Zn)/Fe as well as the large C/Fe ratio observed in
certain EMP stars with [Fe/H] $\lsim -2.5$.
In the present paper, we show that if we introduce higher explosion
energies and mixing-fallback in the core-collapse SN models of $M \sim
20 - 130 M_\odot$, the above abundance features of both typical and
C-rich EMP stars can be much better explained.
We suggest that the abundance patterns of the [Fe/H] $\sim -2.5$ stars
correspond to supernova yields with normal explosion energies, while
those of the carbon un-enhanced ([C/Fe] $< 1$) stars with [Fe/H]
$\simeq -4 \sim - 3$ correspond to high-energy supernova yields. The
abundance patterns of the C-rich ([C/Fe]$\gsim 2$) and low [Fe/H]
($\simeq -5 \sim -3.5$) stars can be explained with the yields of
faint SNe that eject little $^{56}$Ni as observed in SN1997D.  In the
supernova-induced star formation model, we can qualitatively explain
why the EMP stars formed by the faint or energetic supernovae have
lower [Fe/H] than the EMP stars formed by normal supernovae.  We also
examine how the abundance ratios among iron-peak elements depend on
the electron mole fraction $Y_e$, and conclude that a large explosion
energy is still needed to realize the large Co/Fe and Zn/Fe ratios
observed in typical EMP stars with [Fe/H] $\lsim -3.5$.

\end{abstract}


\section{Introduction}

In the early universe, where the metal content of gas is very low, the
enrichment by a single supernova can dominate the pre-existed metal
contents (e.g., Audouze \& Silk 1995).  Low mass stars formed in the
gas survives until today, and observed as extremely metal-poor (EMP)
stars.  Since EMP stars may preserve abundance patterns synthesized by
a single or few supernovae (SNe), the abundance patterns of those
stars may be used to test supernova explosion and nucleosynthesis
theories, and to infer the nature of the first generation stars and
supernovae.

 The abundance patterns of EMP stars show interesting trends below
[Fe/H] $\sim -2.5$; with increasing [Fe/H], [Mn/Fe] and [Cr/Fe]
increase while [Co/Fe] and [Zn/Fe] decrease, where [X/Y]$\equiv
$log(X/Y)$-$log$(X/Y)_\odot$ (McWilliam et al. 1995; Primas et
al. 2000; Spite et al. 2003).  These trends can be explained if the
EMP stars with lower [Fe/H] were enriched by a supernova ejecting
relatively more complete Si-burning matter (e.g., Co, Zn) than
incomplete Si-burning matter (e.g., Mn).  This is realized if the
``mass-cut'', that divides the supernova ejecta and the central
remnant, is relatively deeper (Nakamura et al. 1999).  A question is
why lower metallicity EMP stars are enriched by SNe with relatively
deeper mass-cuts. In this paper (Section 2), we show that the
variation of the explosion energy can nicely explain this relation
(see also Umeda \& Nomoto 2002a,b, 2003 and Nomoto et al. 2003, for
brief explanations).

 The large Zn/Fe ratios typically observed in EMP stars (Primas et
al. 2000; Depagne 2003; Cayrel et al. 2003) have not been explained
with conventional supernova yields, except for the suggestion that Zn
could be produced in the neutrino-powered wind (Hoffman et al. 1996).
Umeda \& Nomoto (2002a; UN02 hereafter) have provided first yields of
the core collapse SN models that have large enough Zn/Fe ratio to be
consistent with the EMP stars if the explosion energies are
sufficiently large (i.e., ``Hypernova'' models; Nomoto et al. 2004).

 One important implication of the hypernova model for the abundances
of EMP stars is that mixing and fall-back are required to take place
in the inner part of the ejecta. The large Zn/Fe is realized only if
the mass-cut is sufficiently deep. If the mass-cut is deep enough to
eject Zn, however, too much Fe is ejected and the ratios between
lighter elements and Fe, such as Mg/Fe, become too small. 

 To solve this problem, UN02 have proposed that mixing-out of Zn and
the subsequent fall-back of sufficient amount of the mixed material
take place in the ejecta.  Mixing due to Rayleigh-Taylor instabilities
is initiated at a steep density gradient when the expanding core is
largely decelerated by the reverse shock generated at the
core-envelope interface (e.g., Ebisuzaki et al. 1989; Arnett et
al. 1989; Hachisu et al. 1990; Kifonidis et al. 2000).  Simultaneously
such a deceleration causes fallback of the mixed material (e.g.,
Chevalier 1989; Herant \& Woosley 1994; see Colgate 1971 for
pioneering work).  Therefore, the extent of mixing and the amount of
fallback both depend on the stellar mass, presupernova density
structure, explosion energy, asphericity, etc., which must be examined
systematically with the multi-dimensional simulations.  Although we
treat the mixing and fallback as free-parameters in the present study,
the resultant abundance pattern in the ejecta in comparison with the
observations may provide interesting constraints on these parameters.
We also note that a similar effect to mixing-fallback also occurs in
the jet-like explosion. In such a model, energetic explosion occurs
only along the jet directions and thus the total Fe mass can be
smaller with enhancement of the complete Si-burning products
(e.g. Maeda \& Nomoto 2003a,b).

 Although Nakamura et al. (1999) have successfully explained the trend
in Fe-peak elements, one may wonder whether the absolute values of the
abundances fit to the observations as well as elements other than
Fe-peak.  Chieffi \& Limongi (2002, CL02 hereafter) compares UN02 and
Woosley \& Weaver 1995 (WW95 hereafter) with the typical EMP
abundances and concluded that none including theirs fit to the
observations well. They proposed as a possible solution that a
progenitor model with a large C/O ratio may solve the discrepancy. In
this paper, instead, we show that the absolute abundance of Co and Mn
are quite sensitive to $Y_e$, and the fit to the observations becomes
significantly improved with a certain choice of reasonable value of
$Y_e$.

 Are the abundances of all EMP stars consistent with hypernova
nucleosynthesis?  There is a sub-class of EMP stars, C-rich EMP stars,
including the most Fe deficient star HE0107-5240 with [Fe/H]$\sim
-5.3$ (Christlieb et al. 2002).  These stars are quite rich in C and
N, typically [C/Fe] $\gsim$ 2, and in some stars Mg as well.  Recently
we have shown that the abundances of these stars are well reproduced
with the yields of core-collapse SNe which undergo small Fe ejection
and mixing-fallback (Umeda \& Nomoto 2003, UN03 hereafter).  We showed
that not all C-rich EMP stars favors high-energy models. In fact the
abundance pattern of HE0107-5240 is well reproduced by a low-energy
($E_{51}\equiv E/10^{51}$ ergs =0.3, where $E$ is the explosion of the
SN) model.  In this paper we compare our models with other C-rich EMP
star abundances and constrains the models from fitting to the data.
We demonstrate that the difference in the degree of the mixing,
fallback and explosion energies may explain both the C-rich and usual
EMP stars.

\section{Nucleosynthesis calculations}

 The calculation method and other assumptions are the same as
described in Umeda et al. (2000), UN02 and UN03.  
The isotopes included in the network for explosive burning
are shown in Table 1. After the
post-process nucleosynthesis calculations, we calculate the final
yields by setting the final mass-cut and the mixing-fallback
parameters. More detailed description and the definition of the
mixing-fallback parameters are given in Section 3.2 (also in UN02). We
treat the mixing and fallback as free-parameters in the present study,
by the reason described in Introduction.

 We take the progenitor mass and the explosion energy as independent
parameters for the following reasons.  For a given progenitor model,
if the explosion mechanism (or the procedure for the artificial
explosion) is specified the remnant mass is uniquely determined as a
function of the explosion energy (e.g., WW95; Limongi \& Chieffi
2003).  However, we do not specify the explosion mechanism, especially
because precise explosion mechanism is unknown for hypernovae.  Also,
the density structure of the progenitors depends complicated on the
initial mass, treatment of convections and uncertainties in the
nuclear reaction rate. Different density structure leads the different
remnant mass with the same explosion energy.  Furthermore, if the
mixing-fallback effect is considered to be a 1D (approximate)
representation of a jet-like explosion, then the energy of the 1D
model no longer matches with the total explosion energy and
independent of the remnant mass.  Note the amount of fallback is not
necessarily smaller for larger $E$ (Herant \& Woosley 1994).

  Since we explode the progenitor model when the central density of 3
$\times 10^{10}$ g cm$^{-3}$ is reached without calculating further
collapse and bounce, our approach may be regarded as simulating the
prompt explosion.  However, the results will not be much different
from the delayed explosion. In the delayed explosion, more materials
fall onto the central remnant, which makes the mass-cut larger and
changes $Y_e$ in the complete Si-burning region. (See Thielemann,
Nomoto, \& Hashimoto 1996 for the study of this effect). This effect
may be important for relatively less massive ($M = 13-15M_\odot$)
progenitors with low explosion energy. On the other hand, in massive
and energetic explosions (i.e., in hypernovae), explosive
nucleosynthesis takes place in outer regions where $Y_e$ is not
sensitive to the location of the mass-cut.  We study the dependence of
nucleosynthesis on $Y_e$ in section 3.1, since $Y_e$ may be modified by
the neutrino process.
 
\section{EMP Stars with Typical Abundance Pattern}

\subsection{Trends in the iron peak elements and hypernovae}

 Umeda \& Nomoto (2002b) has shown that the trends in the abundance
ratios of Fe-peak elements, [(Zn, Co, Mn, Cr)/Fe] vs [Fe/H], can be
understood by the variations of deepness of mass-cut in the explosive
nucleosynthesis of SNe II. We also have suggested that the large Zn/Fe
and Co/Fe ratios in typical EMP stars are well-reproduced by hypernova
nucleosynthesis.  In this section, we describe how closely these facts
are related and how observed trends, and not just variations, can be
explained with the energy of supernovae.

For a larger explosion energy, the supernova shock is stronger and the
temperature after the shock passage is higher. The post-shock region
is radiation dominant, so that the peak temperature is approximately
related to the stellar radius $r$ and the deposited energy $E^*$ as
\begin{equation}
T_9 = (E^*_{51})^{1/4} (r/ 3.16\times 10^4 {\rm km})^{-3/4},
\end{equation}
where $T_9$ is the peak temperature in 10$^9$ K and $E^*_{51}$ is the
deposited energy. Complete Si-burning, which burns Si completely,
occurs for $T_9 > 5$. In this region, elements such as $^{56}$Ni,
$^{64}$Ge (decaying into $^{64}$Zn) and $^{59}$Cu (decaying into
$^{59}$Co) are produced. Incomplete Si-burning occurs for $ 4 < T_9 <
5$. In this region elements such as $^{56}$Ni, $^{52}$Fe (decaying
into $^{52}$Cr) and $^{55}$Co (decaying into $^{55}$Mn) are produced.

\begin{figure}

\vskip -5cm
\vspace {-5cm}

\vspace{-6cm}
\epsscale{1.}
\plottwo{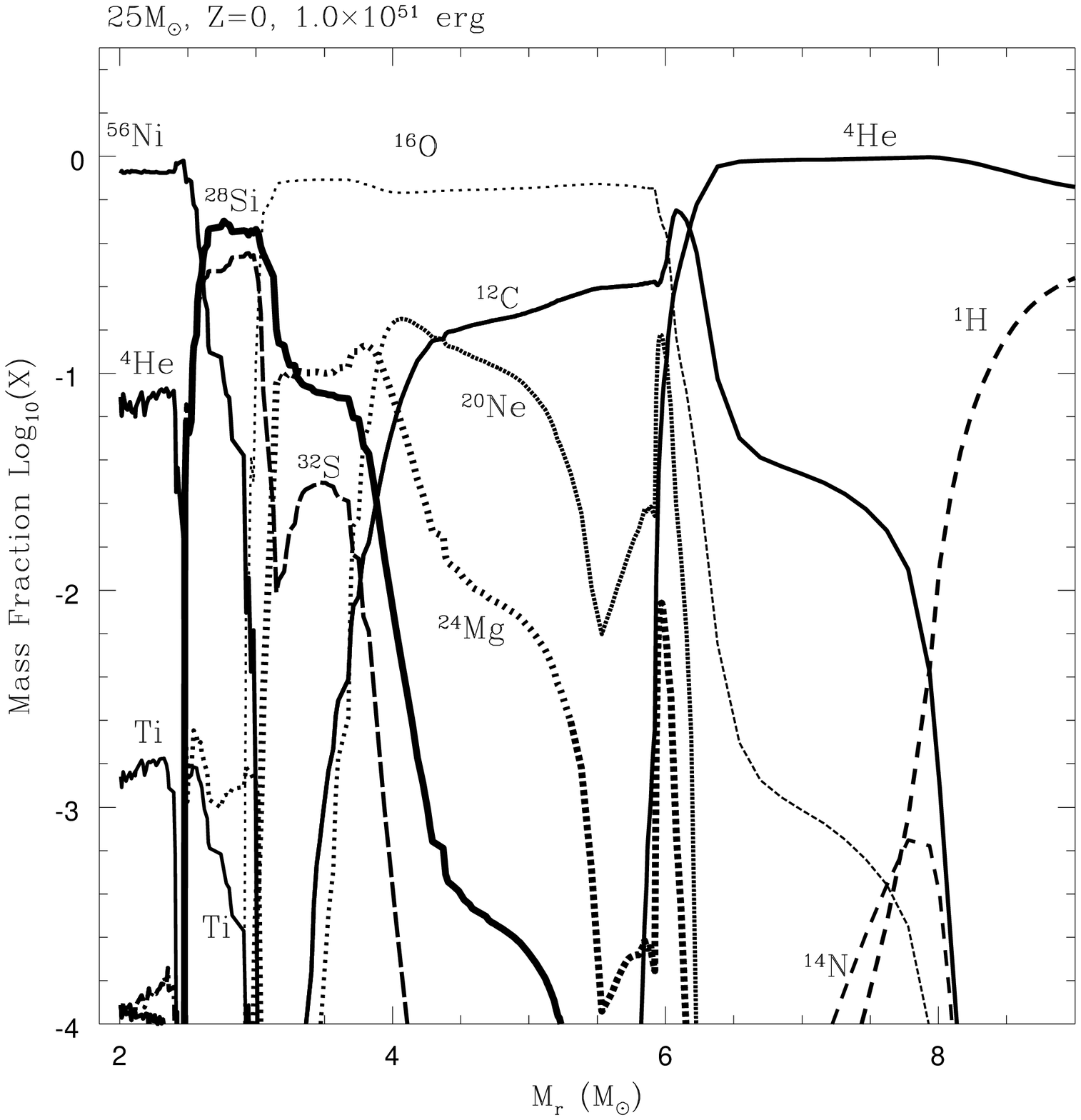}{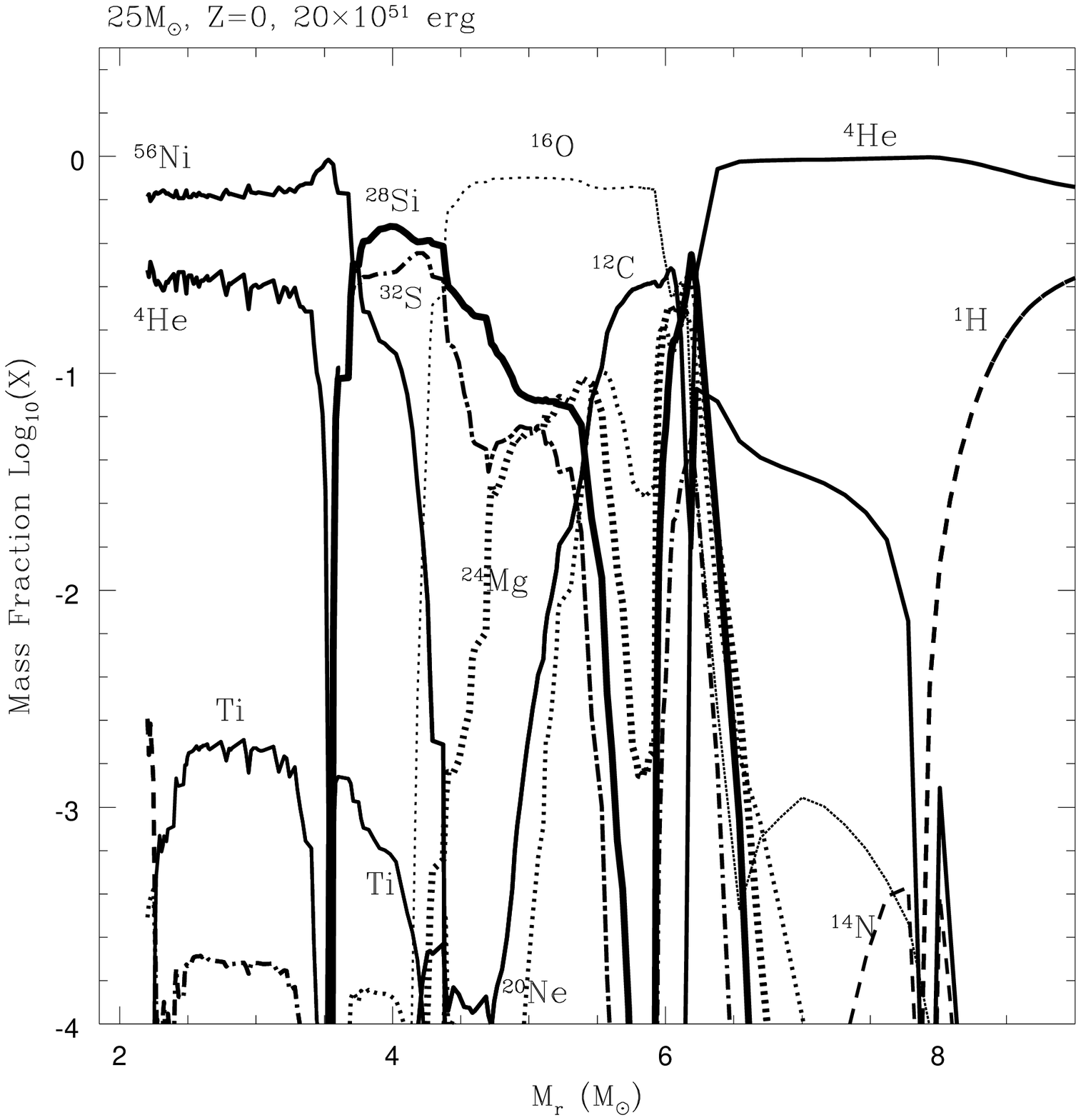}

\caption{Abundance distribution after SN explosion of a 25 $M_\odot$
star with $E_{51}=1$ (left panel) and $E_{51}=20$ (right panel).
Complete Si-burning regions, here it is estimated by X($^{28}$Si) 
$< 10^{-3}$, are $M_r < 2.5M_\odot$ for $E_{51}=1$ and 
$M_r < 3.5M_\odot$ for $E_{51}=20$. Incomplete Si-burning regions,
here their upper edges are estimated by X($^{56}$Ni) 
$< 10^{-3}$, are $ 2.5M_\odot < M_r < 3.0M_\odot$ for $E_{51}=1$ and 
$3.5 M_\odot < M_r < 4.3M_\odot$ for $E_{51}=20$.
For a larger explosion energy, complete Si-burning region is extended
outside. Incomplete Si-burning region is also enlarged, however,
the mass ratio between complete and incomplete Si-burning regions
becomes larger for a larger explosion energy with a fixed mass-cut.
\label{25z0A}}
\end{figure}

For a larger explosion energy, the complete Si-burning region is
enlarged in mass more than the incomplete Si-burning region (see
Figure \ref{25z0A} and its caption).  As a result, the mass ratio
between the complete and incomplete Si-burning regions is larger in a
more energetic explosion if the mass-coordinate of the mass-cut does
not change significantly.  In this sense, increasing the energy
produces similar effects to making the mass-cut deeper without
actually changing the mass coordinate of mass-cut.  In addition,
complete Si burning in the higher energy model has better features
than in the deeper mass cut model in explaining the observed features:
(1) the overproduction of $^{58}$Ni in the latter model (Nakamura et
al. 1999) can be avoided because of larger $Y_e$, and (2) a larger
amount of $^{64}$Zn can be synthesized because of higher entropy, and
thus stronger $\alpha$-rich freezeout in the higher energy model
(Nakamura et al. 2001b; UN03).

 Regarding [Zn/Fe], Zn is quite abundant in a typical EMP star, i.e.,
[Zn/Fe] $\sim$ 0.3 - 0.8 (Primas et al. 2000; Depagne 2003; Cayrel et
al. 2003).  We have shown that such large Zn/Fe ratio is difficult to
produce by SNe with normal explosion energy ($\sim 10^{51}$ erg), but
possible by energetic core collapse SNe with $10^{52}$ erg or more
(UN02).

 We assume that EMP stars are formed in the supernova ejecta mixed
with interstellar matter.  In this ``supernova-induced star formation
model'', the [Fe/H] (or [Mg/H]) of an EMP star is determined by the Fe
(or Mg) mass ejected from a SN, divided by the hydrogen mass in circum
stellar matter swept by the SN shock. It is estimated that the swept
hydrogen mass is roughly proportional to the explosion energy (Ryan,
Norris, \& Beers 1996; Shigeyama \& Tsujimoto 1998).
Thus we may write that
\begin{equation}
 \rm {[Fe/H] \simeq log_{10} (Fe/E_{51})} + C, \quad
 \rm {[Mg/H] \simeq log_{10} (Mg/E_{51})} + C',
\end{equation}
where Fe, Mg and H represent mass fraction of Fe, Mg and H,
respectively; $E_{51}$ is the explosion energy in 10$^{51}$ erg, and
C($'$) are ``constants''.  Here C($'$) may not be exactly constants.
These values depend on how much metal is actually mixed with the
star-forming gas, thus depending on detailed hydrodynamical mixing,
turbulent motion, local inhomogeneities and some other factors.  These
values, therefore, possibly distribute around certain mean values.  In
this case, the EMP stars produced by more energetic SNe are expected
to have smaller [Fe/H] and [Mg/H] in a statistical sense.

 Nakamura et al. (2001b) also systematically studied nucleosynthesis
in hypernovae and showed that Fe/$E_{51}$ and Mg/$E_{51}$ decrease
significantly with increasing $E$; this tendency is consistent with
the observations and models of SN 1998bw ($E_{51}$=50) and SN 1987A
($E_{51}$=1), where Mg/$E_{51}$ and Fe/$E_{51}$ of SN 1998bw are
smaller than SN 1987A by a factor of 20 and 8, respectively (Nakamura
et al. 2001a).

\begin{figure}

\epsscale{1.}
\plotone{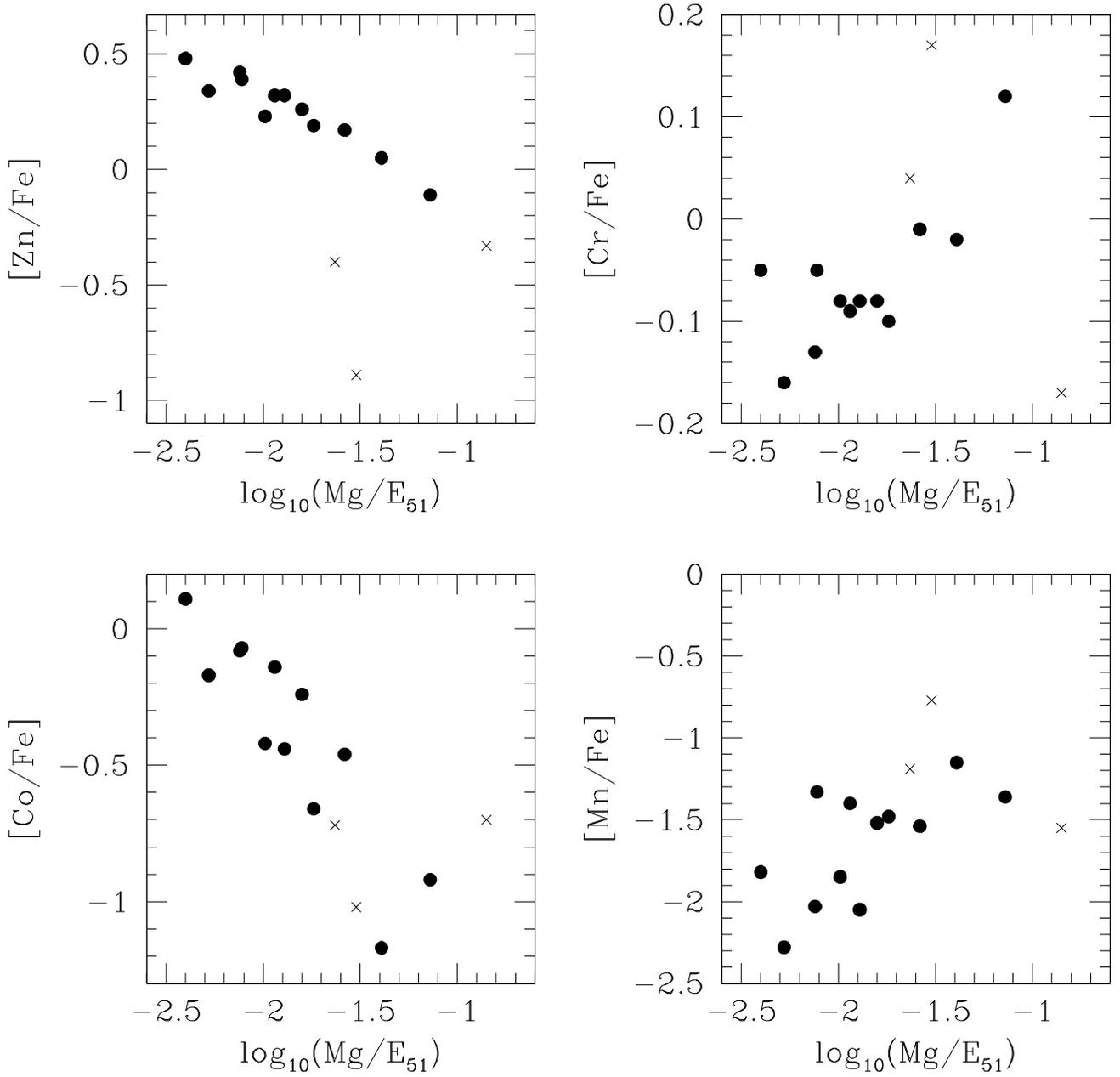}

\caption{Yield ratios 
[(Zn, Co, Cr, Mn)/Fe] of the models in Table \ref{models1}
plotted against log$_{10}$ (Mg/$E_{51}$) $\simeq [Mg/H]$ +constant.  
Here, the crosses are $E_{51}=1$, $M= 13, 15,$ and 25$M_\odot$
models, and solid circles are high energy explosion $(E_{51} \geq 10$)
models.
\label{xmg1}}
\end{figure}

In this explanation, [Fe/H] of EMP stars is almost independent of the
initial metallicity and the age of the SNe progenitor.
We calculate nucleosynthesis for several $Z=0$ models with different
masses and energies as shown in Table \ref{models1},  
and plot the yield
ratios [(Zn, Co, Cr, Mn)/Fe] vs log$_{10}$ (Mg/$E_{51}$) in Figure
\ref{xmg1}. Several quantities related to these models are also
shown in Table 3 and 4.
Here, Mg is adopted for abscissas because the ejected
mass of Mg is less sensitive to the mass-cut than that of Fe.
For the initial He abundance, we adopt $Y=0.247$.  

 Throughout this paper the mass-cut (or the ``initial'' mass-cut in
the mixing-fallback model, see below) is chosen to maximize the Zn/Fe
ratio in the models with the original $Y_e$ distribution, unless
otherwise stated.\footnote{
Zn (the dominant isotope is $^{64}$Zn for Pop III SNe, which is the
decay product of $^{64}$Ge) is mostly produced in the complete
Si-burning region where $Y_e \simeq 0.5$ and the Zn/Fe ratio decreases
for lower $Y_e$ (see UN02 and also Figure 4 of this paper).
Therefore, as the mass-cut decreases, [Zn/Fe] in the ejecta first
increases and then decreases. The ``maximum'' [Zn/Fe] we mention is
the first maximum obtained in this region.  Depending on the
distribution of $Y_e$ in progenitor models, the [Zn/Fe] may show the
second peak for much smaller mass-cut where $Y_e$ is very low (Hoffman
et al. 1996).  However, in this paper, we are not considering such low
$Y_e$ regions because these regions often over-produce too
neutron-rich isotopes together.}
This is because the observed large value of [Zn/Fe] $(= 0.3 - 0.8)$
for typical EMP stars ([Fe/H] $< -3.6$) are almost always
under-produced in our low energy models and also in previously
published other groups' yields.  As mentioned above, [Zn/Fe] is
enhanced in the high-energy models but it is rarely over-produced for
any parameter choices.

As shown below, we often modify $Y_e$ of the progenitors, but then the
location at the Zn/Fe maximum changes. In this paper, we always adopt
the same (initial) mass-cut as the model with the original $Y_e$
distribution.  The observed values of [Co/Fe] in EMP stars also tend
to be underproduced.  Maximizing [Zn/Fe] is also good for making
[Co/Fe] large, because Zn and Co are produced roughly in the same
region. We determine the mass-cut in this way to reduce the parameters
and explore the results in a systematic way.  

\begin{figure}

\vspace{-6cm}
\epsscale{1.}
\plottwo{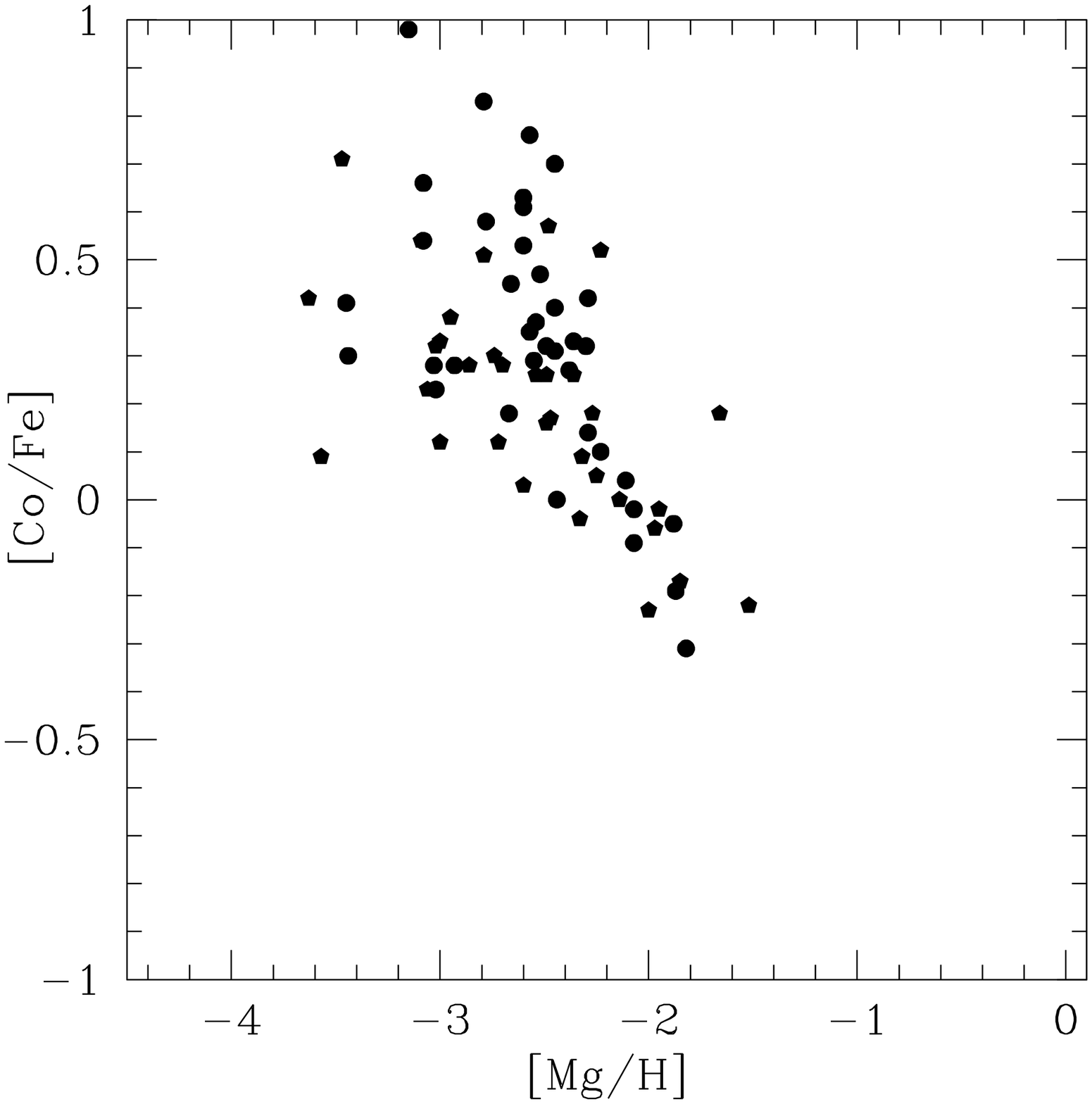}{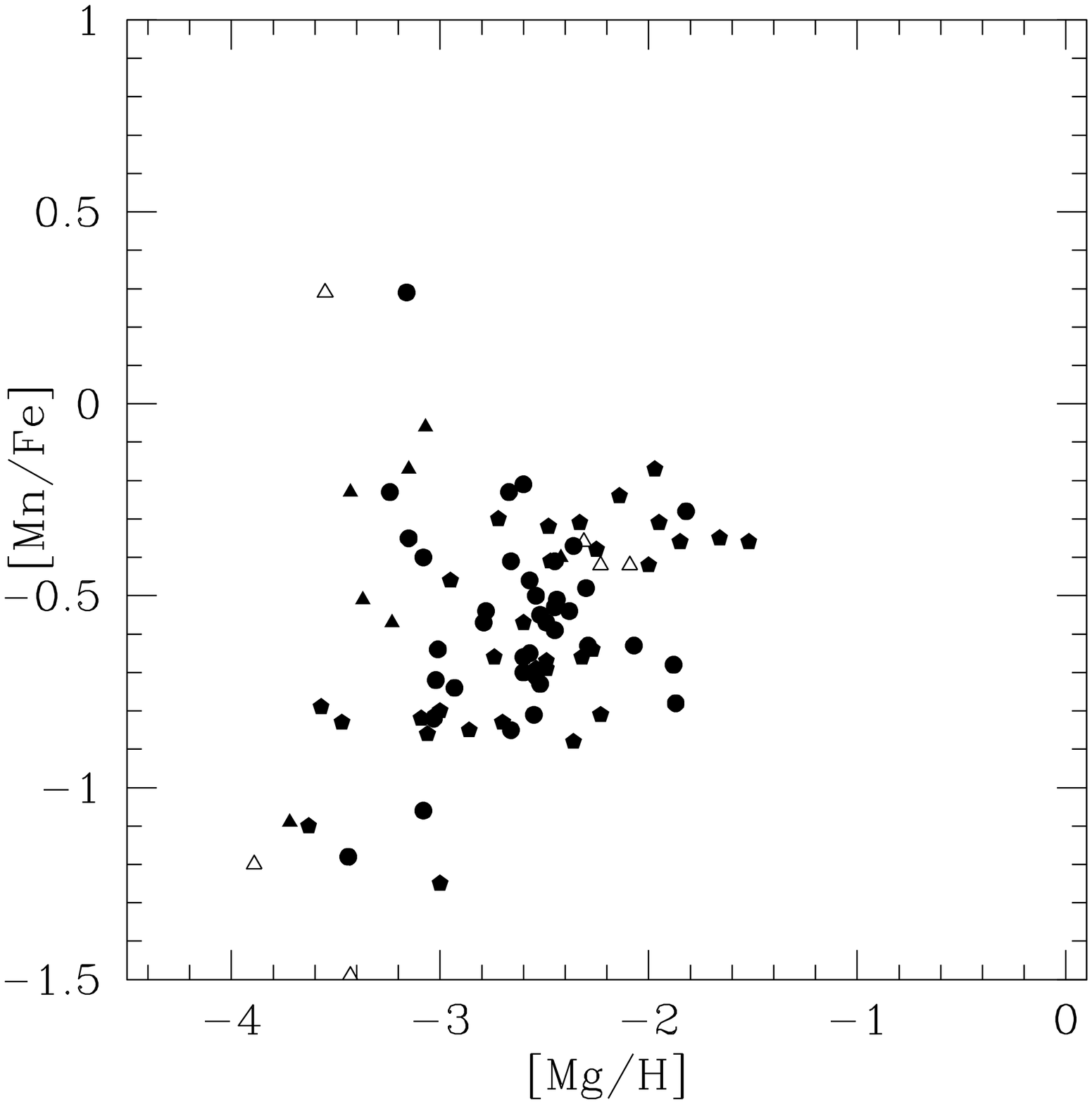}

\caption{Observed abundance ratios of [Co/Fe] and 
[Mn/Fe] vs [Mg/H] in the metal poor stars. The same observed
trends seen in [X/Fe] vs [Fe/H] (see Figure 5) remains 
even if the abscissa
is replaced by [Mg/H]. The reference of the observed points are 
given in Nakamura et al. (1999).
\label{xmgdat}}
\end{figure}

 Figure 2 exhibits that the high-energy models tend to be located at
lower [Mg/H] = log$_{10}$ (Mg/$E_{51})-C'$ (if $C'$ distribute around
a certain peak value), and thus can explain the observed trend. Note
that [Mg/Fe] $\sim 0.3-0.5$ for typical EMP stars and thus the
observed general trends preserve even though the abscissa is changed
from log$_{10}$ (Mg/$E_{51})$ to [Mg/H] as shown in Figure
\ref{xmgdat}.  The observed data vs [Fe/H] are shown later in Figure
5.

 Note that the trends in the Figure 2 are preserved in the
mixing-fallback model described in the subsection 2.2, because all the
quantities in Table 2 are independent of the ejected Fe mass, or the
ejection factor, $f$, as far as the outer boundary of the mixing
region, $M_{\rm mix}$(out), is fixed to be at the outer-boundary of
the incomplete Si-burning region as assumed in UN02.

 Although we could explain the trends, Co/Fe and Mn/Fe ratios in Table
2 are too small to be consistent with observations in the absolute
values.  Here we note that the yields of Co and Mn are sensitive to
the details of the explosion, nuclear reaction rates and $Y_e$.  Among
them, the effect of $Y_e$ change due to the neutrino process during
explosion may be important. In our previous works, we have assumed
that the pre-supernova value of $Y_e$ is preserved during the
explosive burning. In the $Z=0$ models, $Y_e \simeq 0.5000$ above the
pre-supernova oxygen layer and decreases gradually toward the Fe core
(UN02).  However, recent detailed simulations of neutrino transport in
core-collapse SNe show that $Y_e$ may be significantly affected by the
neutrino process during explosion (Liebend\"orfer et al. 2003; Janka,
Buras, \& Rampp 2003). It is interesting that in the deep core $Y_e >
0.5$ may be realized, for which nucleosynthesis has not been
systematically studied before.

The region where large $Y_e$ enhancement occurs due to neutrino
absorption is Rayleigh-Taylor unstable because of associated large
enhancement of entropy due also to neutrino absorption.  The resultant
development of Rayleigh-Taylor instabilities should largely change the
$Y_e$-profile.  Since it is not known how large is the $Y_e$ change
and to which region the $Y_e$ change propagates, we treat the
$Y_e$-profile as a free-parameter and discussed what $Y_e$-profile
would produce a reasonable results.

\begin{figure}

\epsscale{1.}
\plotone{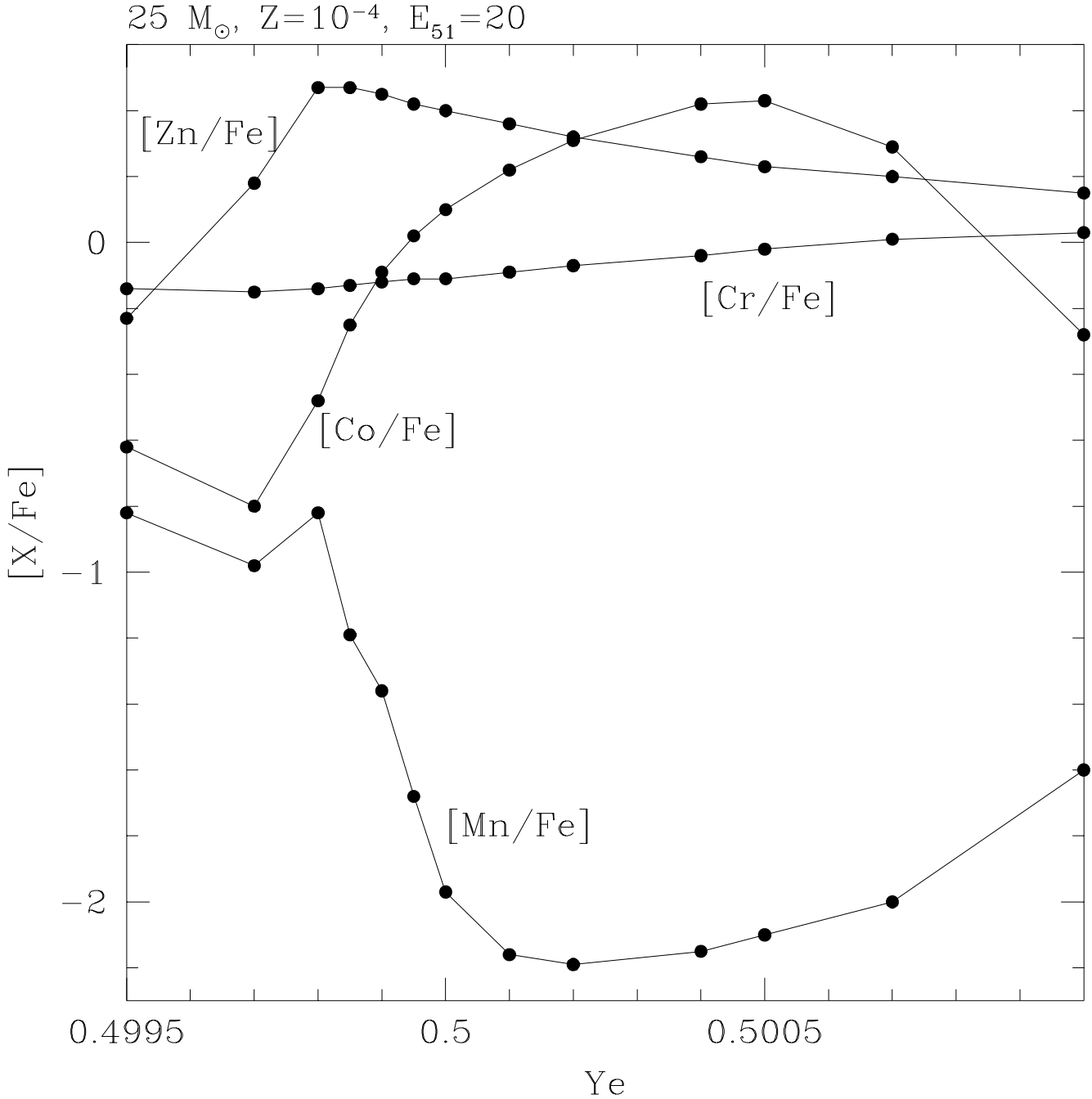}

\caption{The abundance of Fe-peak elements as a function 
of $Y_e$ in the Si-burning region. 
Here, we change the $Y_e$ inside
the incomplete Si-burning region to the value shown in the figure, 
and the mass-cut is chosen to maximize
the Zn/Fe ratio. The supernova model is a $Z=10^{-4}$,
25 $M_\odot$ model with explosion
energy $E_{51}=20$.
\label{yedepend}}
\end{figure}

We show in Figure \ref{yedepend} how the abundances of Fe-peak
elements depend on the value of $Y_e$. Here, we change $Y_e$ inside
the incomplete Si-burning region, and the mass-cut is chosen to
maximize the Zn/Fe ratio.  The adopted supernova model is a 25
$M_\odot$ model with the explosion energy $E_{51}=20$. Production of
Mn is larger for $Y_e < 0.5$ and Co production is significantly
enhanced for $Y_e > 0.5$.  More detailed discussion on the
nucleosynthesis for the $Y_e >0.5$ matter and its implications will be
discussed elsewhere, but here we suggest that the effect of $Y_e >0.5$
may be very important for explaining large Co/Fe ratios observed in
typical EMP stars.

\begin{figure}

\plottwo{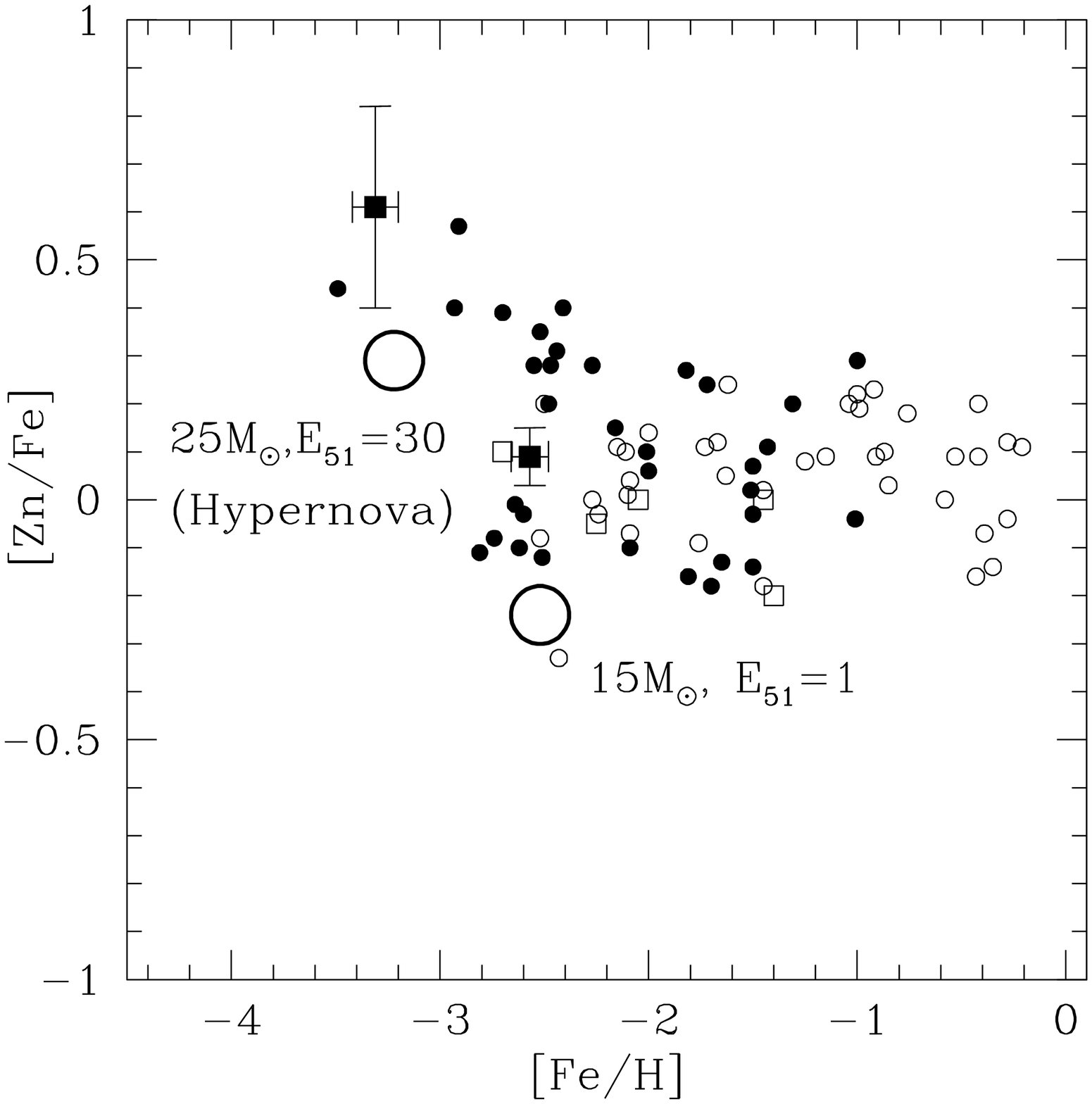}{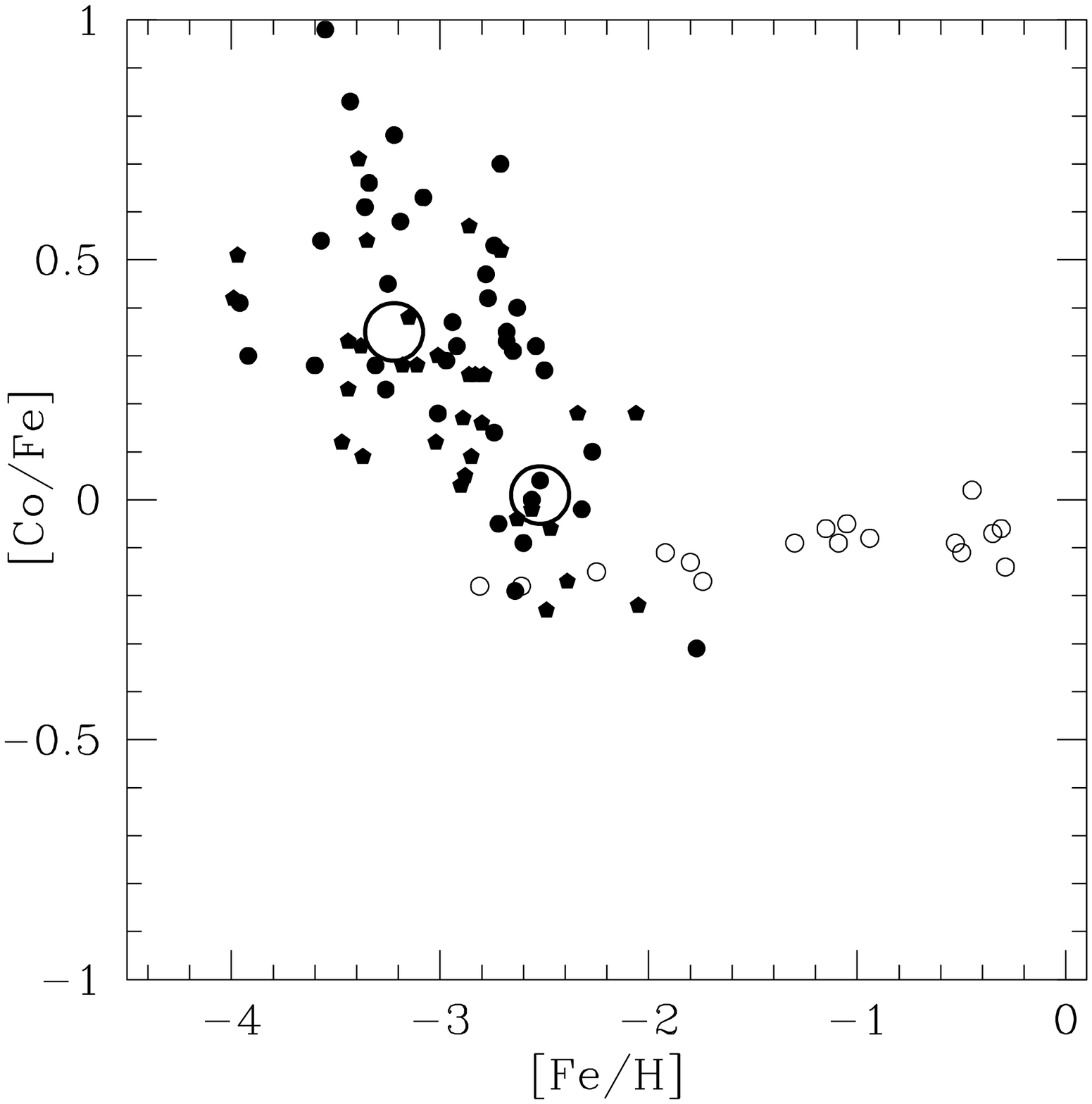}

\plottwo{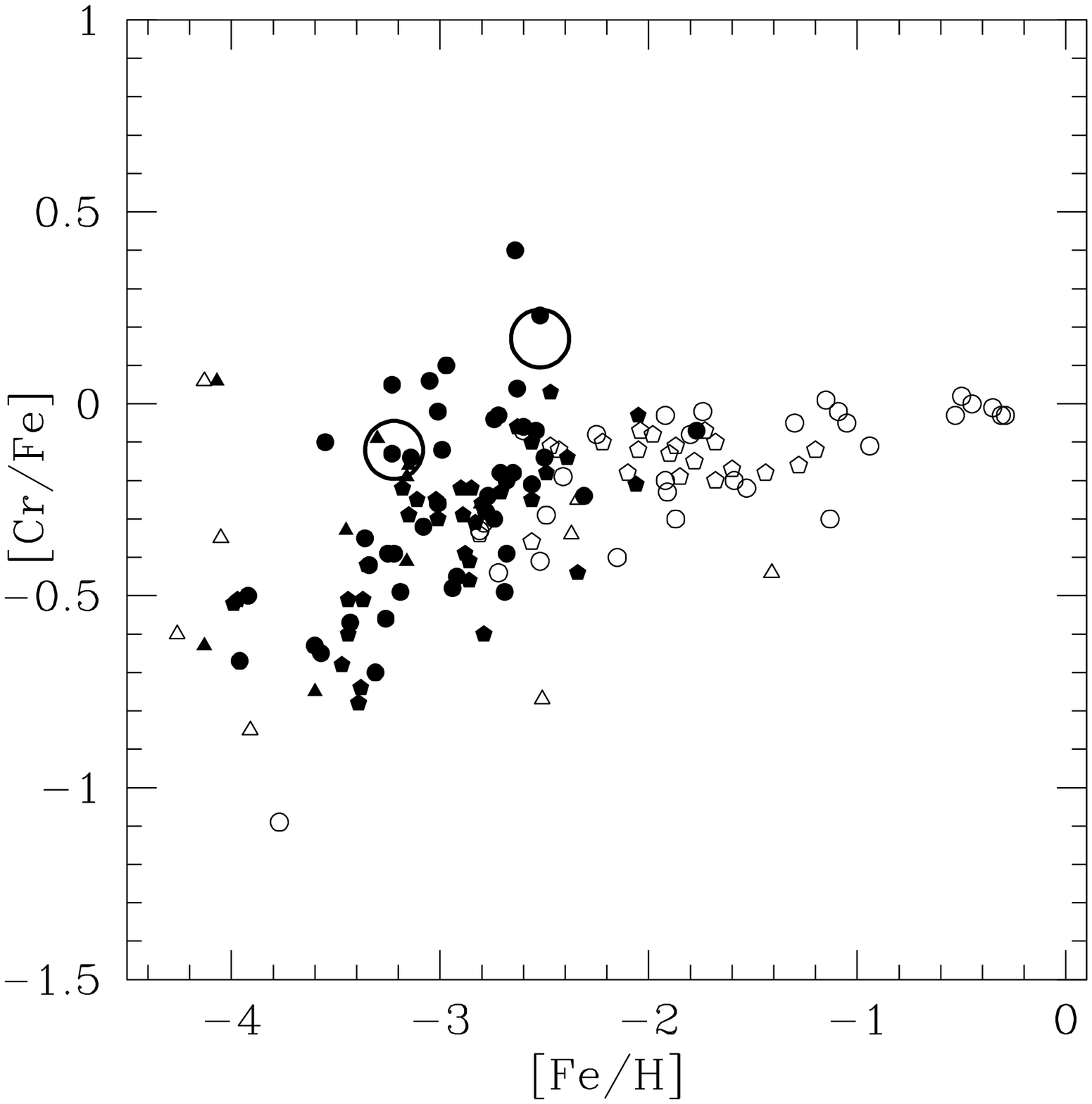}{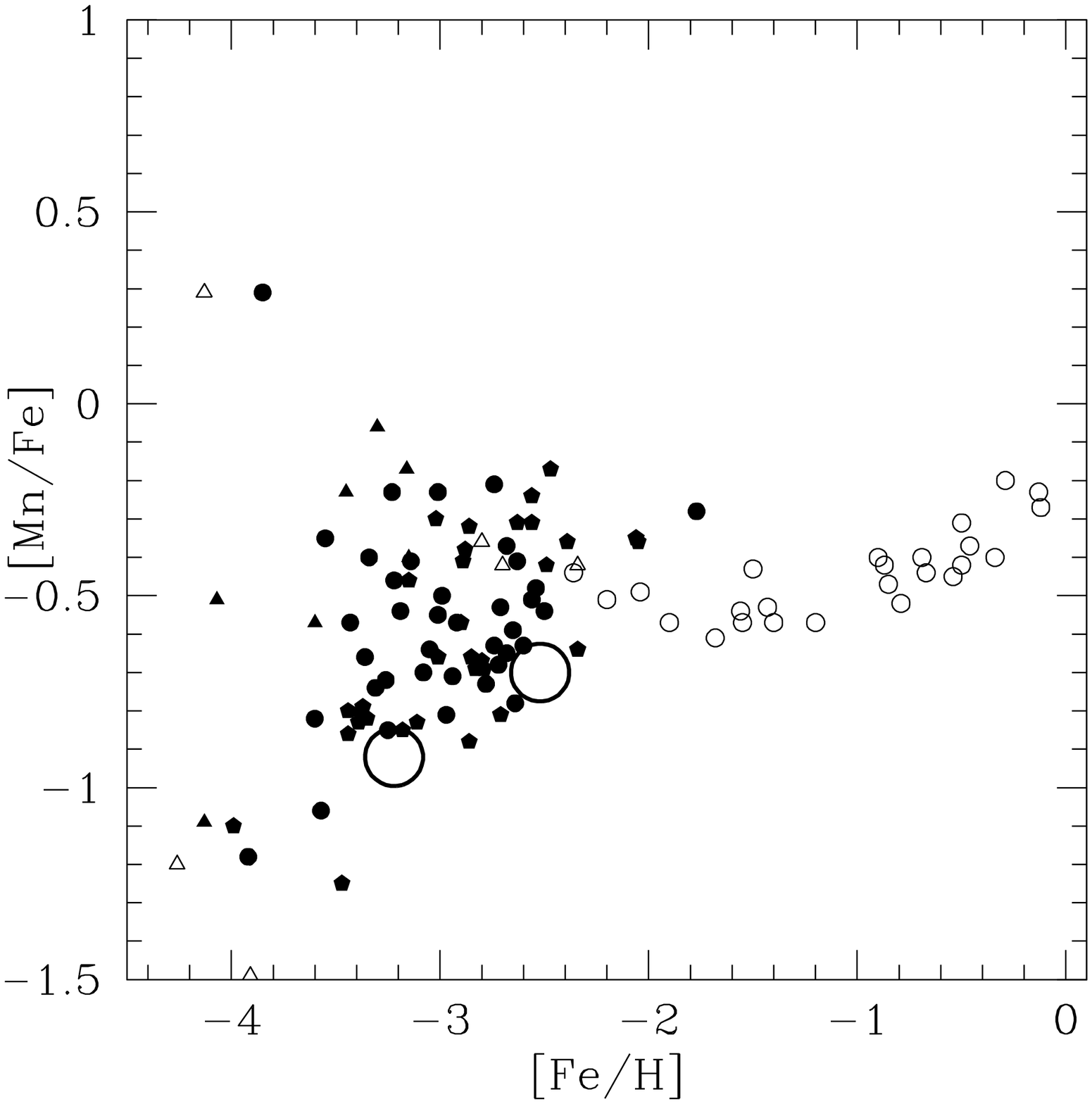}

\caption{Observed abundance ratios of [(Zn, Co, Cr, Mn)/Fe] 
vs [Fe/H] compared with (15$M_\odot$, $E_{51}=1$)
and (25$M_\odot$, $E_{51}$=30) models. In these models, it is assumed that
$Y_e = 0.5001$ in the complete Si-burning region and
$Y_e = 0.4997$ in the incomplete Si-burning region.
\label{xfe1}}
\end{figure}

 As an example, we show in Figure \ref{xfe1} the yields for the
(15$M_\odot$, $E_{51}=1$) and (25$M_\odot$, $E_{51}$=30) models
compared with the observations.  Some data related to these models are
shown in Table 5.  

We determine [Fe/H] of these models by assuming [Fe/H] =
log$_{10}$(Mg/$E_{51}$) + $C$ with $C= -1.0$.  Here we determine
[Fe/H] with log$_{10}$(Mg/$E_{51}$) and not directly with
log$_{10}$(Fe/$E_{51}$) because the ejected mass of Fe is much
uncertain than that of Mg. The ejected Fe mass strongly depends on
mixing-fallback parameters (see Sec. 3.2), but Mg does not as far as
$M_{\rm mix}$(out) is located below Mg-rich region.  This assumption
is not unreasonable because [Mg/Fe] $\sim 0.3-0.5$ for typical EMP
stars and roughly independent of [Fe/H] (e.g., Cayrel et al. 2003).

In these models, we assume that $Y_e = 0.5001$ in the complete
Si-burning region and $Y_e = 0.4997$ in the incomplete Si-burning
region.  We modify $Y_e$ mostly by adjusting the isotope ratios of
silicons.  Modification of $Y_e$ by using other elements is found to
lead to a negligibly small difference in the final result.  For such a
$Y_e$ distribution, both the Co/Fe and Mn/Fe ratios are enhanced and
fits better to the observed values.  This ``inversion'' of $Y_e$ may
be possible according to the most recent explosion calculations.

\subsection{Comparison with individual star}

 In this subsection, we compare the typical abundance pattern of EMP
stars with core-collapse SNe yields. This was, for example, recently
done in CL02. They compared their and other groups' (WW95 and UN02) SN
yields with the observational points given by Norris, Ryan, \& Beers
(2001, NRB01 hereafter).  According to them, all the models including
theirs using the "High" $^{12}$C($\alpha, \gamma$)$^{16}$O rate do not
fit well with observations.  Here the "High" $^{12}$C($\alpha,
\gamma$)$^{16}$O rate is the value given in Caughlan et al. (1985)
which was adopted to explain the solar abundance ratios by SNe II
(e.g., Thielemann, Nomoto, \& Hashimoto 1996; WW95).  According to
them, their low $^{12}$C($\alpha, \gamma$)$^{16}$O rate model (L
model) fits better to the observations, though some elemental ratios
cannot be reproduced. Here the "Low" $^{12}$C($\alpha,
\gamma$)$^{16}$O rate given in Caughlan \& Fowler (1988, CF88
hereafter) is used.  In their L model, Co/Fe and Sc/Fe ratios are
greatly enhanced to be compatible with the observations.

 CL02 explained the reason for these enhancements as follows: The
increase in the carbon abundance leads to a more flattened-out final
mass-radius relation because the contraction of the ONe core is slowed
down by the presence of a very active C-burning shell. As a
consequence, the average density in the regions that experience
complete and incomplete explosive Si burning will be lower as
well. The net result is that the $\alpha$-rich freeze-out is
considerably favored and also that the overall amount of synthesized
$^{56}$Ni is significantly reduced. These effects tend to increase
significantly both [Sc/Fe] and [Co/Fe] (for any chosen mass-cut
location).

 We have investigated from our sample of progenitor models with
relatively high C/O ratios after the central He-burning (C/O$\simeq
0.38 - 0.41$ and $M= 15-18M_\odot$) whether we also obtain such high
Co/Fe ratios. However, so far we have not found such significant
enhancement of the Co/Fe and Sc/Fe ratios (see Table 6). This is
likely because the density structure of the progenitors depends not
only on the C/O ratio but the treatment of convection, and thus the
large C/O ratio may not always lead to significantly low
density. Also, since CL02 used a simple analytic model for the
density-temperature evolution during explosion, this might produce
some differences with our method.  Furthermore, we are not in favor of
the large C/O solution, since it is well known that such large C/O
models overproduce Ne and Na in the solar metallicity models (Woosley
\& Weaver 1993; Nomoto et al.  1997; Imbriani et al. 2001).  CL02
suggested that if the 25$M_\odot$ stars were not the representative of
the solar-metallicity core collapse SNe, this problem may be
avoided. However, then one has to adopt rather unconventional IMF and
such IMF must confront with several observational constraints.

 In this paper, we show a different solution where the fit to the
 observations ([Co/Fe] and [Mn/Fe]) are significantly improved by
 choosing a reasonable value of $Y_e$ but without adopting the high
 C/O; this is because the absolute abundance of Co and Mn are quite
 sensitive to $Y_e$.  

 The large Sc/Fe ratio is still hard to explain by changing $Y_e$. In
 Appendix we show a modified model that enhances Sc/Fe.  The Sc/Fe
 ratio can be largely enhanced if the densities during explosion are
 sufficiently lowered than in the ordinary model by any mechanism,
 e.g., by ejecting small jets. In Appendix, we show in such
 ``low-density'' models, these ratios are significantly enhanced.
 Interestingly, even with the low-density, we still need high
 explosion energy to explain the observed large Zn/Fe ratio together
 with Sc/Fe.

 We will describe more details of the low-density models elsewhere
and in this paper (except the Appendix) only consider the effect of large
$Y_e (\gsim 0.5$) and large explosion energies
to explain the observed Co/Fe ratio.

In Table 7, we summarize several properties of our pre-supernova
progenitor models used in the rest of this paper.  This table shows
the initial stellar mass, metallicity, the adopted $^{12}$C($\alpha,
\gamma$)$^{16}$O rate, the central C/O mass fraction ratio just after
core helium burning, Fe-core mass (defined by $Y_e < 0.49$), O-burning
shell (defined by X(O) $ \simeq 0.1$), C-O core mass (defined by
X(He)$ < 10^{-3}$), and He core mass (defined by X(H)$ < 10^{-3}$).
 
\begin{figure}

\epsscale{.8}

\plottwo{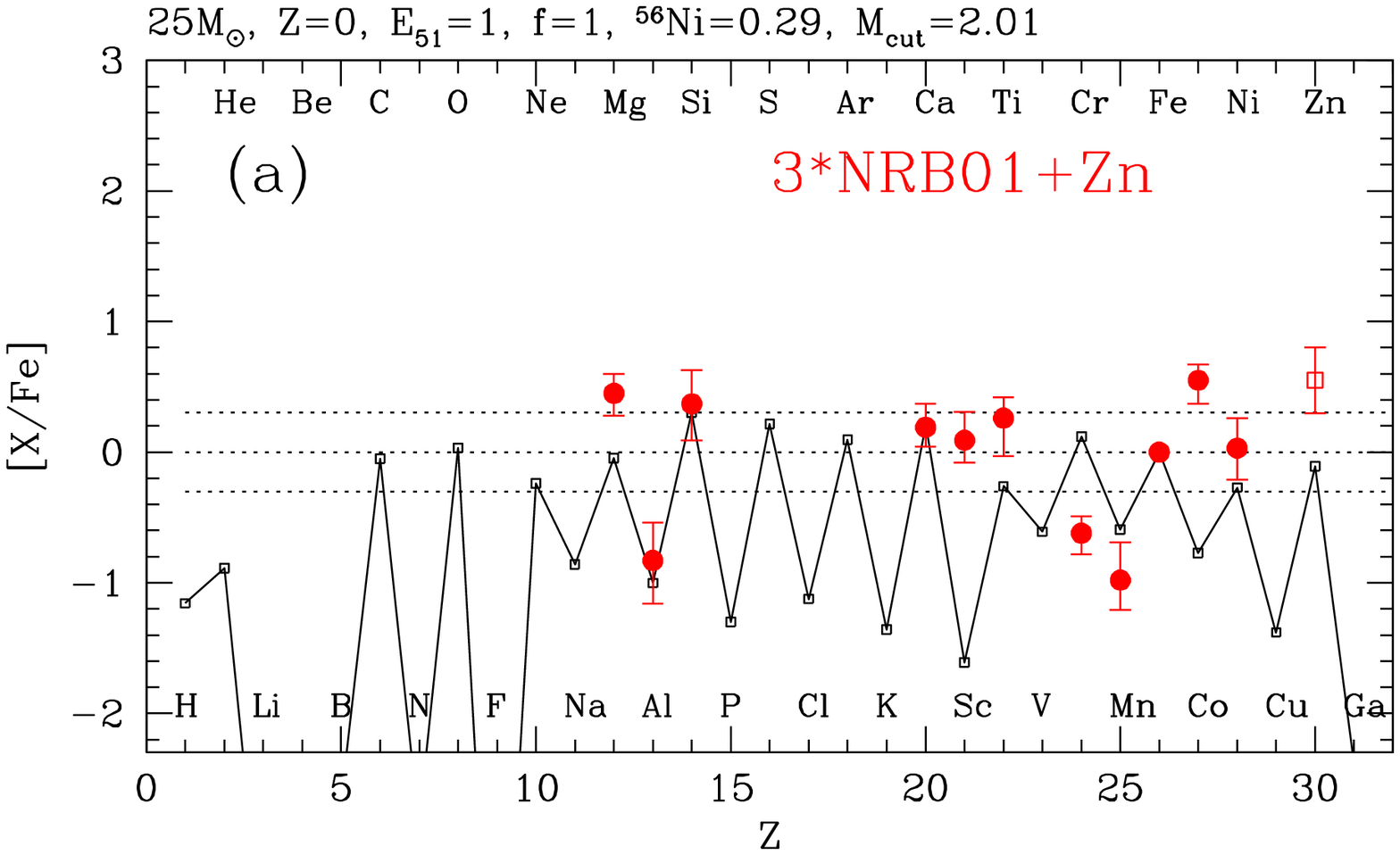}{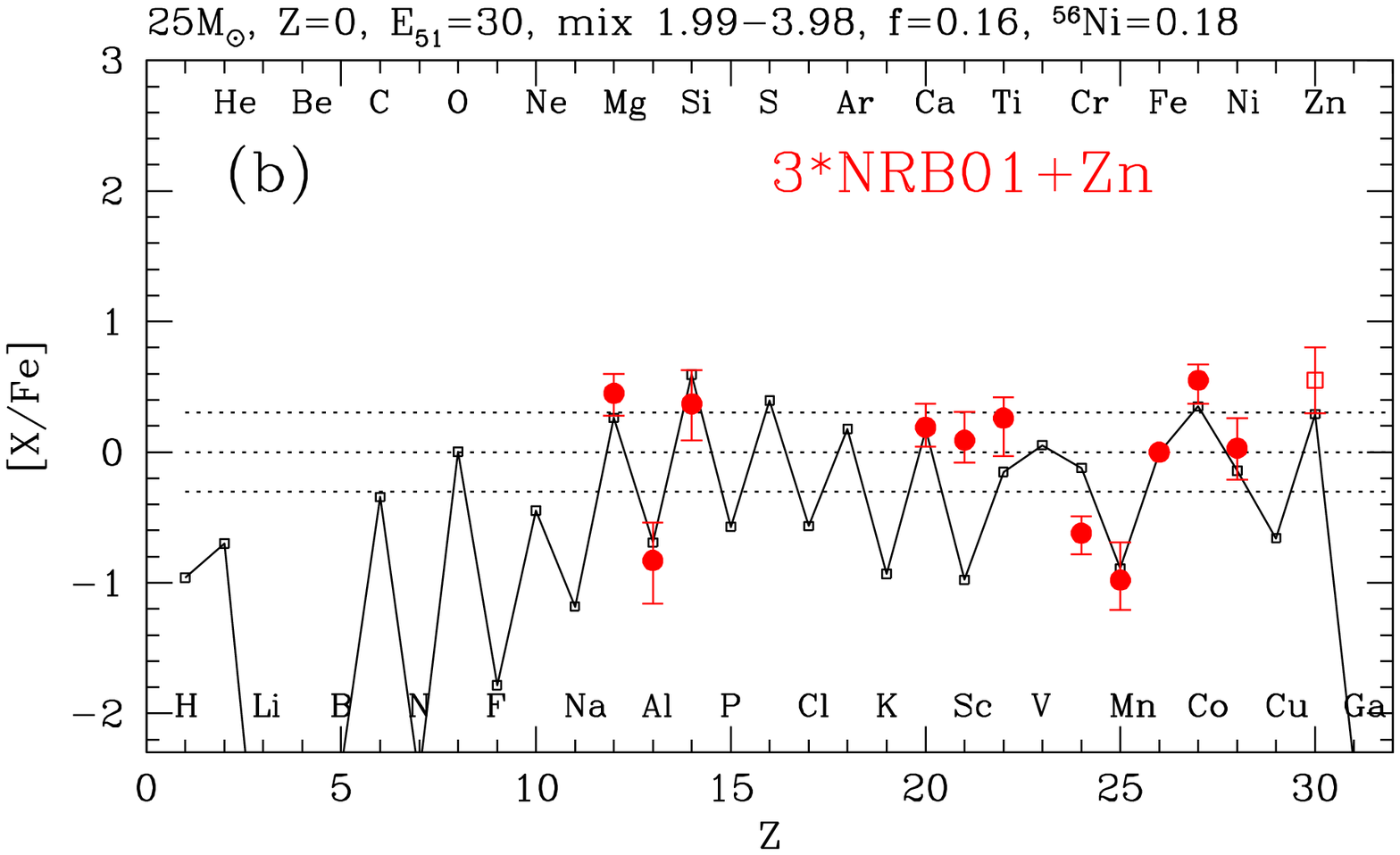}

\plottwo{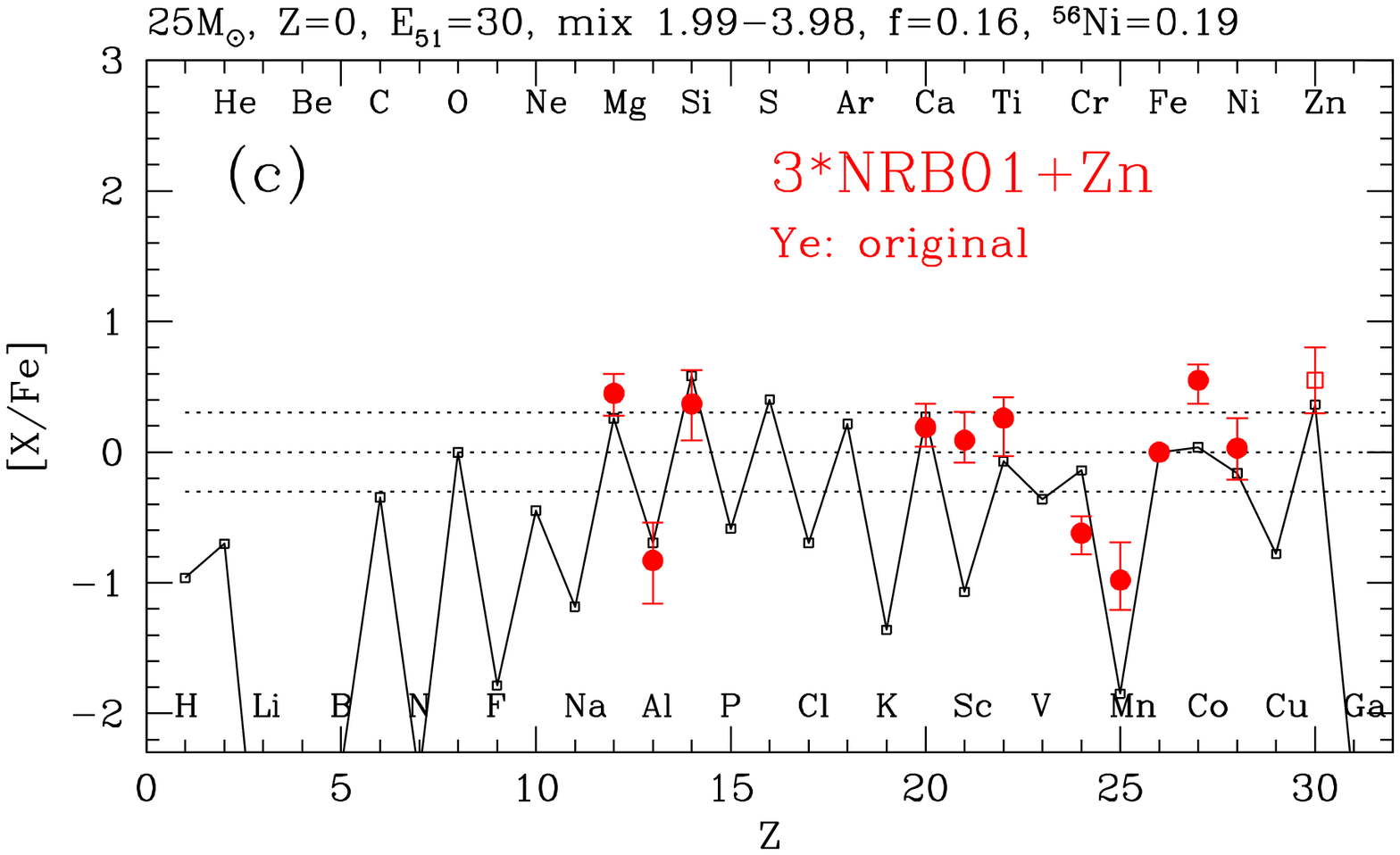}{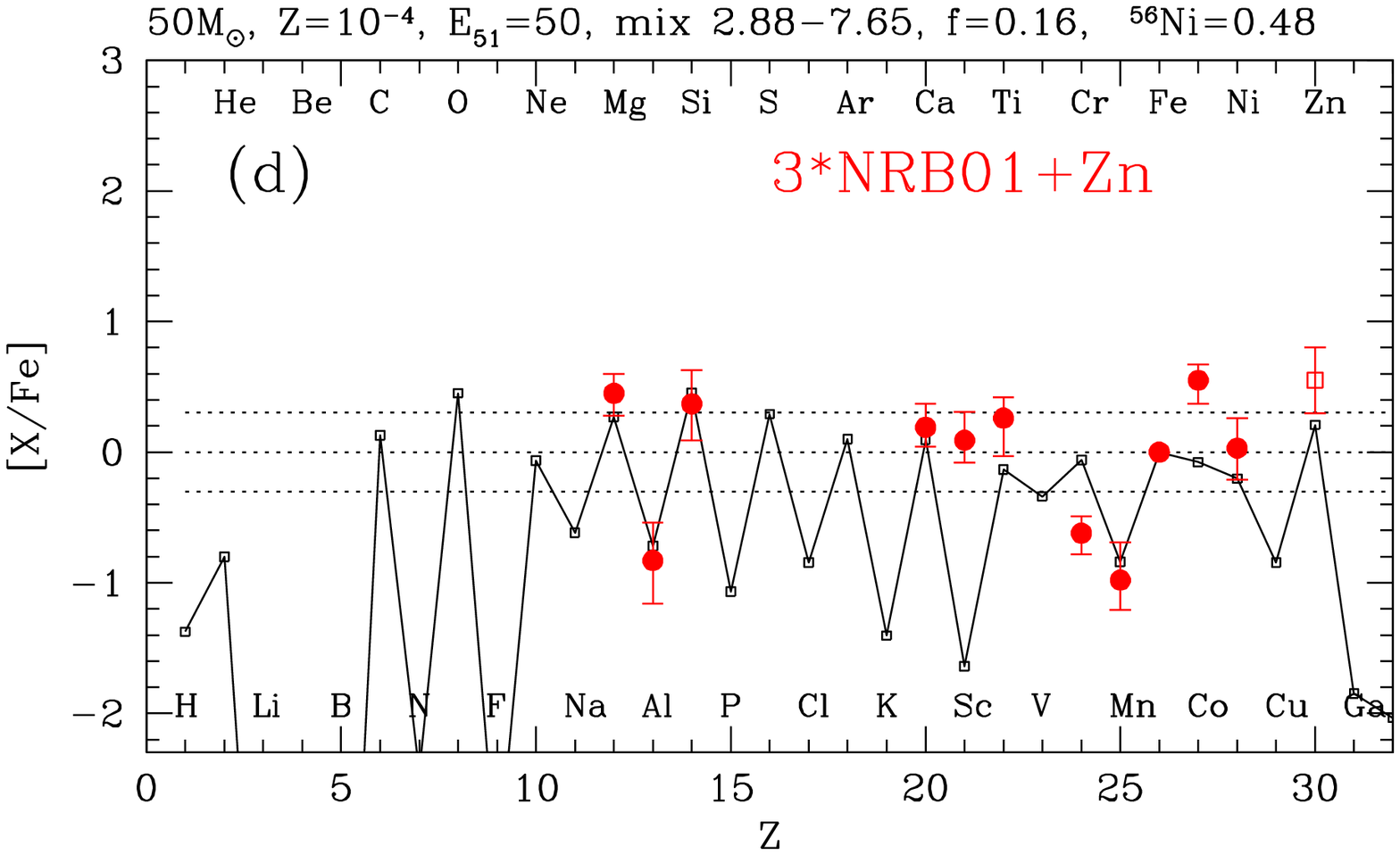}

\caption{Elemental abundances of typical EMP stars at [Fe/H] $\sim$ -3.7
given by NRB01 (solid circles with error bars) 
compared with theoretical supernova yield (solid lines).
The open square with an error bar represents 
the Zn abundance typically observed in EMP stars with [Fe/H]$\sim -3.7$.
In the panel (a), a 25$M_\odot$, $E_{51}$=1 model is shown.
This model does not assume mixing-fallback and the fit to the observation
is not good. In (b:''best'' fitting model) a higher energy with a proper degree of
mixing-fallback is assumed. This fits much better to the observation.
In all models but (c), $Y_e$ during the
explosion is assumed to be $Y_e=0.5001$ in the complete Si-burning
and $Y_e=0.4997$ in the incomplete Si-burning region.
The model in (c) shows the effect of changing the $Y_e$.
Similar goodness of the fitting may be obtained by more massive
more energetic models as shown in (d), though the under-abundance
of [Co/Fe] may suggest that a higher energy model might be better. 
\label{NRB01}}
\end{figure}

 In Figure \ref{NRB01}, we compare our models with typical abundance
patterns of EMP stars.  The observed points are the "averaged"
abundances of three stars CD-38$^\circ$245, CS22172-002 and
CS22885-096 given in NRB01.  These stars have similar abundance
patterns and similar metallicity: the [Fe/H] of these stars are
$-$3.98, $-$3.61, and $-$3.66 respectively.  The solid circles in
Figure \ref{NRB01} represent the averaged abundance of these stars,
and the errorbars represent the range of errorbars of these
stars. Since the variation in [C/Fe] is large and only the upper-limit
is given for CD-38$^o$245, we do not include the [C/Fe] point.  The
NRB01 data do not include the Zn point. However, the overabundance of
Zn is quite common among EMP stars (Primas et al. 2000; Depagne 2003;
Cayrel et al. 2003), [Zn/Fe] is roughly homogeneous for [Fe/H] $<
-3.6$, and the point is crucial to estimate the explosion energy;
therefore, we add the value of [Zn/Fe] in the figure: specifically,
[Zn/Fe]= 0.3 $\sim 0.8$ from Depagne (2003) and Cayrel et al. (2003).
The theoretical yield in Figure \ref{NRB01}(a) is obtained for the
zero metallicity (Pop III) 25$M_\odot$ star after explosion with
$E_{51}$=1.

 The calculation method and other assumptions are the 
same as described in Umeda et al. (2000), UN02 and UN03,
and the progenitor model used in Figure 6(a) is the same as used in UN02.
In this model, the mass-cut is located at mass coordinate 
$M_r=M_{\rm cut}=$ 2.01 $M_\odot$. This mass-cut is chosen to maximize the 
Zn/Fe ratio in the original $Y_e$ models (although $Y_e$ is modified
in this model) as described in Section 2.1.
 The fit in Figure \ref{NRB01}(a) is not very good because of the
underabundances of Mg, Sc, Ti, Co, Zn and overabundance of Cr.\footnote{
 Here, we briefly describe our stance in the comparison between
theoretical models and observations. It is of course best if the
model and the observed points fit exactly. However, most data shown
in this paper does not include unknown systematic errors.
Also in the theoretical models, there should be some uncertainties
not included in this paper. For example, many of nuclear reaction rates
may not be exactly correct and some of them could significantly alter
the results. Therefore, we should always be somewhat tolerate to the
``small'' discrepancies between the models and observations.
At the same time, we should always remind that the matching
between observations and theories might be just an accident and does not
represent the true correctness of the model. In general, we call the
model or a specific parameter choice ``good'' or ``better''
if the model is much closer (even though it is still outside the
shown error-bars) to the observation than our other models.}

 The underabundances of Mg in the model can be resolved if larger
$M_{\rm cut}$ is chosen, for which the ejected mass of Fe (decay
products of $^{56}$Ni) is smaller.  The overabundance of Cr and
underabundances of Co are, on the other hand, improved if $M_{\rm
cut}$ is smaller (e.g., Nakamura et al. 1999). Therefore, these
problems cannot be solved just by changing the mass-cut. In UN02, we
proposed one solution for this problem, which is the mixing-fall back
mechanism.  If the inner part of the ejecta is mixed with the outer
materials, and later some of the mixed materials are fallen-back to
the central remnant, the ratio of the lighter elements, such as Mg and
Al, to Fe increases without changing the abundance ratios in the
Fe-peak elements.
 We also note that the dependence of the yields on the progenitor's
mass is too small to solve the problem of the underabundance of Co/Fe
and Zn/Fe (see e.g., UN02).

 In the mixing-fallback model, we define the following locations
of $M_r$, and a function $f$:
\begin{itemize}

\item $M_{\rm cut}$(ini): initial mass-cut,
which is equal to the internal border of the mixed region.
We called it ``mass-cut'' because the matter above which is ejected
even a fraction.

\item $M_{\rm mix}$(out): outer border of the mixed region.

\item $M_{\rm cut}$(fin): the final mass of the remnant. 

\item $f$: a fraction of matter ejected from the mixed region of
           $M_{\rm cut}$(ini) $\leq M_r \leq$ $M_{\rm mix}$(out).

\end{itemize} 

\noindent
Here, the inner most materials are first mixed in the region between
$M_{\rm cut}$(ini) and $M_{\rm mix}$(out), and some fraction of the
matter, $1-f$, is fallen-back later on to the central remnant.

 As shown in UN02, the Zn/Fe and Co/Fe ratios are significantly
enhanced if the explosion energy is larger. In Figure 6(b), we show
such a high-energy model with $E_{51}$=30.  This is the ``best''
fitted model among the four models shown in Figure 6. We note that
better fitted one may be obtained by slightly changing $Y_e$, $E$ and
mixing-fallback parameters, but finding ``exactly the best'' fitted
model is not the purpose of this paper. What we would like to stress
here is the high-energy models are in much better agreement with
observations than the low-energy models.

 Here the mixing-fallback mechanism is more important than in the low
energy models, because larger amount of Fe (i.e., $^{56}$Ni) is
synthesized so that larger fall-back is necessary for more energetic
explosions. In this model, $M_{\rm cut}$(ini)=2.35 $M_\odot$, $M_{\rm
mix}$(out) = $4.29 M_\odot$, $f=0.1$, and $M_{\rm cut}$(fin)=4.10
$M_\odot$.  $M_{\rm mix}$(out) is chosen to be near the top of the
incomplete-Si burning as in UN02. The ejection factor $f=0.1$ means
that 10\% of the mixed matter is ejected and 90\% is fallen-back. This
model fits much better to the observations than the low energy models.

 Here we explain the meaning of the numbers shown at the top of Figure
6(b) and others. This model uses the initially $25M_\odot$ progenitor
model with initial metallicity $Z=0$ with explosion energy
$E_{51}=30$.  'mix 1.99-3.98' means that $M_{\rm
cut}$(ini)=1.99$M_\odot$ and $M_{\rm mix}$(out)= 3.98$M_\odot$. With
the ejection factor $f=0.16$ one can calculate the final mass-cut or
the remnant mass as $M_{\rm cut}$(fin) $= 3.98-(3.98-1.99)*f = 3.67
M_\odot$. The ejected $^{56}$Ni mass is 0.18 $M_\odot$.

 In the model in Figure 6(b), all elements other than Sc, Ti and Cr
fit well to the observation.  It is known that the Ti abundance is
enhanced if the explosion is aspherical (Nagataki 2000; Maeda \&
Nomoto 2003a,b). For an aspherical explosion Ca, Sc and Zn abundances
are also enhanced, although the explosion energy has to be still
larger than that of the canonical explosion, $E_{51}$=1 (Maeda \&
Nomoto 2003a,b). We also show in Appendix that the ``low-density''
model, in which the density of the progenitor model is artificially
reduced, can also yield significantly enhanced Sc/Fe and Ti/Fe
ratios. In this paper, we do not specify the mechanism to form for the
low-density, but it may be related to the jet-like explosion. If
several discontinuous jets are injected and the inner part of the
progenitor is expanded by weak-jets before a strong jet finally
explodes the star, a low density explosion may be realized.  Although
we have currently no idea about the discrepancy in [Cr/Fe] and has to
consider some modifications to [Sc/Fe] and [Ti/Fe], reasonably good
fits of other elements make us to believe that the typical abundance
pattern of EMP stars can be understood as a result of nucleosynthesis
in energetic core-collapse SNe (or hypernovae).

 For comparison, we show in Figure 6(c) the same model as in Figure
6(b) but with the original $Y_e$ distribution. We find that with the
original $Y_e$ distribution, [Mn/Fe] and [Co/Fe] fit much worse to the
observation than the model in Figure 6(b).

Figure 6(d) shows the comparison with the more massive and more
energetic supernova model ($M = 50 M_\odot$ and $E_{51}$ = 50).
Except for [Co/Fe], the fits of other elements are as good as the
25$M_\odot$ model in Figure 6(b).  This means that from the abundances
of EMP stars, it is difficult to constrain the typical mass of the
progenitor. We can constrain only the set of mass and explosion
energy.  ([Co/Fe] is expected to be larger for models with larger
explosion energy and/or asphericity or lower densities (Appendix),
which will be systematically examined in the future study.)  At least
we can say from the present SNe observations that the progenitors of
energetic core-collapse SNe are more massive than $\sim
20M_\odot$. The upper mass limit is unknown, but should be lower than
$ \sim 140M_\odot$ because above which the stars would explode as
pair-instability SNe and nucleosynthesis patterns are quite different
from EMP stars (e.g., UN02; Heger \& Woosley 2002). Note that the
(newly calculated) progenitor model used in this figure has the
metallicity $Z=10^{-4}=Z_\odot/2000$, but not zero. However, as shown
in our previous work (Umeda, Nomoto, \& Nakamura 2000), the elemental
abundance pattern from SNe II are not much different for $Z=0
-10^{-4}$.  Therefore, the Z=$10^{-4}$ models can be used for the
present purpose.  Detailed metallicity dependent yield will be
published elsewhere.

\section{Abundance Pattern of C-rich EMP Stars }
 
In this section, we compare our core-collapse SNe yields with
abundances of C-rich EMP stars.  Among several C-rich EMP stars we
pick up 5 representative stars, C1:CS22949-037, C2:CS29498-043,
C3:CS22957-027, C4:CS31062-012 and C5:HE0107-5240.  C1 is interesting
because the important elements O and Zn are both observed only for
this star. The abundance pattern of C1 is peculiar because O and Mg
are rich as well as C and N.  The abundances of C2 - C4 are obtained
by Aoki et al. (2002a,b).  C2 is rich in C, Mg, Al and Si more than
C3.  C3 is much more C-rich than C1 \& C2, but not Mg-rich.  C1 - C3
all show no enhancement of s-process elements, while C4 shows
enhancement of [Ba/Fe].  C5 is the most Fe-poor star observed so far.
The model for this star has been discussed in UN03, but in this paper
we modify $Y_e$ in the Si-burning region, and show that the same model
still can explain the observation well.

\subsection{CS22949-037} 

\begin{figure}
\epsscale{1.}

\plottwo{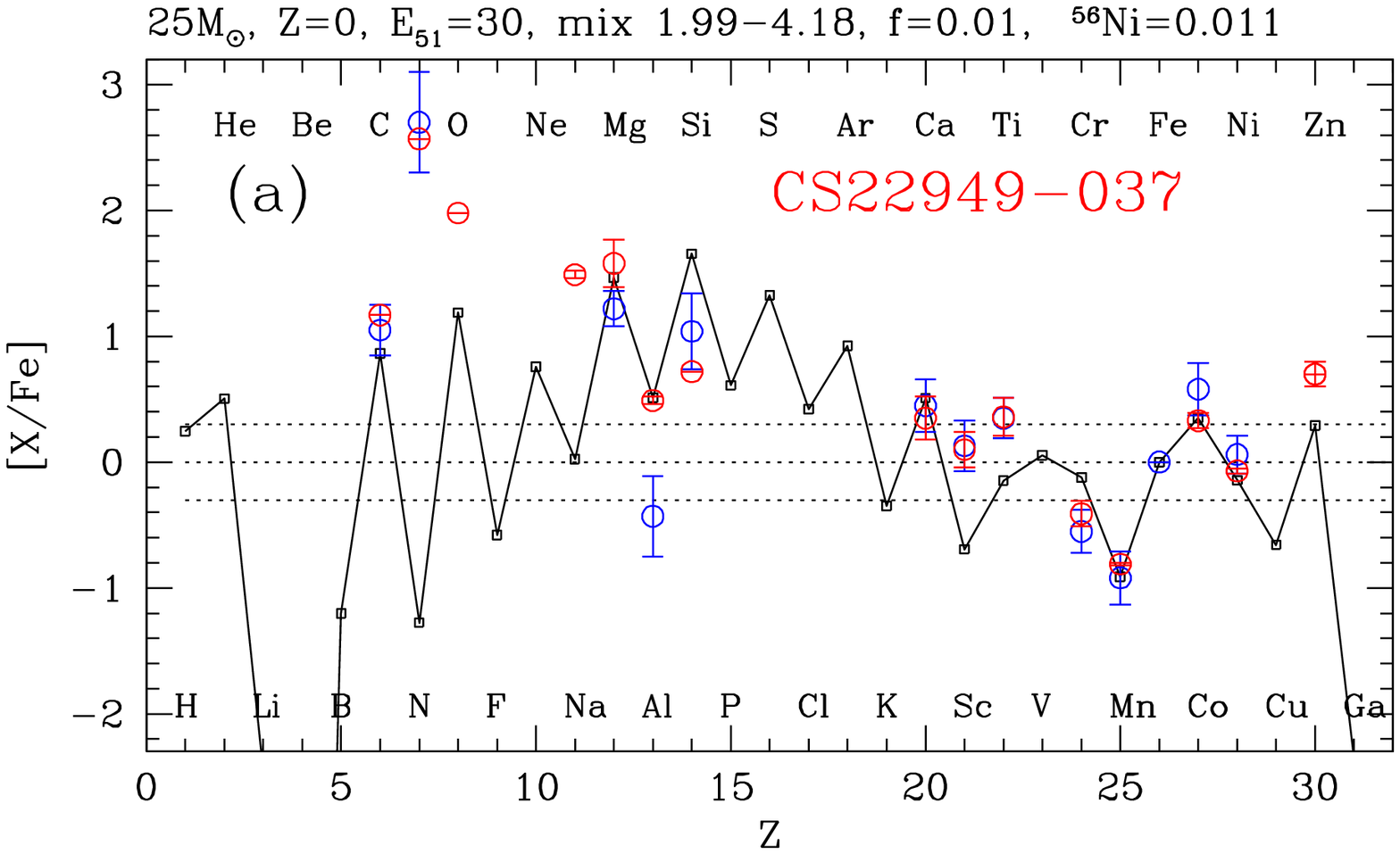}{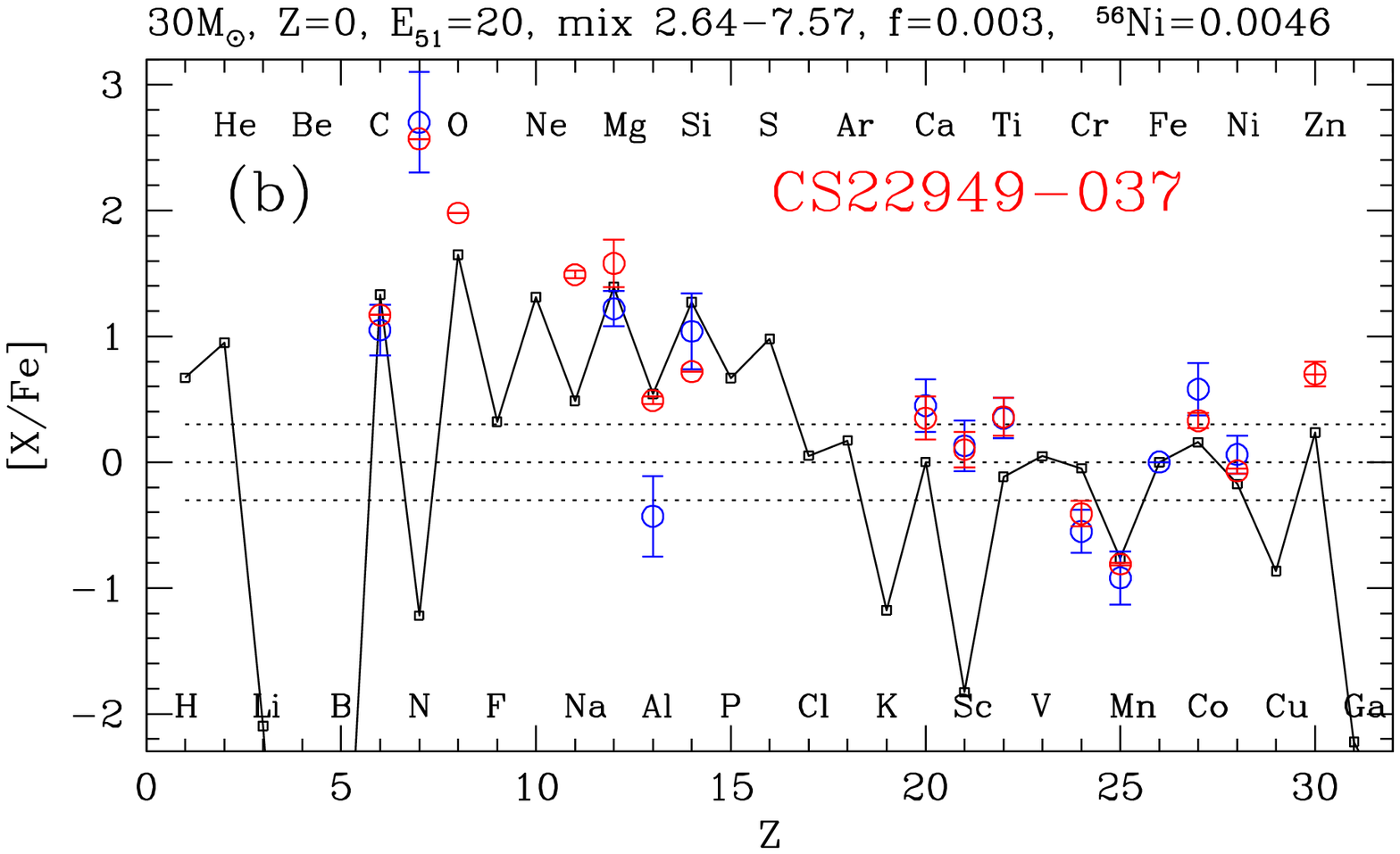}


\caption{Elemental abundances of CS22949-037
compared with theoretical supernova yield (solid lines).
Here the blue circles are data from NRB01  and red
circles are from Depagne et al.(2001). In panel (a),
the model N, O/Fe are underproduced. The model in panel (b)
fits better if N is enhanced by the uncertain mixing mechanism
that may occur in the Pop III progenitors. 
\label{CS22949}}
\end{figure}

CS22949-037 is one of the most Fe-poor giants known ([Fe/H] $\simeq
-4.0$). The detailed abundance pattern of this star was first observed
by NRB01, and then by Depagne et al. (2002).  Those abundances are
shown in Figure \ref{CS22949} by blue circles (NRB01) and by red
circles (Depagne et al. 2002). Two results are mostly consistent, but
Depagne et al. obtained the abundances of Zn and O, which are very
important for constraining the SN models. The Al point in Depagne et
al. is larger than NRB01 because we have adopted NLTE corrections
($\Delta $[Al/H]$\simeq 0.65$) suggested in their paper.  The NLTE
correction is also added to the Na point ($\Delta $[Na/H]$\simeq
-0.6$). For this star, no enhancement of r- and s-process elements is
observed (e.g., [Ba/Fe]= $-$0.6).

 This star shows large Co/Fe and Zn/Fe ratios as in typical EMP stars,
suggesting the enrichment by a high-energy supernova.  For the model
In Figure \ref{CS22949}(a), therefore, we adopt a high explosion
energy model with $M = 25M_\odot$, $E_{51}=30$, $Z=0$ being the same
as in Figure \ref{NRB01} (b), but a smaller value of $f=0.01$ to
reduce the ejected Fe mass (see other parameters on top of the
figures).  For iron-peak elements, (Mn, Co, Ni)/Fe ratios are fit to
the observations but (Ca, Ti, Zn)/Fe ratios are smaller.  With
$f=0.01$, the ratios of (C, Mg, Al)/Fe are consistent with the
observation but O/Fe is too small.  Also the ratio Si/Fe is 
too large and Mg/Si is too small.

 A larger Mg/Si can be resulted from a smaller explosion energy model
(Nakamura et al. 2001; Umeda et al. 2002).  A larger amount of O is
yielded in a more massive progenitor model.  In order to test these
expectations, we construct a model with $M=30M_\odot$, $E_{51}=20$,
and $Z=0$.  (In this model, as well as the models in the rest of this
paper for consistency, $Y_e$ is modified to be 0.5001 and 0.4997 in
the complete and incomplete Si-burning regions, respectively.)  

\begin{figure}
\epsscale{1.}
\plotone{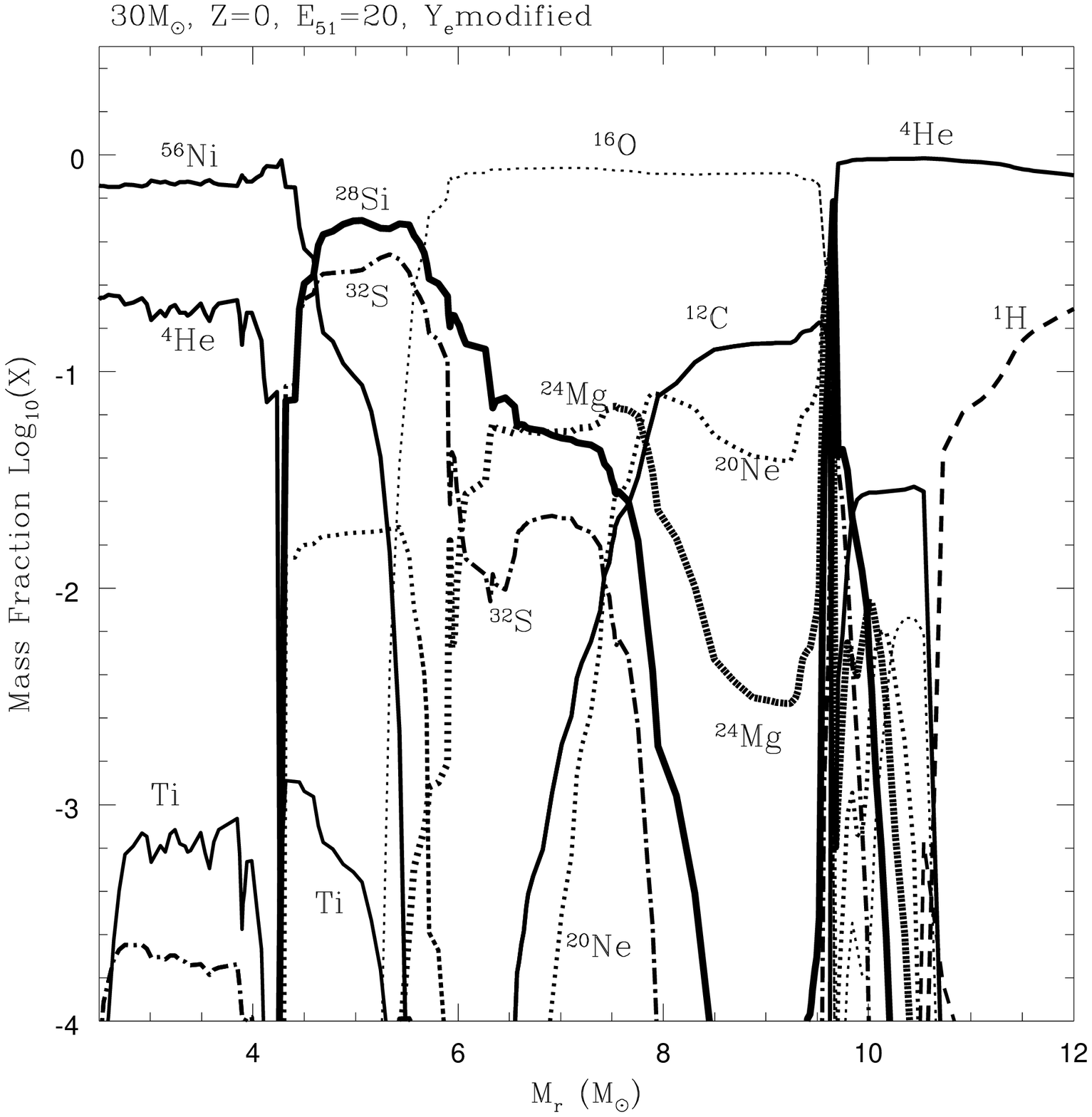} 
\caption{Abundance distribution after SN explosion of a 30 $M_\odot$
star with $E_{51}=20$. $Y_e$ in this model is modified to 0.5001 and
0.4997 in the complete and incomplete Si-burning regions, respectively.
\label{30z0e20ye5001-4997}}
\end{figure}

 As shown in Figure \ref{CS22949}(b), this model indeed yields a
larger O/Fe ratio, thus being closer to the observation than the model
in Figure \ref{CS22949}(a).  Also, because of the smaller explosion
energy (or smaller energy to mass ratio, $E/M$) and larger $M_{\rm
mix}$(out) (= 7.57 $M_\odot$), Mg/Si is as large as observed.
If we adopted smaller $M_{\rm mix}$(out), the amount of Mg
mixed into the fallback material would be smaller so that the Mg/Si
ratio in the ejecta would be even larger than the observed
ratio. (Note that Si is distributed in the deeper layer than Mg as
seen in Figure \ref{30z0e20ye5001-4997}).  The LTE Al/Fe value is
difficult to fit, while the inclusion of the NLTE correction improves
the agreement.

 In order to obtain an overall improvement in the fit between the
model and the observation, further systematic survey of models is
needed for wider parameter space.  For example, the underabundance of
Ti/Fe and Zn/Fe may be improved in the ``low-density'' explosion model
(Appendix).

 The remaining problems to be explained are the N and Na abundances.
In many cases N and Na are underproduced in the metal-poor massive SN
progenitors than those observed.  However, if the surface H is mixed
into the He layer by convection or rotational mixing in the
progenitor, significant amount of N and Na may be produced (e.g.,
WW95).

\subsection{CS29498-043} 

 This star is also an extremely Fe-poor giant with [Fe/H] = $-$3.75.
Its abundance pattern is similar to CS22949-037, being very rich in C,
N, Mg, Al, and Si (Aoki et al. 2002b). Another similarity to
CS22949-037 is no enhancement of s-process elements (e.g., [Ba/Fe] =
$-$0.45).  However, there are some differences: C, Mg, and Al are more
abundant and the Mg/Si ratio is larger in CS29498-043.  For
CS22949-037 we adopt the high energy model because of the large Zn/Fe
and Co/Fe ratios.  For CS29498-043, there are no data of these
elements, so that we first adopt a normal energy model.  We also
assume smaller $f$ (i.e., larger fallback) for CS29498-043 than
CS22949-037 to produce the larger [C/Fe].

\begin{figure}

\epsscale{.8}

\plottwo{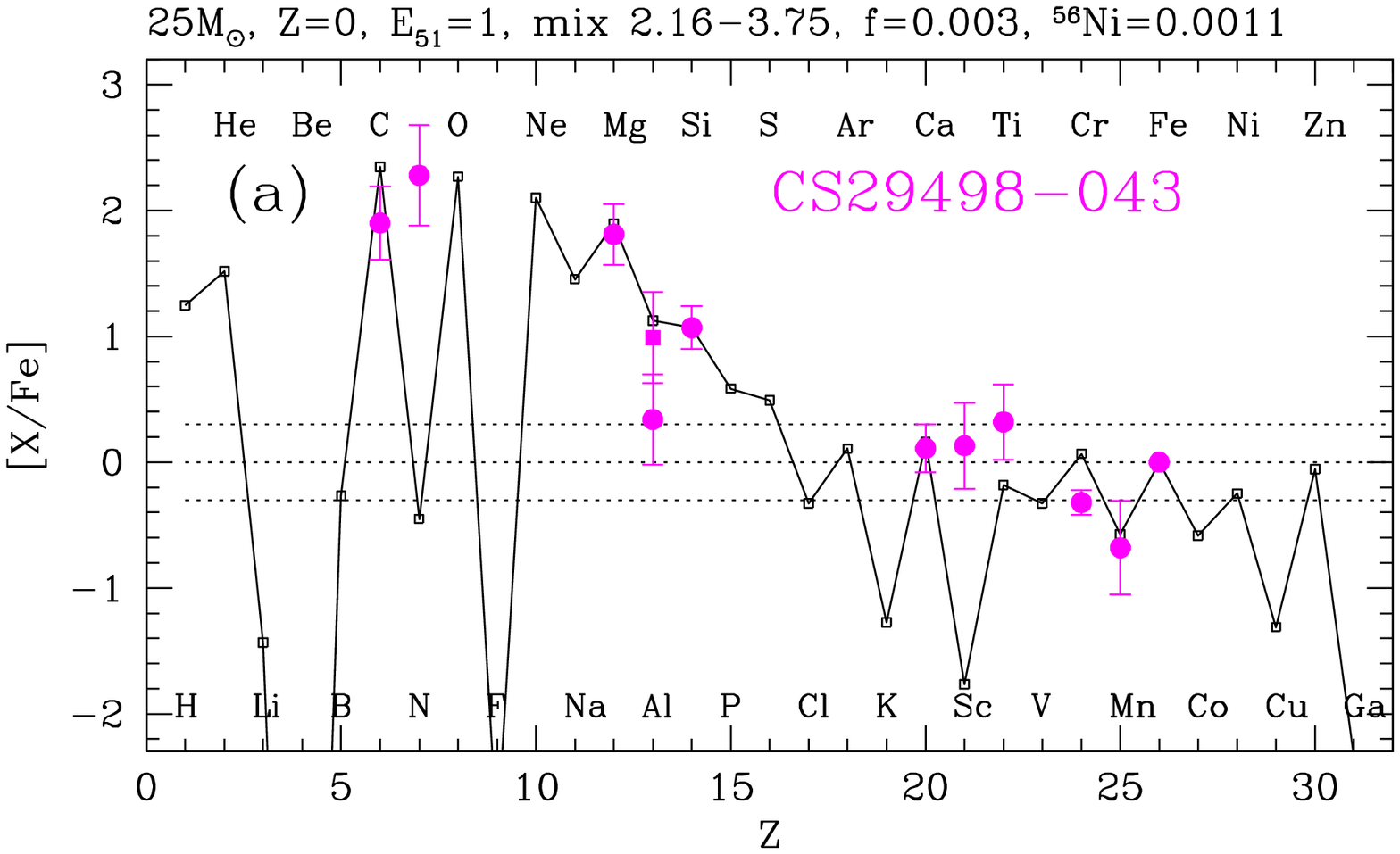}{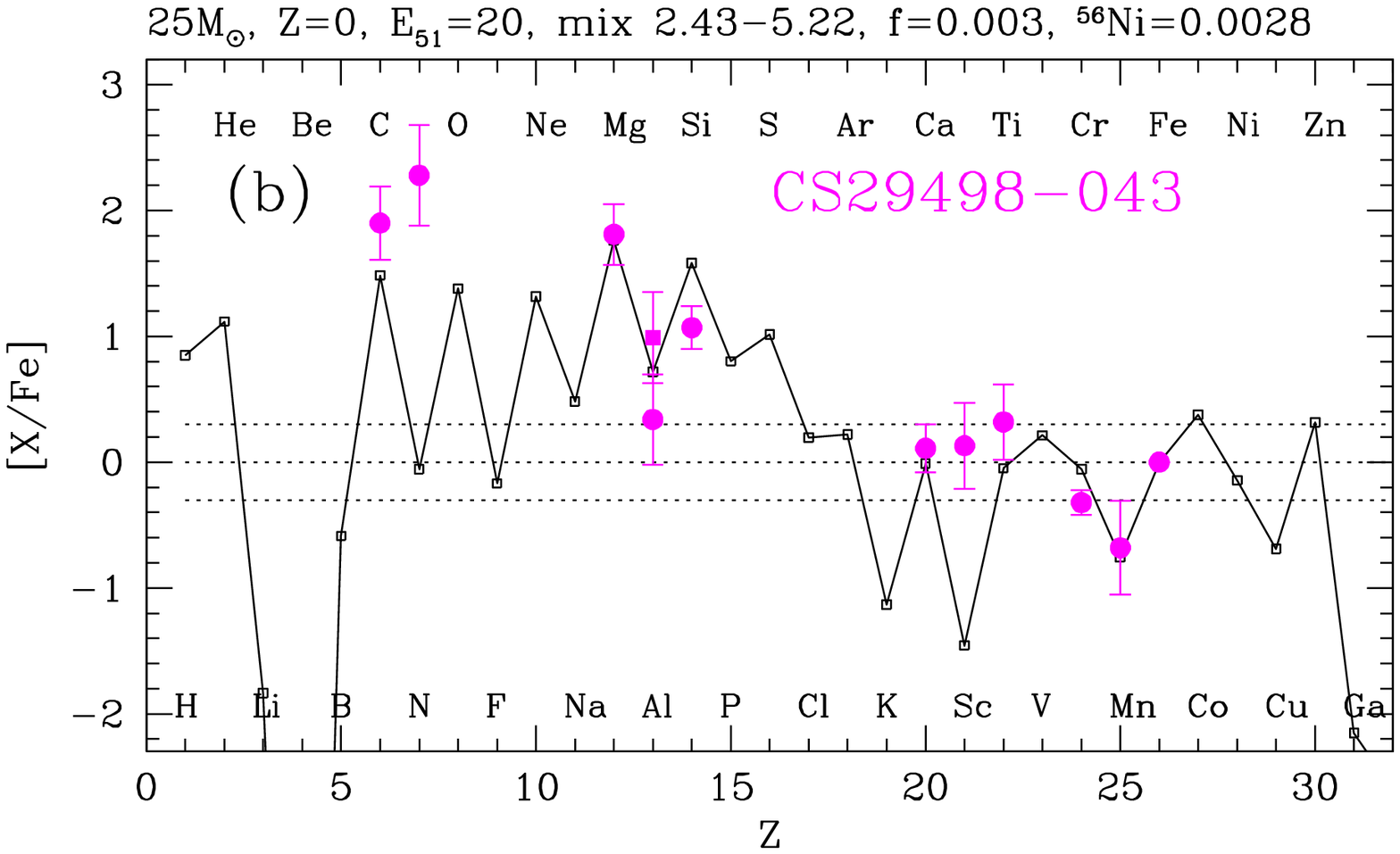}

\plottwo{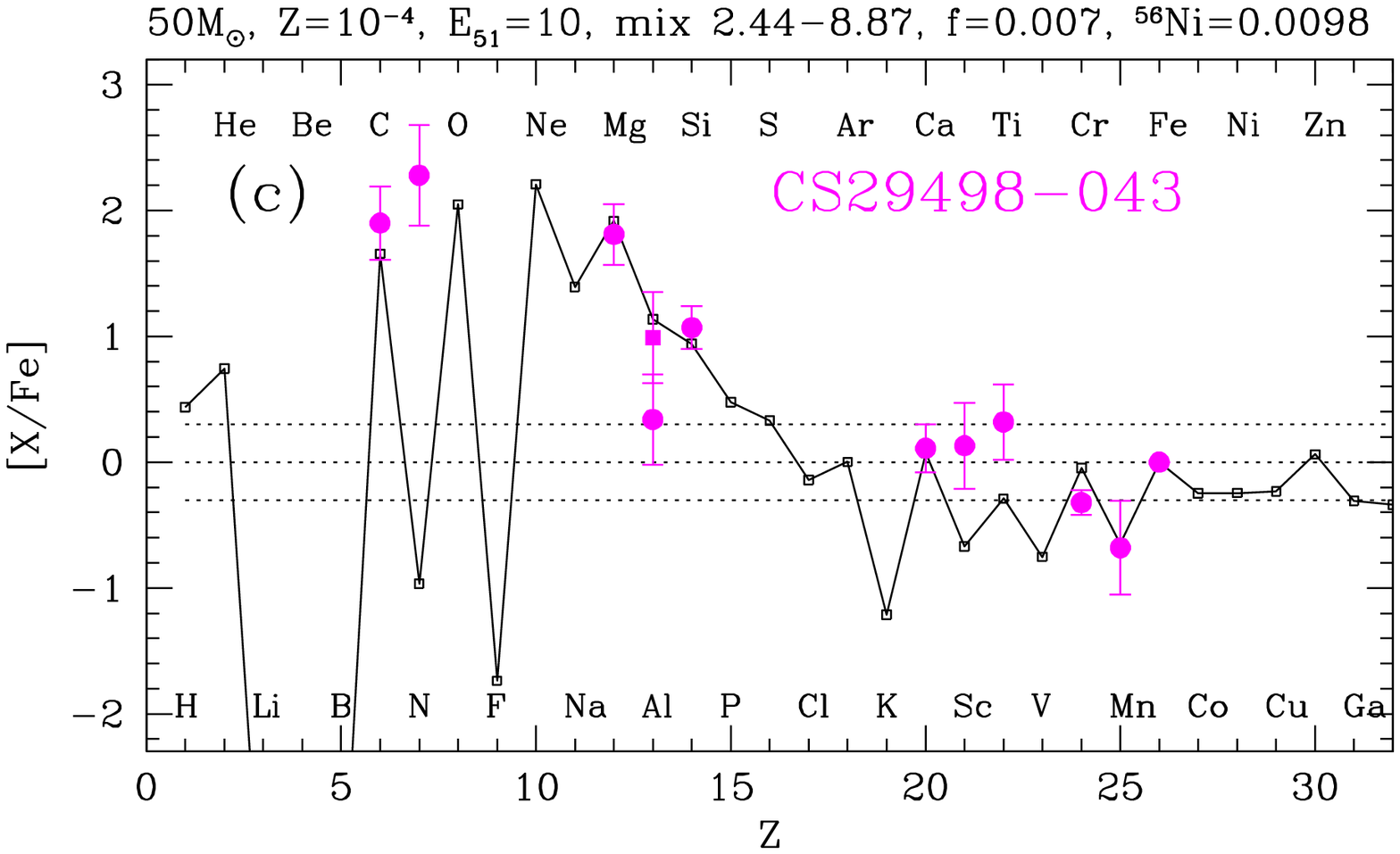}{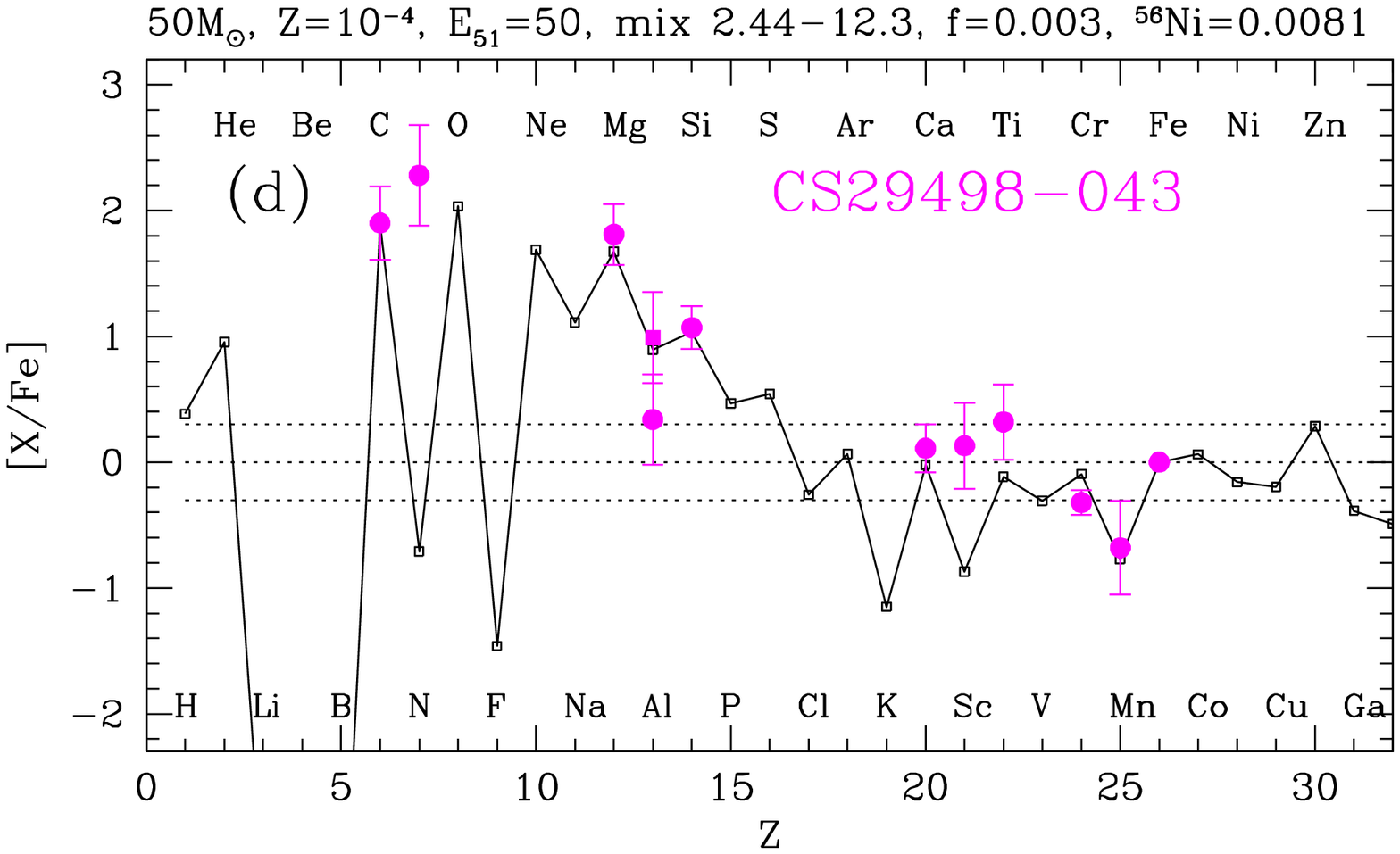}

\caption{Elemental abundances of CS29498-043
compared with theoretical supernova yield (solid lines).
The square and circle points in Al represent
the abundance data with and without the NLTE correction, respectively.
The model (a) with 25$M_\odot$, Z=0, and $E_{51}=1$ fits 
well with the observation except for N, Al, Sc, and Ti, 
while the same progenitor model with higher
explosion energy, $E_{51}=20$, (model (b)) overproduces Si/Fe ratio.  
More massive and more energetic models (c) and (d) also give
relatively good fits (except for N, Al, Si, and Ti).
\label{CS294}}
\end{figure}

 In Figure \ref{CS294}(a) we compare the observed data with the
theoretical model of ($M= 25M_\odot$, $E_{51}=$1, $Z=$0). This model
reproduces the observed abundance ratios for (C, Mg, Si, Ca, Mn)/Fe
within the error bars, shows a factor of $\sim 2$ deviation for (Ti,
Cr)/Fe, and significantly underproduces N, Al, and Sc.  For the Al/Fe
ratio, the possible NLTE correction of $\Delta$[Al/H] $\simeq 0.65$
shown by the solid square in Figure \ref{CS294}(a) could improve the
agreement between the model and observations.  Observational data for
O, Na, Co, and Zn are highly desired to constrain the model.

 Although Co and Zn data are not available, we can still constrain $E$
for given $M$ as follows: The relatively large Mg/Si ratio may
constrain the upper limit of the explosion energy, because the larger
explosion energy yields a smaller Mg/Si ratio.  As mentioned above,
the Mg/Si ratio decreases with increasing $M_{\rm mix}$(out) since Si
is produced in the deeper region than Mg. However, the Mg/Si ratio
does not decrease further if $M_{\rm mix}$(out) increases the Mg-rich
region.  For example, for the (Z=0, 25$M_\odot$) model, the explosion
energy $E_{51}=20$ is too large as shown in Figure \ref{CS294}(b). In
this model, $M_{\rm mix}$(out) is taken sufficiently large to maximize
the Mg/Si ratio, but that is too small compared with the observation.

 With our progenitor models (given in UN02), the large [C/Fe] of this
star is hard to explain with very massive models ($M> 30M_\odot$) with
relatively large $^{12}$C($\alpha,\gamma)^{16}$O rate, such as the
CF85 $^{12}$C($\alpha,\gamma)^{16}$O rate, because typically the C/O
ratio after He burning is smaller for more massive stars.  However as
shown in Figure 7(c), we can construct a model with larger [C/Fe] for
example by adopting a relatively small,
$^{12}$C($\alpha,\gamma)^{16}$O rate, such as the CF88 rate.  An
example is given in Figure \ref{CS294}(c) and (d).  The progenitor
model is the same as used in Figure \ref{CS22949}(c): the 50$M_\odot$
and $Z=10^{-4}$ model with the CF88 $^{12}$C($\alpha,\gamma)^{16}$O
rate.  Models 9(c) and 9(d) have a relatively small energy,
$E_{51}=10$, and a relatively large energy, $E_{51}=50$, respectively.
Here the mixing-fallback parameters are properly chosen to fit to the
C/Fe, Mg/Fe, Al/Fe and Si/Fe points. As mentioned above, a higher
energy model tends to produce larger Si/Mg ratio as well as larger
Ti/Fe, Co/Fe and Zn/Fe ratios.  The Ti/Fe ratio suggests that the
higher energy model $E_{51}=50$ might be better. However, we certainly
need Co and/or Zn data to constrain the energy. Note that, while
$E_{51}=10$ is not small for less massive supernovae, this is not so
strong explosion if the progenitor mass is as large as 50$M_\odot$ in
a sense that $E/M$ (velocity square of the ejecta) is not large.

 These examples show that it is difficult to constrain $M$ and $E$
independently, especially if the $^{12}$C($\alpha,\gamma)^{16}$O rate
is unknown. On the contrary, we may be able to constrain $E$ and $M$
independently if the $^{12}$C($\alpha,\gamma)^{16}$O rate, or the C/O
ratio after the He-burning is known.

 In summary, as far as the currently available data is used, the
abundance of CS29498-043 is better explained with nucleosynthesis in
moderately energetic SNe with mixing and large amount of fallback.
The mixing and fallback are necessary.  If the large [C/Fe] is
realized by a larger mass cut without the mixing and fallback, the
elements from Si to Ca are overproduced.  Smaller mass and less
energetic models, such as ($Z$=0, 20$M_\odot$, $E_{51}$=1), also show
a similar level of fit.  In order to estimate the explosion energy,
determinations of the Zn/Fe and Co/Fe ratios will be important for
this star.

 Tsujimoto \& Shigeyama (2003) also considered a supernova model with
small iron ejection to explain the abundance patterns of CS22949-037
and CS29498-043. The idea seems similar to ours, but they concluded
that the abundances of these stars cannot be explained by mixing (and
fallback) mechanism.  This conclusion, however, is based on the
assumption that these stars have the yields of low energy SNe and hence
the (Cr, Mn, Co)/Fe ratios are different from normal supernovae.
However, the small $^{56}$Ni mass (and thus large fallback mass) does
not necessarily mean the small explosion energy if the ejecta mass is
larger, because the mass of fallback matter is determined by the
balance between the explosion energy and gravity (e.g., Herant \&
Woosley 1994).  In fact, the large abundances of Zn and Co of
CS22949-037 cannot be produced in the low energy model.  Even for the
high energy model, the mixing of (Co, Zn) is necessary to enhance
(Co, Zn)/Fe in the ejecta and the fallback is necessary to reduce the
ejected $^{56}$Ni mass.

\subsection{CS22957-027}

This giant star is very C-rich ([C/Fe]=2.39) though the metallicty
[Fe/H]= $-$3.11 is not the smallest (Aoki et al. 2002a).  Compared
with two stars considered above, the Mg/Fe and Al/Fe ratios are
smaller. The s-process elements are not enhanced ([Ba/Fe]=$-$1.23) as
in the above two stars.

Without the Zn/Fe or Si/Fe data, it is hard to infer the explosion
energy. However, from the relatively large [Mn/Fe], we assume a
relatively small explosion energy. The most important character of
this star is the large [C/Fe] and the large [C/Mg].  These two large
ratios require relatively large scale mixing and a large amount of
fallback.  We adopt a model with ($M=25M_\odot$, $Z=0$, $E_{51}$=1)
with mixing in the region of $M_r=2.1$ to 4.8$M_\odot$ and the
ejection factor $f$=0.003.

 As shown in Figure 10(a), this model reproduces the observed
abundance ratios (C, Al, Mg, Ca, Mn, Ni)/Fe within the error bars
(being as good as in Figure \ref{CS294}(a)), shows a factor of $\sim
2$ deviation for (Ti, Cr)/Fe, and significantly underproduces N.
Further observational data for O, Na, Si, Co, and Zn are highly
desired to constrain the model.

If the explosion energy is increased, the synthesized Mg mass
generally increases, which makes the agreement with the observations
worse.  However, if the progenitor mass is increased as well as the
explosion energy, a similarly good fit can be made (e.g. Figure 10(b)
and 10(c) for $Z=10^{-4}$, 50$M_\odot$, $E_{51}$=10 and $E_{51}=50$
models), though a large scale mixing and a quite small ejection factor
($f$=0.0005) is required.

\subsection{CS31062-012}

There is a subclass of C-rich EMP stars which show some enhancement of
the s-process abundances (e.g., Ryan 2002). CS31062-012 is one of such
examples (Aoki et al. 2002c), which has [Ba/Fe] = 1.98 being more than
two orders of magnitude larger than that of CS22957-027 and
CS29498-043.  Although Ba is also produced in the r-process, the
abundance pattern of other neutron capture elements suggests that this
Ba is s-process origin. In Aoki et al. (2002c) the abundance of
another star of this type, CS22898-027, is given and its abundance
pattern is similar to CS31062-012.

\begin{figure}
\epsscale{.5}
\plotone{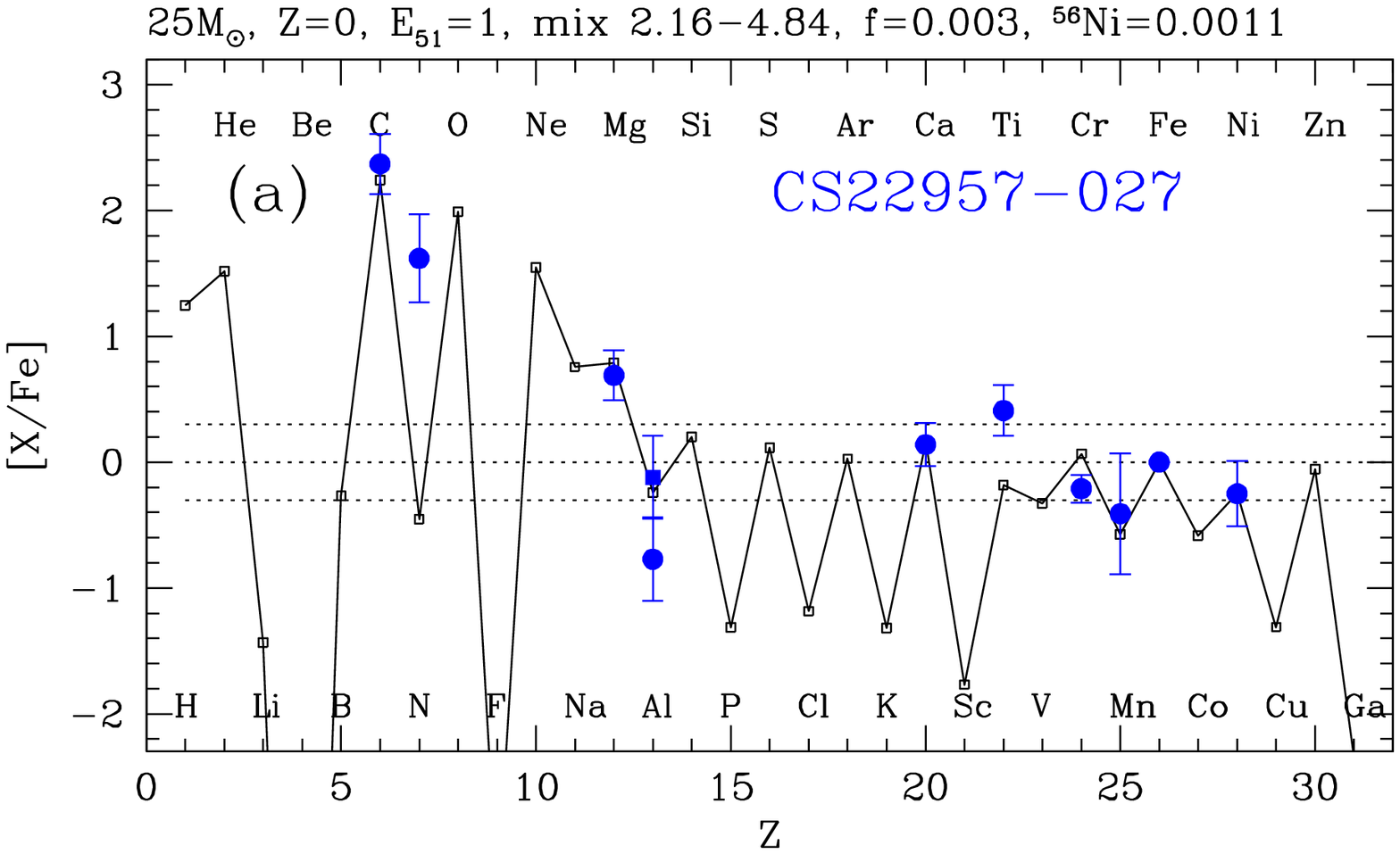}

\epsscale{1.}
\plottwo{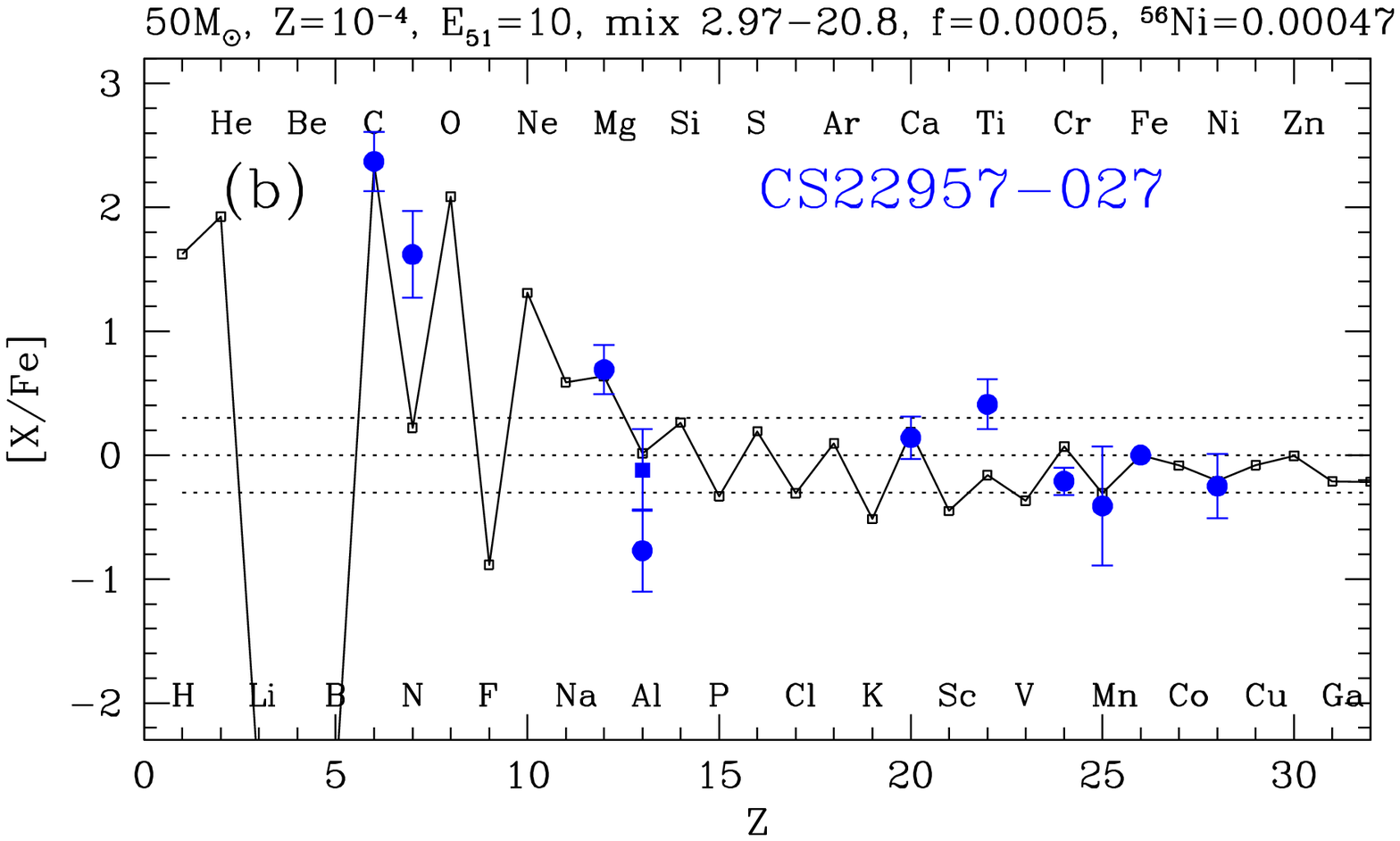}{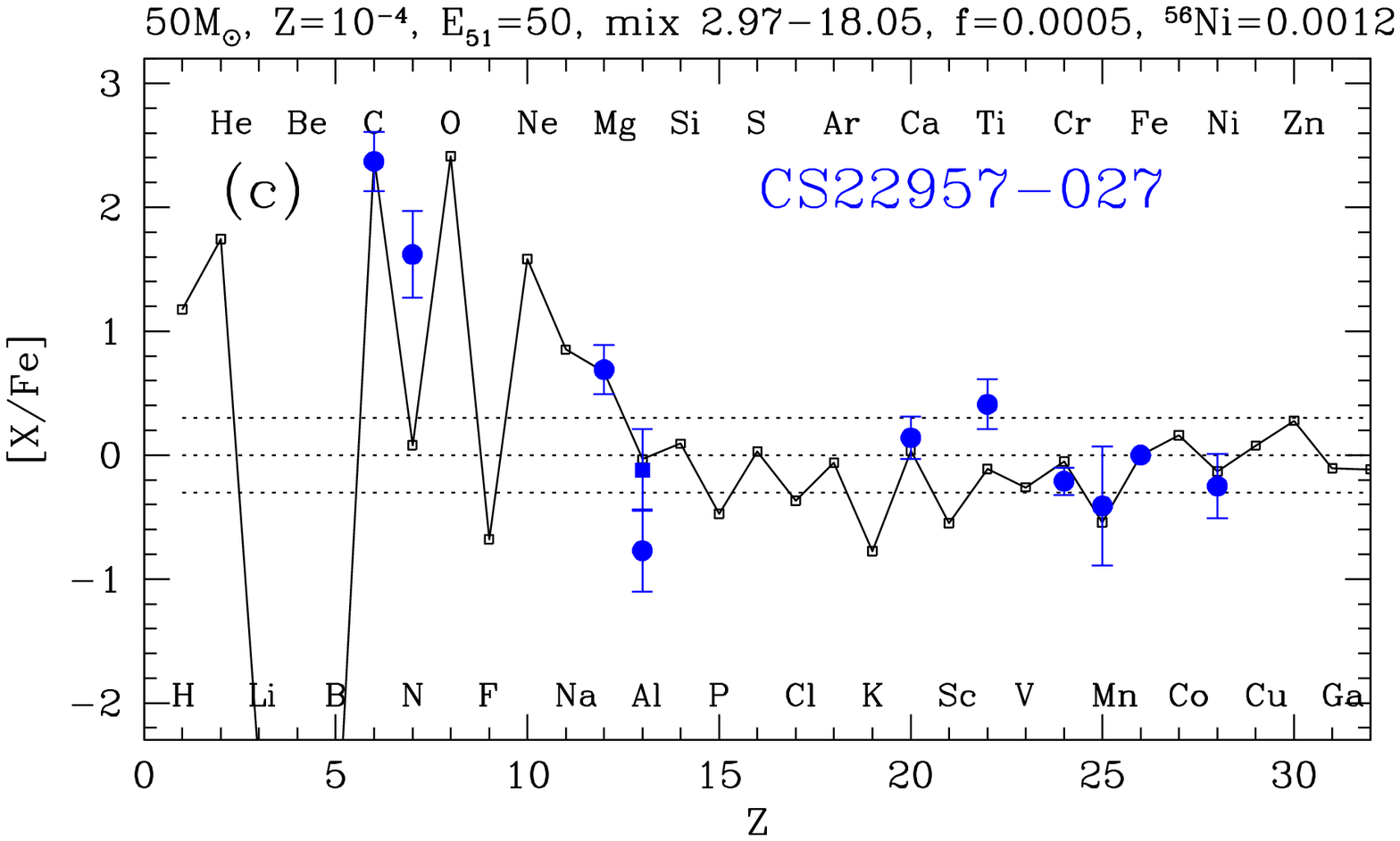}

\caption{Elemental abundances of CS22957-027
compared with theoretical supernova yield (solid lines).
The model (a), 25$M_\odot$, Z=0, with $E_{51}=1$, and
more massive and more
energetic models (b) and (c) all give similar results.
\label{CS22957}}
\end{figure}

\begin{figure}
\epsscale{.6}
\plotone{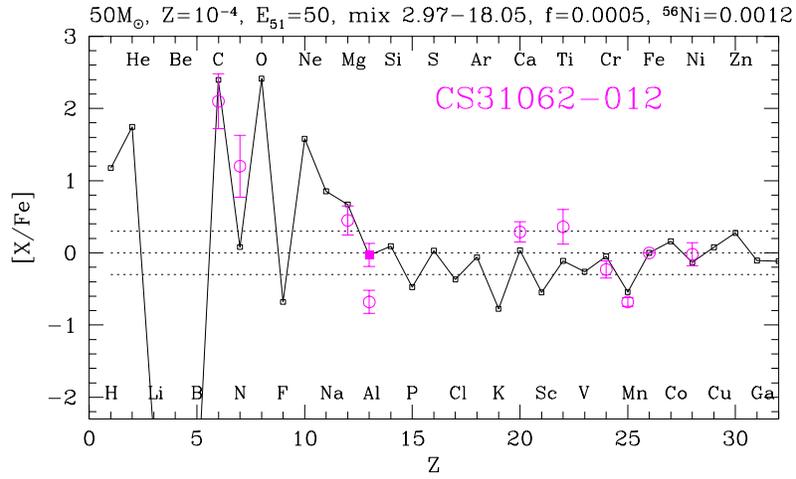}

\caption{Elemental abundances of CS31062-012
compared with theoretical supernova yield (solid lines).
The model is 50$M_\odot$, Z=10$^{-4}$, with $E_{51}=50$.
This is the same model for CS22957-027
in Figure 10(c). Interestingly, this star has the signature
of the s-process elements, while CS22957-027 shows no enhancement
of the s-process elements. 
\label{CS310}}
\end{figure}

In Figure \ref{CS310}, we compare the observed abundance pattern of
CS31062-012 with the same theoretical model as in Figure
\ref{CS22957}(b). This shows that the same model fits equally-well to
both CS22957-027 and CS31062-012 for the abundance ratios of (C, Al,
Mg, Ca, Mn, Ni)/Fe.

Then the question arises what makes the difference in the abundance of
Ba?

 One possible source of the s-process elements is the mass transfer
from an AGB companion star. However, as mentioned in Introduction,
some of these kinds of stars have no indication of binary companions,
although these companion stars once transferred masses might have been
departed to the un-observed distances (Ryan 2002).  Another
possibility is the s-process during the pre-SN evolution. We note that
CS31062-012 and other Ba-rich stars are somewhat more metal-rich than
Ba-normal stars.  For example, [Fe/H] of Ba-rich metal-poor stars,
CS31062-012, CS22898-027, LP625-44 and LP706-7 are [Fe/H]= $-2.55,
-2.26, -2.71$ and $-2.74$, respectively. So the question is whether a
SN from such a metal-poor as [Fe/H] $\sim -3$ can have an ejecta with
[Ba/Fe] $\sim$ 2.

Observations show that the [Fe/H] $\sim -3$ stars typically have
[Ba/Fe] = $-1.5 \sim -0.5$ (e.g., NRB01). Then in order for the SN
ejecta to have [Ba/Fe]$\sim$2, the enhancement factor of Ba, which is
the ratio of initial (pre-stellar evolution) to final (post-SN) Ba
masses in the ejecta, has to be larger than $10^{2} - 10^{3.5}$. Note
that in our model for CS31062-012, the mass of ejected Fe is so small
that [Fe/H] of the ejecta mixed with circumstellar matter changes only
a little after the SN explosion. We need s-process nucleosynthesis
calculations for such metal-poor stars to judge if the enhancement
factor of this amount is possible.

\subsection{HE 0107-5240}

This star has the lowest [Fe/H] ($\simeq -5.3$) among the observed EMP
stars (Christlieb et al. 2002). Understanding the origin of this star
has special importance, because it has been argued that low mass star
formation is prohibited below a certain metallicity (e.g., below
[Fe/H] $\sim -4$; Schneider et al. 2002) due to inefficient gas
cooling.

 In UN03 we discussed that this star is the second generation star,
whose abundance pattern can be understood by the enrichment of
population III core-collapse supernovae as is similar to other EMP
stars discussed in this paper.  For HE0107-5240, the ejecta is Fe-poor
but C-rich (see below), then the low mass star formation can be
possible with the C, N, O cooling.  More detailed implication about
the formation of this star is given in UN03 (also Schneider et
al. 2003; Bonifacio et al. 2003; Shigeyama et al 2003; Limongi et al.
2003).  Here we briefly explain how the abundance of this star can be
explained in our model.

 This star has extremely high C/Fe ratio, [C/Fe] $\simeq 4$, which
requires very small $^{56}$Ni ejection, e.g., $M_{^{56}\rm Ni} \simeq
8 \times 10^{-6}M_\odot$ in the 25$M_\odot$ SN model.  Contrary to the
large C/Fe ratio, the Mg/Fe ratio is almost solar. This requires that
the mixing region is extended to the entire He-core, and only tiny
fraction of the matter, 0.002\%, is ejected from this region.  The
explosion energy of this SN model is assumed to be relatively low,
$E_{51}=0.3$, which is necessary to reproduce the subsolar ratios of
[Ti/Fe] $\simeq -0.4$ and [Ni/Fe] $\simeq -0.4$.  Since the large
fallback mass does not necessarily require a small explosion energy,
and constraints from Ti and Ni on the energy are not as strong as Zn,
it is important to measure Zn/Fe to see if it is as small as predicted
by the small energy model and to obtain a stronger constraint on the
explosion energy.  As in other models described in this paper, the
abundances of N and Na may be enhanced during the EMP star evolution.

\begin{figure}
\epsscale{.6}
\vskip -3cm
\plotone{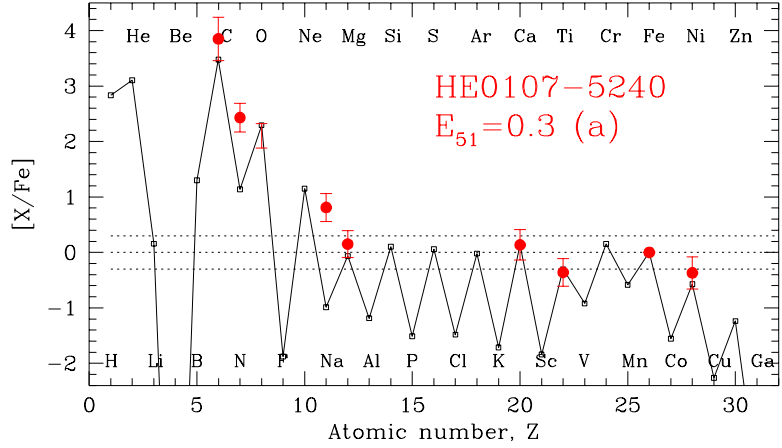}
\vskip -2cm
\plotone{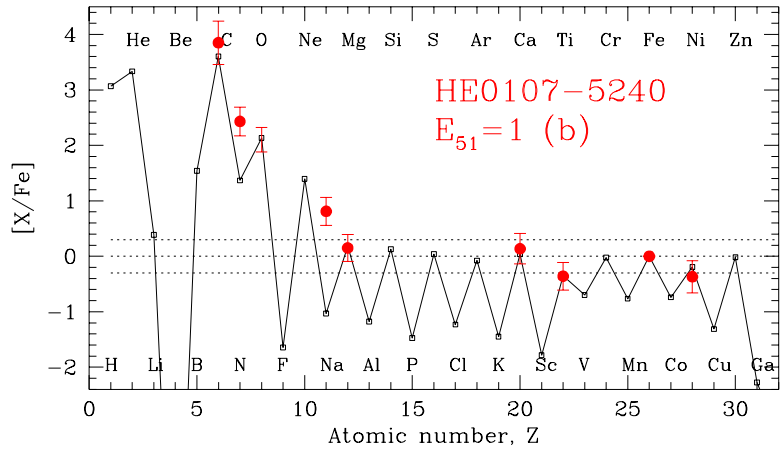}

\caption{
Elemental abundances of HE0107-5240, compared with a theoretical
supernova yield. HE0107-5240 is the most
Fe-deficient, C-rich star yet observed, with [Fe/H]$= -5.3$ and very
large ratios of [C/Fe] and [N/Fe].  
The observed data shown by the filled circles with error bars are 
taken from Christlieb et al. (2004).
For [O/Fe], we show the range suggested by Bessel et al. (2004). 
Here the supernova models are the
population III 25$M_\odot$ core collapse SNe, with explosion
energies $E_{51}$ = 0.3 (a) and $E_{51}$ =1.0 (b).  
In these models, only a small fraction
of the materials in the mixed region, 0.007\% (a) and 0.004\% (b), are ejected.
The ejected Fe (or $^{56}$Ni) masses, 
$2.5 \times 10^{-5}M_\odot$ (a) and $1.7 \times 10^{-5}M_\odot$ (b), 
are so small that the large C/Fe ratio can be realized.
\label{HE0107}}
\end{figure}

 In Figure \ref{HE0107}(a) we compare the model with the updated observed
abundance pattern (Christlieb et al. 2004).  The model is basically
the same as adopted in UN03, but the following two points: (1) $Y_e$
in the complete and incomplete Si-burning regions are modified to keep
the consistency in this paper, and (2) mixing-fallback parameters are
slightly changed to reduce the [O/Fe].  The effect of (1) is marginal
and the results are essentially identical to the original $Y_e$ model:
the Co abundance is a little larger with this modification, but other
yields are roughly identical to the model in UN03.  We try to reduce
the [O/Fe] because the newly obtained observed data for [O/Fe]
(Bessel, Christlieb, \& Gustafsson 2004) seems slightly smaller than
the value of the model in UN03. The value in UN03 was [O/Fe]= +2.9,
and the value in Bessel et al. (2004) is [O/Fe]= +2.3 for a
plane-parallel LTE model atomosphere. They also estimated that
systematic errors due to 3D effects may reduce [O/Fe] by 0.3 to 0.4
dex or even more.  In Figure 12 (a), we reduce the [O/Fe] value by simply 
increasing the ejection factor to $f= 7\times 10^{-5}$ 
(it was $2\times 10^{-5}$ 
in UN03). Other parameters are $M_{\rm cut}$(ini)=1.90$M_\odot$ and
$M_{\rm mix}$(out)=6.01$M_\odot$.  These parameter changes 
lead also to smaller [C/Fe]
but it is within the error-bar given in Christlieb et al. (2004). Now
the new values are [C/Fe]=+3.5, [N/Fe]=+1.1, and [O/Fe]= +2.3. These values in
UN03 were [C/Fe]=+4.0, [N/Fe]=+1.7, and [O/Fe]=+2.9.
If the C/O ratio is larger, it is better reproduced by a slightly higher 
explosin energy model as shown by the model with $E_{51}=1$ in 
Figure 12(b).
For this case, [C/Fe]=+3.6, [N/Fe]=+1.4, and [O/Fe]=+2.1.
As described above, $E$ cannot be well-constrained without
the Co/Fe and Zn/Fe data. 

\section{Summary and Discussion}

 We have compared the abundances of EMP stars with nucleosynthesis
yields of individual supernovae, and obtained constraints on the model
parameters from the comparisons.  Previously CL02 showed that
theoretical yields (with the ``High'' $^{12}$C($\alpha,
\gamma$)$^{16}$O rate like ours) and observations do not match well,
especially for the (Ti, Cr, Co, Ni)/Fe and (Si, Ca, Al)/Mg ratios.
However, CL02 did not consider high energy explosion models and the
mixing-fallback process.

 By using the ``High'' $^{12}$C($\alpha, \gamma$)$^{16}$O rate,
however, we can reproduce the observed (Si, Ca, Al)/Mg ratios,
probably because of the effects of high-energies and the
mixing-fallback in the explosion.

\subsection{Zn, Co, Mn and Ni}

\subsubsection{Production in Hypernova models}

 The underabudances of Zn and Co in ordinary SN models are
significantly improved in the large explosion energy models as shown
in UN02. The Co abundance was, however, still lower than the
observation. In this paper we point out that the abundance of Co is
significantly enhanced for $Y_e \gsim 0.5$. The abundance of Mn is
also sensitive to $Y_e$ and in our model $Y_e \simeq 0.4995-0.4997$
gives a relatively good fit to the observations. Since Mn is mostly
produced in the incomplete Si-burning region and Co is produced in the
complete Si-burning, best fit to the observations may be with the
inversion of $Y_e$ in the Si-burning region. This may sound
unrealistic, but the most recent simulations of core-collapse SNe
indeed predict such inversion of $Y_e$ by the effect of neutrino
processes (e.g., Liebend\"orfer et al. 2003; Janka, Buras, \& Rampp
2003).  We note that Co can be enhanced by the $Y_e$ effect but the
large energy is still necessary to explain the observed large Co/Fe
ratios observed in EMP stars.

 CL02 found underproduction of the Ni/Fe ratio in all their models.
However, we have not found difficulties in producing large Ni/Fe
ratios; Ni is enhanced in relatively deeper with smaller $Y_e$
region. In summary, the previous bad fits of these elements to the
observations can be significantly improved by considering the
high-energy models and the variation of $Y_e$.

\subsubsection{Other possibilities for $^{64}$Zn production ?}

 Hoffman et al. (1996) have proposed another site to produce
$^{64}$Zn, which is the neutrino-powered wind just after the shock is
launched in the deepest layers of the star.  However, the total yield
for this model has not been given, so that it is not clear if the
proper amount of $^{64}$Zn is ejected without overproducing unwanted
elements such as $^{56-62}$Ni, especially because the Zn production
site of this model has very low $Y_e$.  In the model of Hoffman et
al. (1996), Zn is mostly produced as the neutron-rich isotope
$^{64}$Zn, while in our model, the dominant Zn is the decay of a
(neutron-proton) symmetric isotope $^{64}$Ge.

 The observed features may be in favor of our models.  First, relative
homogeneity of [Zn/Fe] in EMP stars (e.g., Cayrel et al. 2003) suggest
that the production site of Zn and r-process elements are different,
because only in few EMP stars, r-process elements are enhanced.
Second, it appears that the Zn/Fe enhanced stars are also Co/Fe
enhanced.  This is easily understood in our models, because Zn and Co
are produced almost in the same region in our hypernova model.  On the
other hand, in the model of Hoffman et al. (1996) the most Zn-rich
region is different from the most Co-rich region.

\subsection{N, Na, Ti, Cr and Sc}

 Of course, our supernova yields do not fit to the observations
perfectly for all elements.  N, Na, Ti, Cr and Sc are the examples for
which the discrepancies between theory and observations are relatively
large. Among them, the N abundance can be explained rather easily by
the CN-cycle and first dredge-up in the EMP star. Also, N and Na may
be synthesized either in the supernova progenitors or in low-mass EMP
stars, by the mixing of hydrogen into the He shell-burning layer
through an extra mixing process (Iwamoto, Umeda, \& Nomoto 2003).  The
underabundance of Sc and Ti may be enhanced in the ``low-density
model'' (see Appendix) or aspherical or jet like explosions (Maeda \&
Nomoto 2003a; Nagataki 2000).  At present we do not have explanations
about the overproduction of Cr.

\subsection{Why [Fe/H] and abundance of Fe-peak elements are related? } 

 We have shown that the observed trend in the abundance of Fe-peak
elements with [Fe/H] can be understood in the supernova induced star
formation model, in which [Fe/H] is estimated by equation (2), since
the EMP stars enriched by high-energy supernovae tend to have lower
[Fe/H]. The [Zn/Fe] and [Co/Fe] increase with increasing $E$, while
[Mn/Fe] and [Cr/Fe] decrease with $E$. As a result [Zn/Fe] and [Co/Fe]
increase with decreasing [Fe/H], while [Mn/Fe] and [Cr/Fe] decrease.
This success supports the idea of the supernova induced star formation
model, and the idea that EMP star abundances are mostly determined by
a single SN.

\subsection{Mixing-fallback and the mass ratios between heavy and light elements} 

 With the mixing-fallback mechanism, the abundance ratios between
relatively light and heavy elements, such as Mg/Fe and Al/Fe, can be
smaller to fit to the observation. Not only the fallback but the
mixing during the explosion is necessary; without mixing complete
Si-burning products such as Zn and Co are not ejected, leading too
small [Zn/Mg] and [Co/Mg].  The fit to the abundance of typical
[Fe/H]$\sim -3.7$ stars (NRB01) is significantly improved when the
matter below the Si-burning region is mixed and only 10\% of the
matter ejected from this region.  We note that similar effect occurs
for the aspherical or jet-like explosion (Maede \& Nomoto 2003a,b). In
reality it may be the combined effects of mixing, fall-back and
asphericity.

\subsection{Variations in the C-rich EMP star abundance; 
degree of Mixing-fallback and explosion energy} 

 We have shown that the abundance patterns of not only the typical EMP
stars but also C, N-rich EMP stars can be explained by the
core-collapse supernova yields with the different explosion energy and
the degree of mixing-fallback.  In general, C, N-rich EMP stars can be
formed in the ejecta of ``faint'' supernova that eject little Fe
because of the large amount of fallback.  Such SNe are not
hypothetical, but have been observed (Nomoto et al. 2002 for a
review).  The prototype is SN1997D, which was modeled as a low energy
($E_{51}=0.4$) explosion of a 25$M_\odot$ star (Turatto et al. 1998).

 As progenitors of the C-rich EMP stars, some of these ``faint''
supernovae are likely to have low energies (e.g., HE0107-5240) but
some might have high-energies (e.g., CS22949-037).  To estimate the
explosion energy, the observations of the Co/Fe and Zn/Fe ratios are
highly desirable.  In the low energy models, Zn/Fe and Co/Fe are
always underproduced for any parameter choices we considered.  Only in
the high-energy models, Zn/Fe and Co/Fe can be large in our models.
As shown in the appendix, the Co/Fe ratio may be enhanced in the
low-density model, but large explosion energy is still required to
explain the observed ratios. We note that models in the set L of CL02
produce large enough [Co/Fe] but [Ni/Fe] is significantly
underproduced and [Zn/Fe] is not given.  If [Zn/Fe] or [Co/Fe] are not
observed, some constraints on the energy can be obtained from the
Si/Mg ratio.

 There are variety of the abundance pattern in C, Mg, Al, Si and S
among the C-rich EMP stars.  We have shown that these varieties can be
explained with different $M_{\rm mix}$(out). This is one advantage of
the mixing-fallback model compared with other explanations. For
example, in the mass transfer model from the companion AGB stars, it
is difficult to explain the overabundance of Si.  For the
mixing-fallback model, what causes the difference of $M_{\rm
mix}$(out)? Since we do not know how ``hypernovae'' explode and how
much asphericity and rotation exist, it is currently not possible to
answer this question. However there are some observational
suggestions:
 As shown in Section 2, the abundance of typical EMP stars can be
explained by high-energy SNe with relatively small $M_{\rm mix}$(out).
On the other hand, the abundance of HE0107-5240 can be explained by
low energy SNe with relatively large $M_{\rm mix}$(out). These facts
suggest that the relative position of $M_{\rm mix}$(out) decreases
with the explosion energy. This is consistent with the intuition that
for a smaller explosion energy the velocity of the ejecta is lower,
thus leading a larger amount of fallback (larger $M_{\rm mix}$(out)).
However, we do not expect one to one correspondence between the energy
and the amount of fallback, because geometry of the explosion and the
rotational speed should affect the amount of fall-back.

 One may wonder how sensitive of our results to the mixing parameters,
$M_{\rm cut}^{\rm ini}$ and $M_{\rm mix}$(out).  In this paper we
choose $M_{\rm cut}$(out) to maximize the Zn/Fe ratio (see Section 3.1
for more detail).  Variations of $M_{\rm cut}^{\rm ini}$ changes the
Zn/Fe ratio, but this is not so sensitive to the ratio as long as
$M_{\rm cut}^{\rm ini}$ is located deep inside of the complete
Si-burning region.  The variation of $M_{\rm mix}$(out) changes the
abundance ratios between various elements. Using the model for
CS29498-043 (Figure 9, and its abundance distribution is shown in
Figure 13), a 50$M_\odot$ and $E_{51}=50$ model, we show in Figure 14,
how [C/Mg] and [Mg/Si] varies as a function of $M_{\rm mix}$(out). We
also show in this figure that the ranges corresponding to the observed
error bars of these abundance ratios (red lines for [Mg/Si] and blue
lines for [C/Mg]).  The region in which both the observations are
satisfied is shown by the region labeled `Allowed Region'.  For this
model, the allowed region of $M_{\rm mix}$(out) is about $\Delta M =
3.5 M_\odot$ and thus we do not need fine tuning of the parameter to
satisfy the observation. In most cases also, we do not need fine
tuning of $M_{\rm mix}$(out).

\subsection{Ejected mass of Mg and mixing-fallback}

 We have shown that the observed C-rich EMP stars have various
[Mg/Fe], but it can be explained with the mixing-fallback model. In
our model the observed large [C/Fe] is realized by the small ejected
Fe mass, corresponding to a faint SN. If the mixing-fallback region
does not extend beyond the Mg layers, the Mg/Fe ratio is larger for
smaller ejected Fe mass. On the other hand, if the mixing-fallback
region extends beyond the Mg layer, the Mg/Fe is not necessary large
for a little Fe-ejection because Mg ejection mass can be also small.

 Shigeyama, Tsujimoto \& Yoshii (2003) claimed that the abundance
pattern of C-rich EMP stars cannot be explained with the faint SN
model, unless [Mg/Fe] is also large.  This is because they assumed
that the ejected Mg mass is only the function of the main-sequence
mass of the SNe (e.g., Shigeyama \& Tsujimoto 1998).  In other words,
they implicitly assume that the mixing-fallback region does not reach
the Mg layers.  However, the ejected Mg mass is not the only function
of $M$.  Even for stars with the same mass, the Mg mass can defer
depending on the explosion energy, mass-cut and the degree of
mixing-fallback.  Thus the diversity of [Mg/Fe] can be explained with
the mixing-fallback model.

\subsection{Degeneracy in the $E$ and $M$: Can we constrain $M$?}

  If observations provide the Co/Fe or Zn/Fe ratios, we can infer
whether the explosion was of relatively high or low energy. However,
we can constrain only a set of the explosion energy, $E$, and the
progenitor's mass, $M$, to fit the observations, because more massive
and larger energy models give similar abundance ratios.  Typically, in
a more massive star the C/O ratio after the He-buring is
smaller. Since the final yields depends on the C/O ratio (e.g., Weaver
\& Woosley 1993; Nomoto \& Hashimoto 1998; Imbriani et al. 2001), one
may be able to constrain the progenitor mass by closely comparing
observations with theoretical yields.  Unfortunately, the C/O ratio is
quite sensitive to the uncertain $^{12}$C($\alpha, \gamma$)$^{16}$O
rate and the treatment of convection, so at this moment we are not
able to determine $M$ alone.  Further detailed modeling of individual
and chemical evolution of galaxies, and comparison with the
observations are necessary to determine $M$ and $E$ independently.

 Although $E$ and $M$ degenerate, we can still set the upper limit to
$M$ as $ M \lsim 130M_\odot$, because above which the stars become
pair-instability supernovae (PISNe) and the yields do not fit to the
EMP star abundances (see the next subsection).

\subsection {Pair-Instability Supernovae}

 We have shown that the ejecta of core-collapse supernova explosions
of $20-130 M_\odot$ stars can well account for the abundance pattern
of EMP stars. In contrast, the observed abundance patterns cannot be
explained by the explosions of more massive, $130 - 300 M_\odot$
stars. These stars undergo PISNe and are disrupted completely (e.g.,
UN02; Heger \& Woosley 2002), which cannot be consistent with the
large C/Fe observed in HE0107-5240 and other C-rich EMP stars.  The
abundance ratios of iron-peak elements ([Zn/Fe] $< -0.8$ and [Co/Fe]
$< -0.2$) in the PISN ejecta (Figure 15; UN02; Heger \& Woosley 2002)
cannot explain the large Zn/Fe and Co/Fe in the typical EMP stars
(McWilliam et al. 1995; Primas et al. 2000; Norris et al. 2001) and
CS22949-037.  Therefore the supernova progenitors that are responsible
for the formation of EMP stars are most likely in the range of $M \sim
20 - 130 M_\odot$, but not more massive than 130 $M_\odot$.

\bigskip

 Yield tables for some of the models are shown in
http://supernova.astron.s.u-tokyo.ac.jp/\textasciitilde umeda/data.html 
Other yields and related information can be uploaded upon requests.

\acknowledgments

 We would like to thank N. Iwamoto, K. Maeda and M. Y. Fujimoto for
useful discussion.  This work has been supported in part by the
grant-in-Aid for Scientific Research (15204010, 16042201, 16540229)
and the 21st Century COE Program of the Ministry of Education,
Science, Culture, Sports, and Technology in Japan.


\bigskip

\begin{center}
{\bf Appendix: A low density model}
\end{center}

 In this appendix, we present the abundance patterns of the
``low-density models''.  In this model, the density during explosive
burning is lower than the original model, thus enhancing the
$\alpha$-rich freeze-out. As a result, Sc/Fe, Ti/Fe, Mg/Fe and also
Ca/Fe, Co/Fe, and Zn/Fe ratios are enhanced to be in a better
agreement with the observed ratios.  Here, we artificially reduce the
density of the progenitor model by a factor of three.  In this paper,
we do not specify how such a low density is realized, but we propose
one possibility: Recent studies on SNe have revealed that a certain
class of SNe explode very energetically (hypernovae) leaving a black
hole behind.  Suppose that such a hypernova explosion is induced by
the jets perpendicular to the accretion disk around the black hole. If
a relatively weak jet expands the interior of the progenitor before a
strong jet forms a strong shock to explode the star, major explosive
burning takes place in lower density than in the original progenitor.

 Figures 16(a) shows a low-density model compared with the elemental
abundances of typical EMP stars at [Fe/H]$\sim -3.7$ used in Figure 6.
In this model, the density of the presupernova progenitor is reduced
to 1/3 without changing the total stellar mass. $Y_e$ of this model is
not modified from the original value. Compared with the original
density model shown in Figure 16(b), we find that in the low-density
model the Sc/Fe ratio is significantly larger, also, Ca, Ti, V, Mn,
Co, Ni and Zn to Fe ratios are larger, being in a better agreement
with the observational points. Slight deviations in [Mn/Fe] and
[Co/Fe] from the observations may be resolved by varying $Y_e$. Then,
except [Cr/Fe], all the data shown here can be fitted in this model.

 In Figure 16(c) we show the abundance pattern when the density for
the post-process is artificially reduced to 1/3. In this case, also
Sc/Fe, Ti/Fe and Co/Fe ratios are larger than in original models.
This example demonstrates that the density during the explosion is
important for the abundance ratios of Sc/Fe and Co/Fe and thus a
realistic explosion simulation is important for these elements.

 An important question is, since the Zn/Fe ratio is larger in the
low-density model, whether high-energy explosions are required to
explain the observed large Zn/Fe (UN02) for the low-density models. As
shown in Figure 16(d), the low-density alone is difficult to realize
the large [Zn/Fe] even for $E_{51}=10$ for the 25$M_\odot$ model. This
model realizes the large [Co/Fe] but over-produces [Ca/Fe] and
[Ni/Fe], thus being much worse agreement with observation than the
higher-energy models.

 Low-density models also give better fits to the C-rich EMP star data,
because Ca, Sc, Ti, Co and Zn to Fe ratios are enhanced without
changing the abundance of lighter elements. In Figure 17, we show a
$25M_\odot$ model with a reduced progenitor density, compared with the
CS22949-037 data. This is the low-density version of Figure 7(a), and
we find again that Sc/Fe and Ti/Fe are enhanced and Co/Fe and Zn/Fe
are close to the observed value even with the original value of $Y_e$.



\vskip -10cm

\clearpage

\clearpage

\clearpage

\clearpage

\clearpage

\clearpage

\clearpage

\clearpage

\clearpage

\begin{figure}
\epsscale{1.}
\plotone{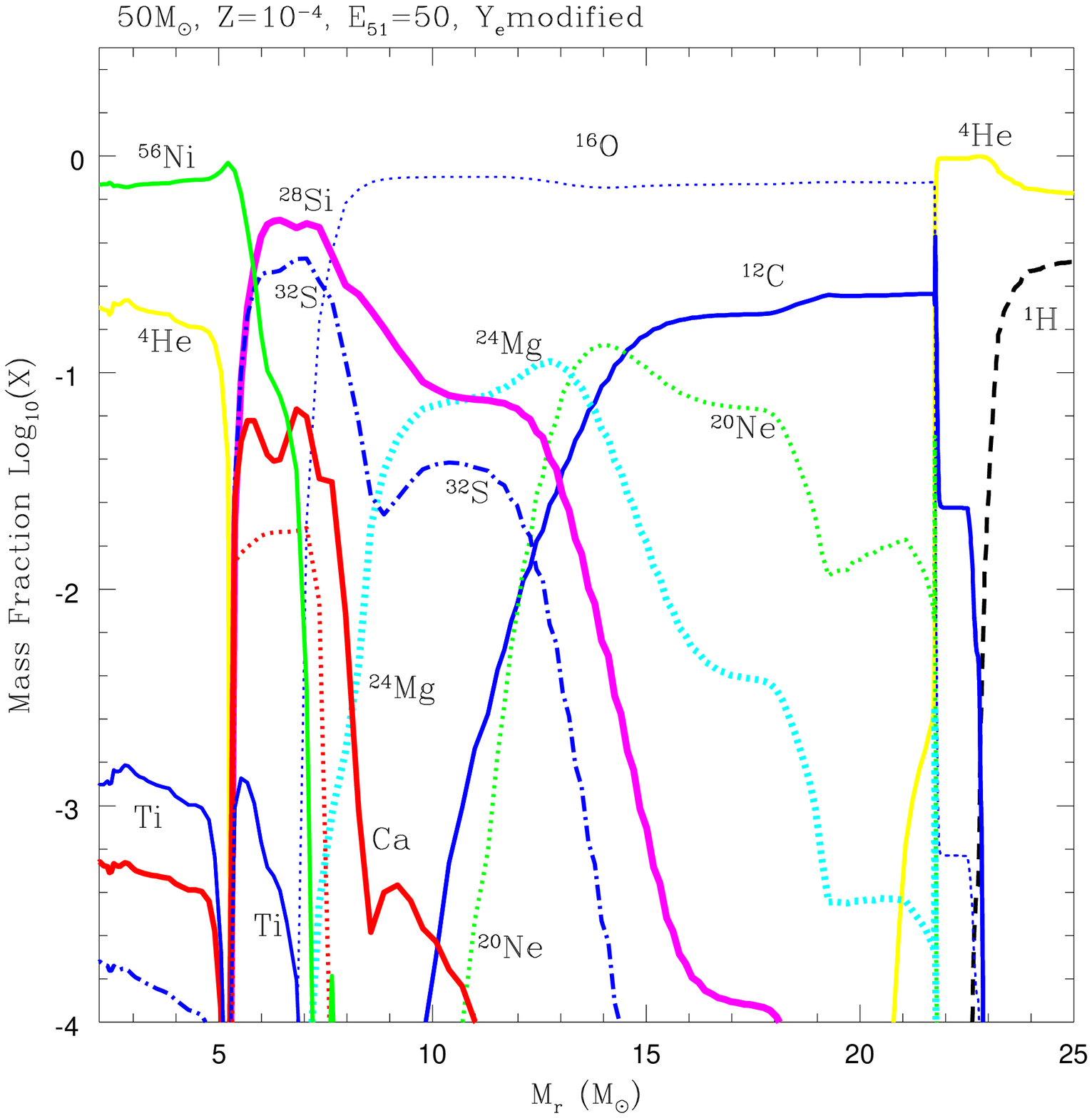} 

\caption{Abundance distribution after SN explosion of a 50 $M_\odot$
star with $E_{51}=50$. $Y_e$ in this model is modified to 0.5001 and
0.4997 in the complete and incomplete Si-burning regions, respectively.
\label{50z-4e50ye5001-4997}}
\end{figure}

\clearpage

\begin{figure}
\epsscale{1.}
\plotone{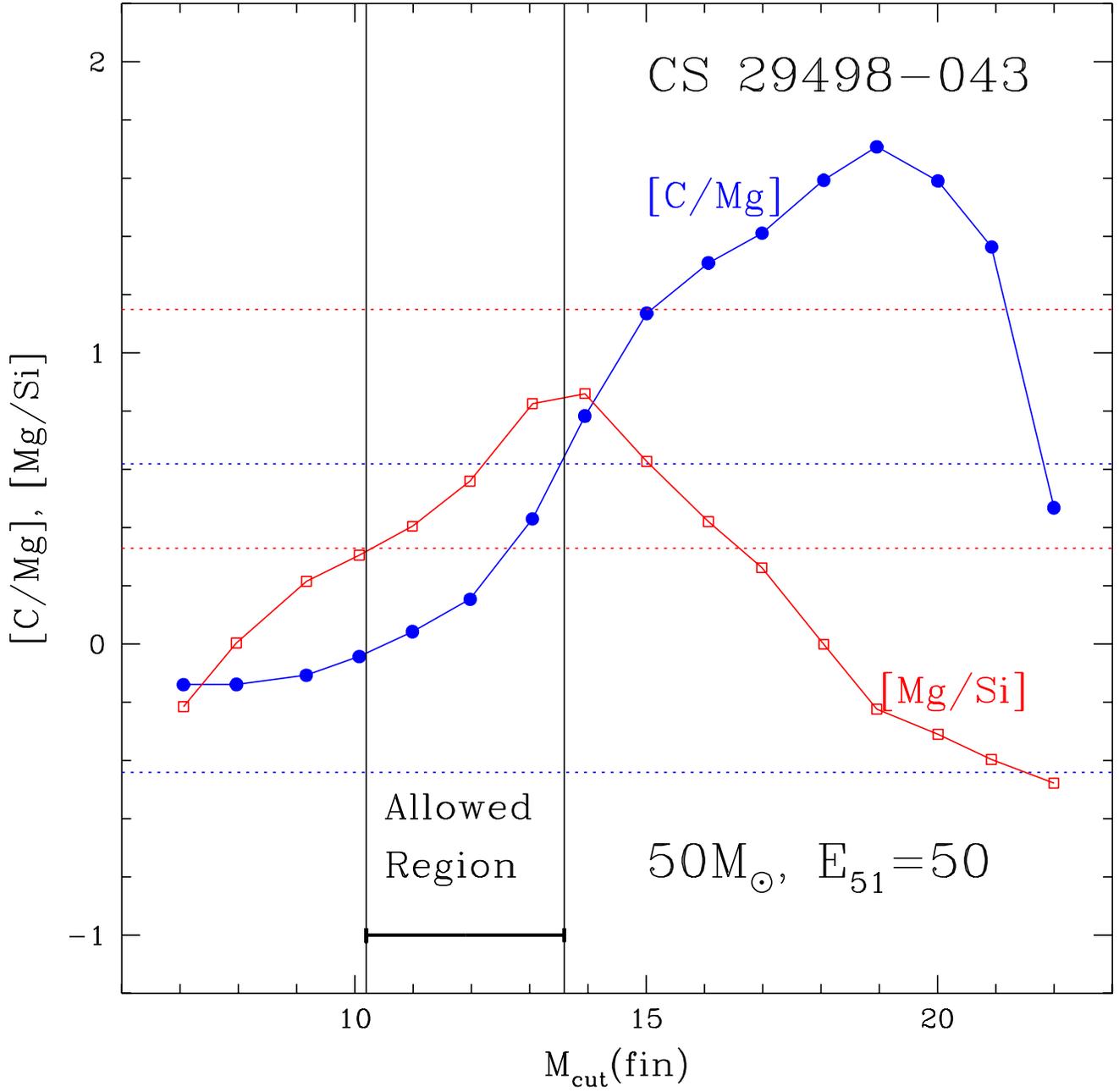} 

\caption{The abundance ratios
[C/Mg] and [Mg/Si] of the model in Fig.\ref{50z-4e50ye5001-4997}
(50$M_\odot$, Z=$10^{-4}$, $E_{51}=50$)
as a function of final mass-cut $M_{\rm cut}^{\rm fin}$.
The initial mass-cut and the ejection factor
are fixed to $M_{\rm cut}^{\rm fin}=2.44M_\odot$
and $f=0.003$, respectively. These ratios are compared with the
observation of CS29498-043, that are shown by the region
between two blue dashed-lines ([C/Mg]) and red dashed-lines ([Mg/Si]).
The range of $M_{\rm cut}^{\rm fin}$ in which both the observational
points are satisfied is $M_{\rm cut}^{\rm fin} \simeq 10.2 - 13.6
M_\odot$ and indicated as the ``Allowed Region''.
\label{cmgsi}}
\end{figure}

\clearpage

\begin{figure}
\epsscale{.6}
\plotone{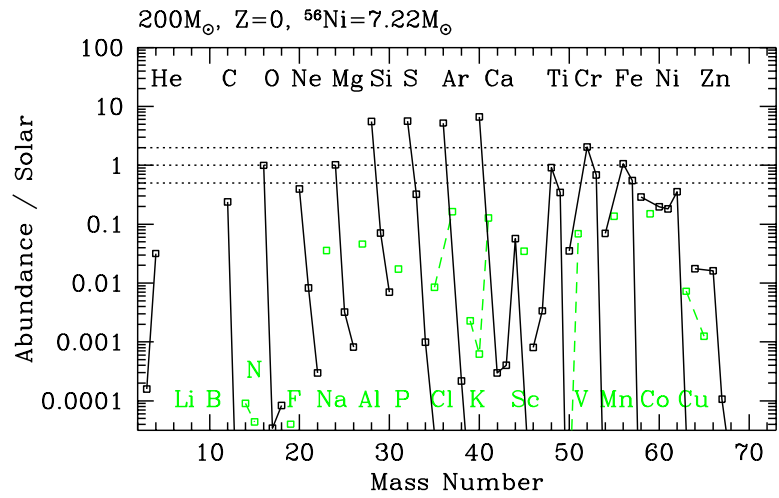}

\caption{Yields of a pair-instability
supernova from the 200 $M_\odot$ star (UN02).
}
\end{figure}

\begin{figure}
\epsscale{1.}

\epsscale{.8}

\plottwo{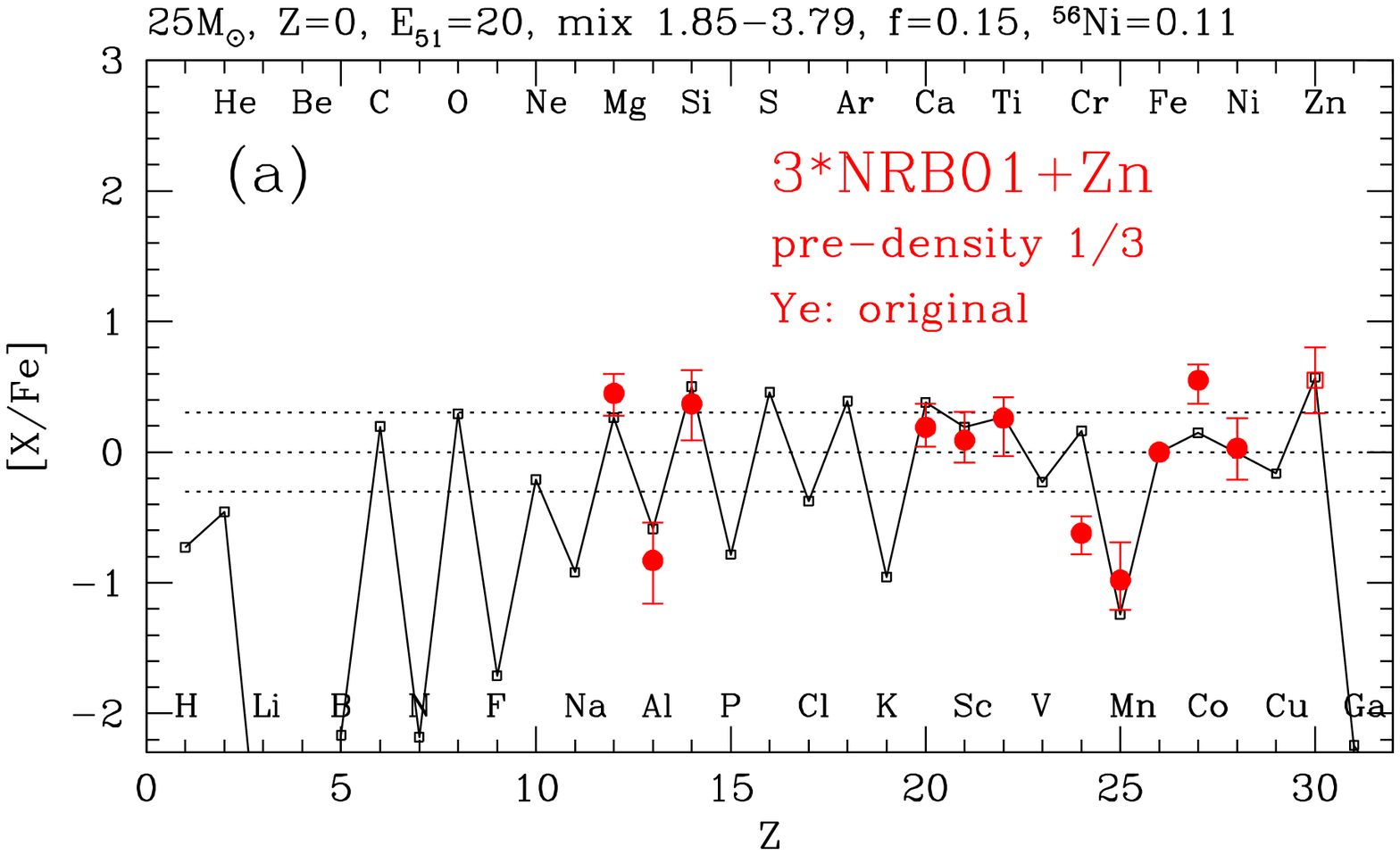}{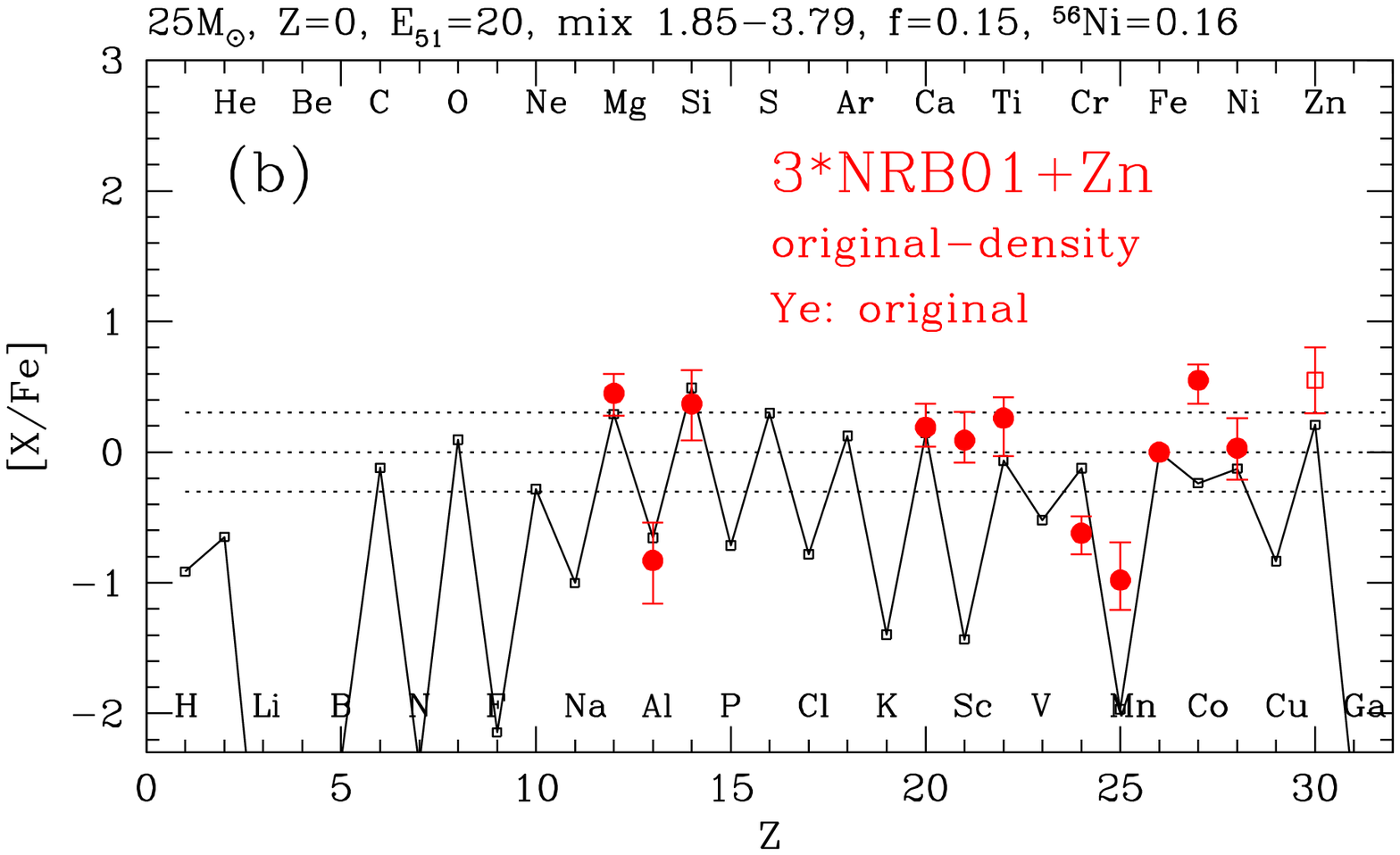}

\plottwo{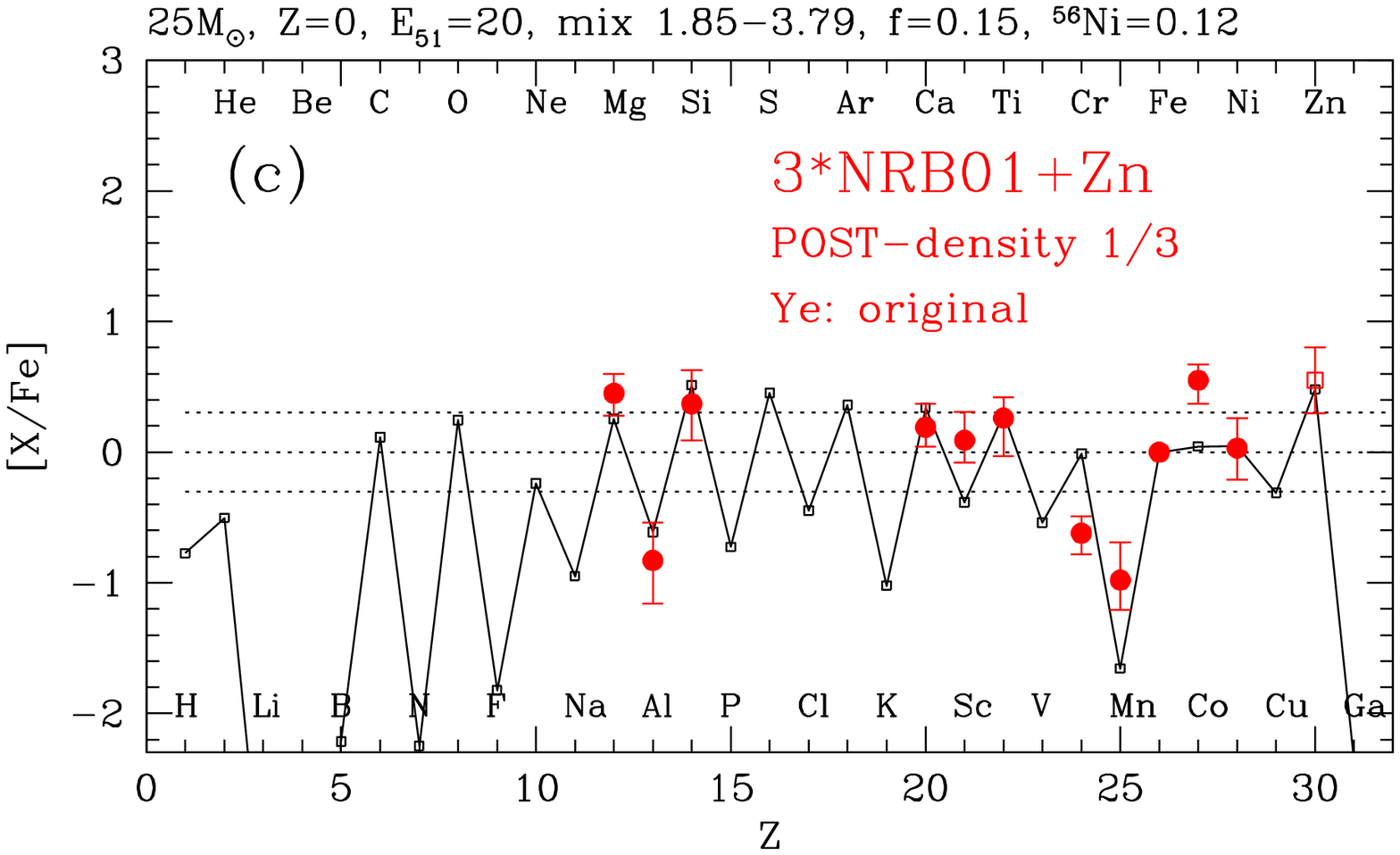}{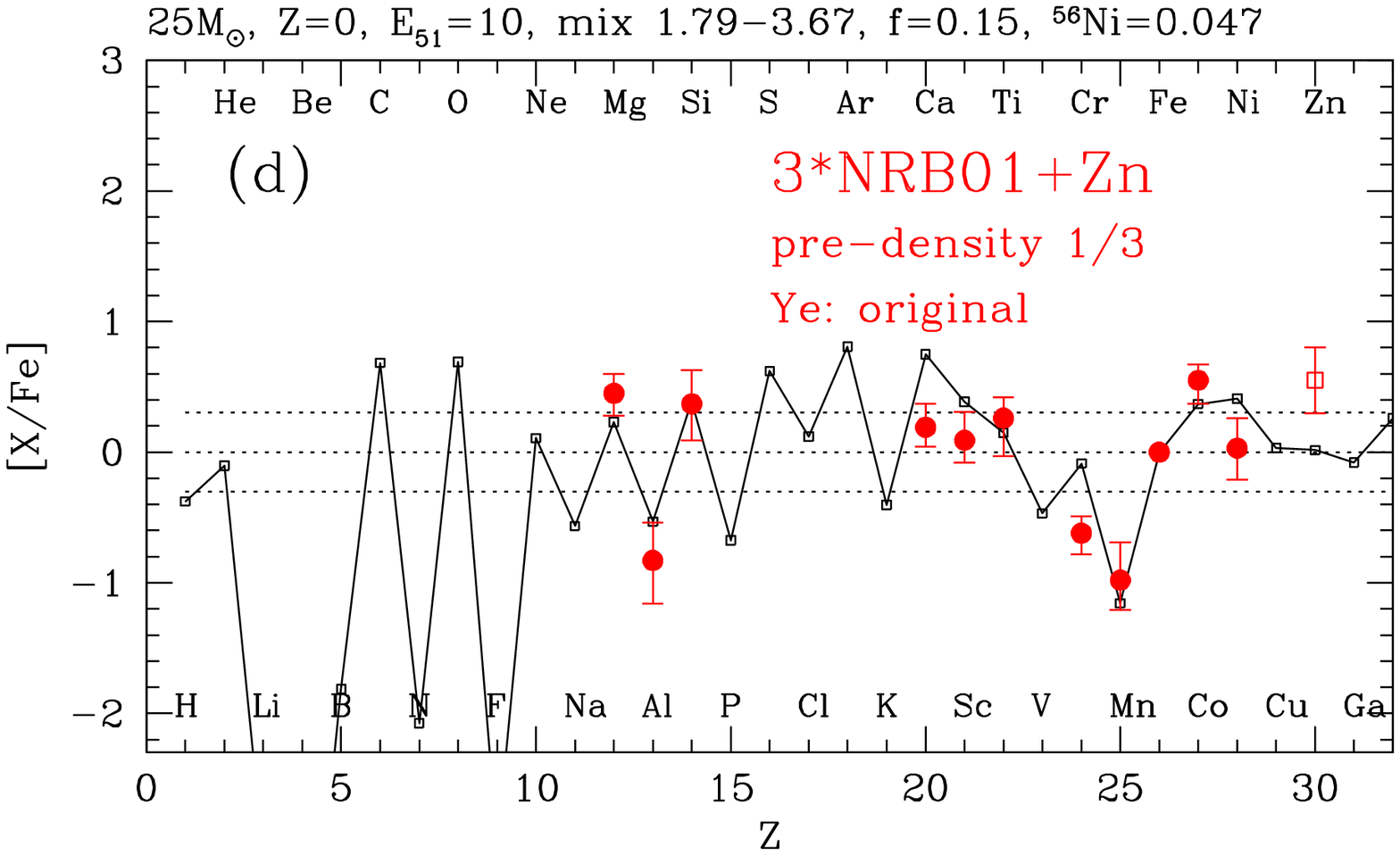}

\caption{Elemental abundances of typical EMP stars at [Fe/H] $\sim$ -3.7
given by NRB01 (solid circles with error bars) 
compared with yield (solid lines) of low-density models.
In the panel (a) the density of the pre-supernova progenitor 
is reduced to 1/3 with keeping total stellar mass. $Y_e$ of this model
The panel (b) shows a model with the original density for comparison.
In (c), the density for the post-process calculations are reduced to 1/3
from the original value obtained by the hydrodynamical calculations.
This figure shows that a realistic hydrodynamical calculations
is important especially for Sc/Fe, Ti/Fe and Co/Fe ratios.
The panel (d) shows a low-density model with a lower energy.
As shown in this figure in the low-density models, [Zn/Fe] is enhanced
but still a high-energy explosion is required to fit to the observed data.
\label{LOW1}}
\end{figure}

\begin{figure}
\epsscale{1.}
\plotone{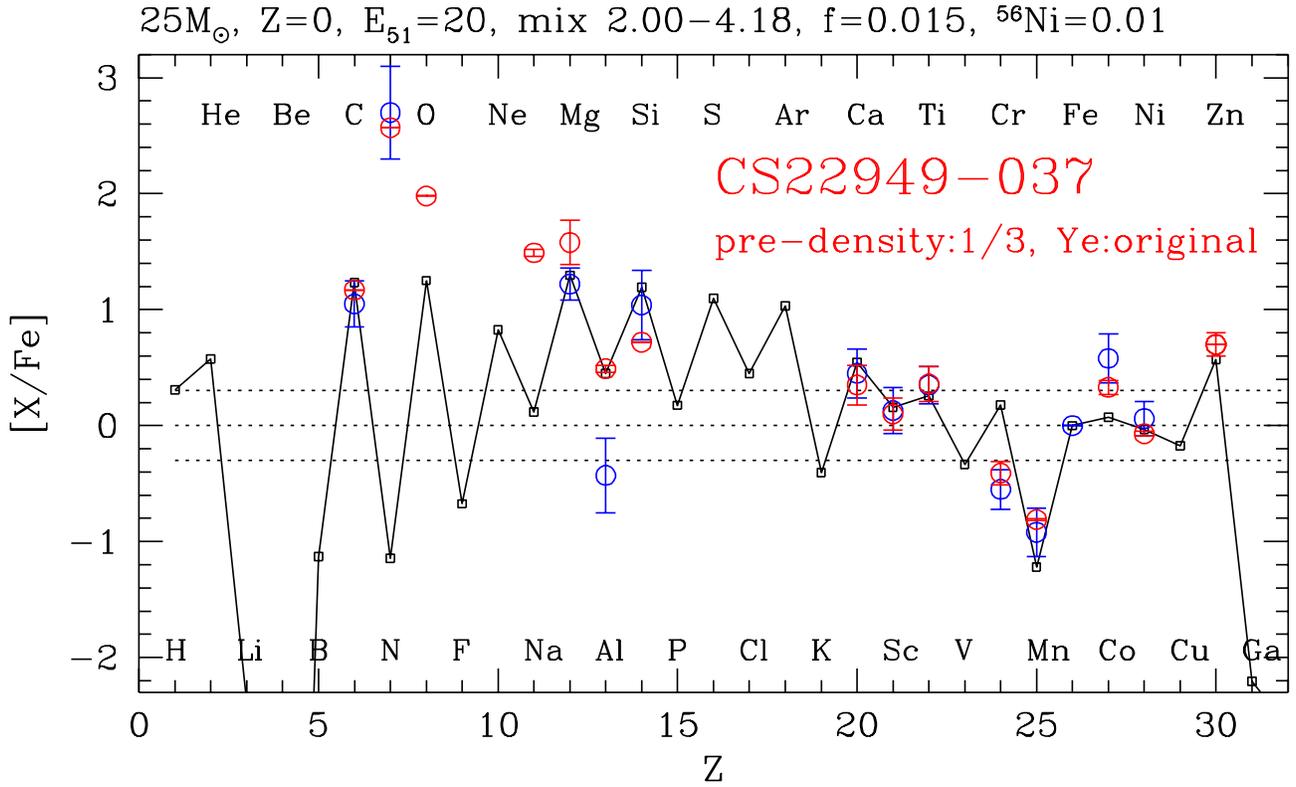}

\caption{Elemental abundances of  CS22949-037
compared with a low-density model. In this model,
[Ca/Fe], [Sc/Fe] and [Ti/Fe] are well reproduced. 
Co/Fe and Zn/Fe ratios are also enhanced from the original
density model, giving a better fit.
\label{LOW2}}
\end{figure}


\clearpage

\begin{deluxetable}{lclc}
\tablecaption{Isotopes included in the network for explosive
 burning \label{tbl-2}}
\tablewidth{0pt}
\tablehead{
\colhead{Isotope } & \colhead{A} &  \colhead{Isotope} &  
\colhead{A} 
}  
\startdata 
n --- & 1 & H --- & 1-3  \cr 
He --- & 3-4 & Li --- & 6-7  \cr 
Be --- & 7-9 & B --- & 8-11  \cr 
C --- & 11-13 & N --- & 13-15  \cr 
O --- & 14-18 & F --- & 17-19  \cr 
Ne --- & 18-22 & Na --- & 21-23  \cr 
Mg --- & 22-27 & Al --- & 25-29  \cr 
Si --- & 26-32 & P --- & 27-34  \cr 
S --- & 30-37 & Cl --- & 32-38  \cr 
Ar--- & 34-43 & K --- & 36-45  \cr 
Ca --- & 38-48 & Sc --- & 40-49  \cr 
Ti --- & 42-51 & V --- & 44-53  \cr 
Cr --- & 46-55 & Mn --- & 48-57  \cr 
Fe --- & 50-61 & Co --- & 51-62  \cr 
Ni --- & 54-66 & Cu --- & 56-68  \cr 
Zn --- & 59-71 & Ga --- & 61-73  \cr 
Ge --- & 63-75 & As --- & 65-76  \cr 
Se --- & 67-78 & Br --- & 69-79  \cr 
\enddata
\end{deluxetable}

\clearpage

\begin{deluxetable}{cccccccc}
\tablecaption{Yield ratios [(Zn, Co, Cr, Mn)/Fe] as a function of $M$ and $E_{51}$
\label{models1}}
\tablewidth{0pt}
\tablehead{ \colhead{model} &
\colhead{(M, E$_{51}$)} & \colhead{[Zn/Fe]} &  \colhead{[Co/Fe]} &  
\colhead{[Cr/Fe]} &  \colhead{[Mn/Fe]} & \colhead{Mg} & 
\colhead{log$_{10}$(Mg/E$_{51}$) }}  
\startdata 
A & (13, 1) & -0.40 & -0.72  & 0.04  & -1.19 & 0.023& -1.63  \cr 
B & (15, 1) & -0.89 & -1.02  & 0.17  & -0.77 & 0.030& -1.52  \cr 
C & (20, 1) & -1.66 & -0.62  & -0.22  & -0.89 & 0.020& -1.70  \cr 
D & (25, 1) & -0.56 & -1.21  & 0.11  & -0.74  & 0.14& -0.85  \cr 
E & (25, 10) & 0.32 & -0.44  & -0.08  & -2.05 & 0.13& -1.89  \cr 
F & (25, 20) & 0.42 & -0.08  & -0.13  & -2.03 & 0.15& -2.12  \cr 
G & (25, 30) & 0.48 & 0.11  & -0.05  & -1.82 & 0.12& -2.40  \cr 
H & (30, 1) & -0.21 & -0.89  & -0.06  & -0.89 & 0.40& -0.40  \cr 
I & (30, 10) & 0.05 & -1.17  & -0.02  & -1.15 & 0.41& -1.39  \cr 
J & (30, 20) & 0.19 & -0.66  & -0.10  & -1.48 & 0.36& -1.74  \cr 
K & (30, 30) & 0.23 & -0.42  & -0.08  & -1.85 & 0.31& -1.99  \cr 
L & (30, 50) & 0.34 & -0.17  & -0.16  & -2.28 & 0.26& -2.28  \cr 
M & (50, 1) & -0.13 & -1.89  & 0.26  & -1.72 & 0.75& -0.12  \cr 
N & (50, 10) & -0.11 & -0.92  & 0.12  & -1.36 & 0.73& -1.14  \cr 
O & (50, 30) & 0.17 & -0.46  & -0.01  & -1.54 & 0.79& -1.58  \cr 
P & (50, 50) & 0.26 & -0.24  & -0.08  & -1.52 & 0.80& -1.80  \cr 
Q & (50, 70) & 0.32 & -0.14  & -0.09  & -1.40 & 0.80& -1.94  \cr 
R & (50, 100) & 0.39 & -0.07  & -0.05  & -1.33 & 0.78& -2.11  \cr 
\hline

\enddata
\tablecomments{These are $Z=0$ models.
The ejected Mg mass in $M_\odot$ and log$_{10}$(Mg/E$_{51})$ are 
also shown. Note that in these models, 
the $Y_e$ during the explosion is unmodified.
This is because [Mn/Fe] and [Co/Fe] are systematically underproduced 
compared with the observations. We note in the mixing-fallback model
all the quantities shown here are independent of the ejected Fe mass,
or the ejection factor, $f$,
as far as the outer boundary of the mixing region, $M_{\rm mix}$(out),
is fixed to be at the outer-boundary of the incomplete Si-burning region.
}
\end{deluxetable}

\begin{deluxetable}{ccccc}
\tablecaption{$Y_e$ at $M_{\rm cut}$(ini) and 
$M_{\rm Si-burn}$ for the selected models in Table 2.
}
\tablewidth{0pt}
\tablehead{ \colhead{model} &
\colhead{ $M_{\rm cut}$(ini)} & \colhead{$M_{\rm Si-burn}$} & 
\colhead{$Y_e$($M_{\rm cut}$(ini))} &  
\colhead{$Y_e$($M_{\rm Si-burn}$)}   
}  
\startdata 
A & 1.53 & 1.61  & 0.4996  & 0.4996  \cr 
B & 1.62 & 1.90  & 0.4995  & 0.4998  \cr 
D & 2.20 & 2.91  & 0.4998  & 0.5000  \cr 
F & 2.32 & 4.22  & 0.4998  & 0.4998  \cr 
G & 2.50 & 4.52  & 0.4998  & 0.4998  \cr 
M & 2.80 & 3.99  & 0.4999  & 0.5000  \cr 
N & 2.97 & 5.06  & 0.5000  & 0.5000  \cr 
O & 2.80 & 7.90  & 0.4999  & 0.4999  \cr 
P & 3.06 & 8.86  & 0.5000  & 0.4999  \cr 
\hline

\enddata
\tablecomments{$M_{\rm cut}$(ini) is the mass-cut (or the initial
mass-cut in the mixing-fallback model) and $M_{\rm Si-burn}$
is the mass-coordinate of the outer-boundary of explosive incomplete
Si-burning region which is defined by X($^{56}$Ni) $=10^{-3}$.
}
\end{deluxetable}

\clearpage

\begin{deluxetable}{cccccc}
\tablecaption{Mass-coordinate, $M_{\rm r}$, and some
other quantities where peak-temperatures during explosion,
$T_{\rm peak}$, reach to the specific temperatures 
for the selected models in Table 2.
}
\tablewidth{0pt}
\tablehead{ \colhead{model} &
\colhead{$T_{\rm peak,9}$} & \colhead{$M_{\rm r}$} & 
\colhead{$\rho_{\rm preSN,6}$} &  
\colhead{$\rho_{\rm peak,6}$} &  
\colhead{$Y_e$}   
}  
\startdata 
A & 7    & 1.50  & 2.9     & 17.4    & 0.4995  \cr 
  & 5    & 1.56  & 1.3     & 4.6     & 0.4999  \cr 
  & 4    & 1.63  & 0.46    & 2.1     & 0.5000  \cr 
  & 3.2  & 1.68  & 0.24    & 1.2     & 0.5000  \cr 
\hline
D & 7    & 2.05  & 4.5     & 27.6    & 0.4996  \cr 
  & 5    & 2.49  & 1.5     & 4.3     & 0.4999  \cr 
  & 4    & 2.81  & 0.72    & 1.8     & 0.5000  \cr 
  & 3.2  & 3.13  & 0.45    & 1.2     & 0.5000  \cr 
\hline
F & 7    & 2.76  & 0.78    & 5.9     & 0.5000  \cr 
  & 5    & 3.49  & 0.28    & 1.4     & 0.5000  \cr 
  & 4    & 4.07  & 0.14    & 0.66    & 0.5000  \cr 
  & 3.2  & 4.71  & 0.065   & 0.39    & 0.5000  \cr 
\hline
G & 7    & 2.96  & 0.58    & 3.5     & 0.5000  \cr 
  & 5    & 3.75  & 0.28    & 1.1     & 0.5000  \cr 
  & 4    & 4.41  & 0.096   & 0.48    & 0.5000  \cr 
  & 3.2  & 5.01  & 0.045   & 0.26    & 0.5000  \cr 
\hline

\enddata
\tablecomments{$T_{\rm peak,9}$ is the peak-temperature
during explosion at $M_r$ in the units of $10^9$ (K).
$\rho_{\rm preSN,6}$ and $\rho_{\rm peak,6}$ are the
pre-supernova and peak-density at $M_r$ in the units of 
$10^6$ (g cm$^{-3}$). $Y_e$ is the pre-supernova value
of $Y_e$ at $M_r$.
}
\end{deluxetable}

\begin{deluxetable}{cccccccc}
\tablecaption{Some related data for the models in Figure 5.
\label{models2}}
\tablewidth{0pt}
\tablehead{ 
\colhead{(M, E$_{51}$)} & \colhead{[Zn/Fe]} &  \colhead{[Co/Fe]} &  
\colhead{[Cr/Fe]} &  \colhead{[Mn/Fe]} & \colhead{Mg} & 
\colhead{log$_{10}$(Mg/E$_{51}$) }}  
\startdata 
 (15, 1) & -0.24 & 0.01  & 0.17  & -0.70 & 0.030& -1.52  \cr 
 (25, 30) & 0.29 & 0.35  & -0.12  & -0.92 & 0.18& -2.22  \cr 
\enddata
\tablecomments{The numbers shown here are different from those in Table 2,
because in these models the $Y_e$ during the explosion is modified:
$Y_e = 0.5001$ in the complete Si-burning region and
$Y_e = 0.4997$ in the incomplete Si-burning region.
In Figure 5, [Fe/H] is determined by [Fe/H]=  log$_{10}$(Mg/$E_{51}$) + 
$C$ with $C= -1.0$ due to the reason described in the text. Note that
all these numbers are roughly independent of mixing-fallback parameters
as far as  $M_{\rm mix}$(out) is not too large.}
\end{deluxetable}

\clearpage
\begin{deluxetable}{cccc}
\tablecaption{The relation between the C/O ratio 
after the He-burning and [Co/Fe] in the ejecta of our models
}
\tablewidth{0pt}
\tablehead{ 
\colhead{Model} & \colhead{C/O} &  
\colhead{[Co/Fe]$'$(for $^{56}$Ni=0.07$M_\odot$)} &  
\colhead{[Co/Fe]$'$(for Zn/Fe max)}   
}  
\startdata 
 13A & .29/.70 & -0.72  & -0.72  \cr 
 15A & .27/.75 & -3.57  & -1.02  \cr 
 20A & .22/.78 & -3.53  & -0.62  \cr 
 25A & .25/.74 & -3.37  & -1.21  \cr 
\hline
 13B & .35/.65 & -0.29  & -0.23  \cr 
 13C & .46/.52 & -2.80  & -0.80  \cr 
 15B & .33/.67 & -3.29  & -1.37  \cr 
 15C & .41/.56 & -2.50  & -0.81  \cr 
 20C & .37/.60 & -1.11  & -0.45  \cr 
 20D & .38/.62 & -0.42  & -0.33  \cr 
 25C & .29/.68 & -1.07  & -0.87  \cr 
\enddata
\tablecomments{
The number in the model name is the initial mass of the
progenitor. The models 'A' are the ones used in UN02
and have relatively low central C/O mass fraction ratios
just after the helium burning. These models adopt the
$^{12}$C($\alpha, \gamma$)$^{16}$O rate, 1.4 times the
value of Caughlan \& Fowler 1988 (CF88). The models B, C, D, 
which have relatively larger C/O values than the models A,
are calucated with the $^{12}$C($\alpha, \gamma$)$^{16}$O rate
1.3 times the value of CF88 and also 
assumes faster convective mixing than 
models A (detail of these models are presented elsewhere). 
The models B, C and D have initial metallicity 
Z=0, 0.02 and 0.001, respectively. [Co/Fe]$'$ are the
usual [Co/Fe] for Z=0 models, but for other metallicity
models they are integrated only for the matter in the 
Si-burning regions. Then we could compare the Co/Fe ratios
of the explosively produced matter even though they have
different initial metallicities. We show the [Co/Fe] values
for two different mass-cuts. One is determined by the ejected mass of
$^{56}$Ni=$0.07M_\odot$ and the other makes the
Zn/Fe ratio maximum. The latter mass-cut is usually smaller,
giving a larger [Co/Fe]. These models, both ``small''
and ``larger'' C/O models, significantly underproduce the
large [Co/Fe] values observed in typical EMP stars. 
}
\end{deluxetable}

\clearpage

\begin{deluxetable}{ccccccccc}
\tablecaption{Progenitor models 
\label{tbl-1}}
\tablewidth{0pt}
\tablehead{
\colhead{(M, Z)} & \colhead{CF88} &  \colhead{C/O} &  
\colhead{Fe} &  \colhead{O} &  \colhead{CO} & \colhead{He}
& \colhead{$Y_e$(Si)}  & \colhead{$Y_e$(O)}}  
\startdata 
(25, 0) & 1.4 & .25/.74          & 1.70  & 2.2 & 5.7 &7.8  &0.4997 & 0.4998\cr 
(30, 0) & 1.4 & .19/.78          & 1.78  & 2.6 & 9.3 &10.7 &0.4998 
& 0.4998 \cr 
(30, $10^{-4}$) & 1.0 & .29/.70  & 1.86  & 2.9 & 11.4 &13.0&0.4998 
& 0.4999  \cr 
(50, $10^{-4}$) & 1.0 & .16/.79  & 2.21  & 3.6 & 19.3 &21.8&0.4998 
& 0.5000  \cr 
\enddata
\tablecomments{Some data on the progenitor models used for the comparison with
individual SN. The numbers shown are the the initial stellar mass, metallicity, 
the adopted $^{12}$C($\alpha, \gamma$)$^{16}$O rate, the central C/O mass fraction
ratio just after the helium burning, Fe-core mass (defined by $Y_e < 0.49$),
O-burning shell (defined by X(O) $ \simeq 0.1$),
C-O core mass (defined by X(He)$ < 10^{-3}$), He core mass
(defined by X(H)$ < 10^{-3}$), $Y_e$ at the O-burning shell, and typical 
$Y_e$ at the convective Si-burning layer, respectively. Here, for the 
$^{12}$C($\alpha, \gamma$)$^{16}$O rate, we multiply a constant number 
shown in the table to the value given in CF88. }
\end{deluxetable}


\begin{thebibliography}{}

\bibitem{}
Aoki, W., Ryan, S. G., Beers, T. C. \& Ando, H. 2002a, ApJ, 567, 1166

\bibitem{}
Aoki, W., Norris, J. E., Ryan, S. G., Beers, T. C. \& Ando, H. 2002b,
ApJ, 576, L141

\bibitem{}
Aoki, W., Norris, J. E., Ryan, S. G., Beers, T. C. \& Ando, H. 2002c, 
Publ. Astron. Soc. Japan, 54, 933

\bibitem{}
Arnett, W.D., Bahcall, J.N., Kirshner, R.P., \& Woosley, S.E. 1989,
ARAA, 27, 629

\bibitem{}
Audouze, J. \& Silk, J. 1995, ApJ, 451, L49

\bibitem{}
Bessell, M.S., Christlieb, N., \& Gustafsson, B. 2004, ApJ, 612, L61

\bibitem{}
Bonifacio et al. 2003, Nature, 422, 834

\bibitem{}
Caughlan, G., \& Fowler, W. 1988, Atomic Data Nucl. Data Tables, 40,
283 (CF88)


\bibitem{}
Caughlan, G., Fowler, W., Harris, M. J., \& Zimmermann, B. A. 1985,
Atomic Data Nucl. Data Tables, 32, 197

\bibitem{}
Chieffi, A., \& Limongi, M. 2002, ApJ, 577, 281 (CL02)

\bibitem{}
Chevalier, R. 1989, ApJ 346, 847

\bibitem{}
Christlieb, N., et al. 2002, Nature, 419, 904

\bibitem{}
Christlieb, N., Gustafsson, B., Korn, A.J., Barklem, P.S., Beers, T.C.,
 Bessell, M.S., Karlsson, T., \& Mizuno-Wiedner, M. 2004, ApJ, 603, 708

\bibitem{}
Colgate, S.A. 1971, ApJ 163, 221

\bibitem{}
Depagne, E., et al. 2002, A\&A, 390, 187

\bibitem{}
Ebisuzaki, T., Shigeyama, T., \& Nomoto, K. 1989, ApJ, 344, L65

\bibitem{}
Hachisu, I., Matsuda, T., Nomoto, K., \& Shigeyama, T. 1990,
ApJ, 358, L57

\bibitem{}
Heger, A., \& Woosley, E. 2002, ApJ, 567

\bibitem{}
Herant, M., \& Woosley, S.E. 1994, ApJ 425, 814

\bibitem{}
Hoffman, R. D., Woosley, E., Fuller, G. M.,
\& Meyer, B. S. 1996, ApJ, 460, 478

\bibitem{}
Janka, H.-Th., Buras, R., \& Rampp, M. 2003, Nucl. Phys. A, 718, 269

\bibitem{}
Kifonidis, K., Plewa, T., Janka, H.-Th. \& M\"uller, E. 2000,
ApJ, 531, L123

\bibitem{}
Liebend\"orfer, M., Mezzacappa, A., Messer, O. E. B., 
Martinez-Pinedo, G., Hix, W. R., \& Thielemann, F.-K. 2003, 
Nucl. Phys. A., 719, 144

\bibitem{}
Limongi, M. \& Chieffi, A. 2003, ApJ, 592, 404 

\bibitem{}
Limongi, M. et al. 2003, ApJ, 594, 123

\bibitem{}
Maeda, K., \& Nomoto, K. 2003a, ApJ, 598, 1163

\bibitem{}
Maeda, K., \& Nomoto, K. 2003b, Prog. Theor. Phys. Suppl.,  
151, 211

\bibitem{}
McWilliam, A., Preston, G. W., Sneden, C., \& Searle, L. 1995, AJ,
109, 2757

\bibitem{}
Nagataki, S. 2000, ApJS, 127, 141

\bibitem{}
Nakamura, T., Umeda, H., Nomoto, K., Thielemann, F.-K.,
\& Burrows, A. 1999, ApJ, 517, 193

\bibitem{}
Nakamura, T., Mazzali, P., Nomoto, K., \& Iwamoto, K.
2001a, ApJ, 550, 991

\bibitem{}
Nakamura, T., Umeda, H., Iwamoto, K., Nomoto, K., Hashimoto, M., Hix,
W. R., \& Thielemann, F.-K. 2001b, ApJ, 555, 880

\bibitem{}
Nomoto, K., et al. 2002, in IAU Symposium 212, A Massive Star Odyssey:
from Main Sequence to Supernova, ed. K A. van der Hucht, A. Herrero,
\& C. Esteban, 395 (astro-ph/0209064)

\bibitem{}
Nomoto, K., Maeda, K., Umeda, H., Tominaga, N., Ohkubo, T., Deng, J.,
\& Mazzali, P. A. 2003, in Carnegie Observatories Astrophysics Series,
Vol. 4: Origin and Evolution of the Elements, ed. A.McWilliam \&
M.Rauch (Pasadena: Carnegie Observatories)
(http://www.ociw.edu/ociw/symposia/series/symposium4/
proceedings.html)

\bibitem{}
Nomoto, K., Maeda, K.,
Mazzali, P. A., Umeda, H., Deng, J., \& Iwamoto, K. 2004, in
Stellar Collapse, ed. C. L. Fryer (Dordrecht: Kluwer), 277
(astro-ph/0308136)

\bibitem{}
Norris, J. E., Ryan, S. G., \& Beers, T. C. 2001, ApJ, 561, 1034
(NRB01)

\bibitem{}
Primas, F., Brugamyer, E., Sneden, C., King, J. R., Beers, T. C.,
Boesgaard, A. M., \& Deliyannis, C. P.  2000, in The First Stars,
ed. A. Weiss, T. Abel, \& V. Hill (Berlin: Springer), 51

\bibitem{}
Ryan, S. G. 2002, in CNO in the Universe, ed.  C.Charbonnel,
D.Schaerer, \& G.Meynet (in press) (astro-ph/0211608)

\bibitem{}
Ryan, S. G., Norris, J. E., \& Beers, T. C. 1996, ApJ, 471, 254

\bibitem{}
Schneider, R., Ferrara, A., Natarajan, P. \& Omukai, K. 2002,
ApJ, 571, 30

\bibitem{}
Shigeyama, T. \& Tsujimoto, T. 1998, ApJ, 507, L135

\bibitem{}
Shigeyama, T. et al. 2003, ApJ, 586, 57

\bibitem{}
Spite, M., Depagne, E., Cayrel, R., Hill, V., Francois, P., Spite, F.,
Nordstr\"om, B. et al. 2003, in Carnegie Observatories Astrophysics
Series, Vol. 4: Origin and Evolution of the Elements, ed. A.McWilliam
\& M.Rauch (Pasadena: Carnegie Observatories)
(http://www.ociw.edu/ociw/symposia/series/symposium4/
proceedings.html)

\bibitem{}
Thielemann, F.-K., Nomoto, K., \& Hashimoto, M. 1996, ApJ, 460, 408

\bibitem{}
Tsujimoto, T., \& Shigeyama, T. 2003, ApJ, 584, L87 

\bibitem{}
Turatto, M. et al. 1998, ApJ, 498, L122

\bibitem{}
Umeda, H., \& Nomoto, K. 2002a, ApJ, 565, 385 (UN02)

\bibitem{}
Umeda, H., \& Nomoto, K. 2002b, in proc. of the 11th Workshop on
Nuclear Astrophysics, ed. W.Hillebrandt \& E.M\"uller 
(Max-Planck-Institut f\"r Astrophysik), 164 (astro-ph/0205365) (UN02b)

\bibitem{}
Umeda, H., Nomoto, K., Tsuru, T.G., \& Matsumoto, H. 2002, ApJ, 578,
855

\bibitem{}
Umeda, H., \& Nomoto, K. 2003, Nature, 422, 871 (UN03)

\bibitem{}
Umeda, H., Nomoto, K., \& Nakamura, T. 2000, in The First Stars,
ed. A. Weiss, T. Abel, \& V. Hill (Berlin: Springer), 150
(astro-ph/9912248)

\bibitem{}
Woosley, S.E., \& Weaver, T.A. 1995, ApJS, 101, 181 (WW95)

\end{thebibliography}
\end{document}